\documentclass[10pt]{article}
\usepackage{geometry}
\usepackage[super]{cite}
\usepackage{amsmath}
\usepackage{mathrsfs}
\usepackage{amsfonts}
\usepackage{amssymb}
\usepackage{graphicx}
\usepackage{multirow}
\usepackage{url}
\usepackage{slashbox}
\usepackage{xcolor}
\usepackage{setspace}
\usepackage[font=footnotesize]{subcaption}
\usepackage[font={footnotesize, stretch=1.25}]{caption}
\usepackage[normalem]{ulem}
\usepackage{soul}
\allowdisplaybreaks[4]
\newcommand*{\US}{U_{k_1}S_{k_2}}

\newcommand*{\UI}{U_{k_1}I_{k_2}}
\newcommand*{\UR}{U_{k_1}R_{k_2}}
\newcommand*{\AS}{A_{k_1}S_{k_2}}
\newcommand*{\AI}{A_{k_1}I_{k_2}}
\newcommand*{\AR}{A_{k_1}R_{k_2}}
\newcommand*{\PS}{P_{k_1}S_{k_2}}
\newcommand*{\PI}{P_{k_1}I_{k_2}}
\newcommand*{\PR}{P_{k_1}R_{k_2}}
\newcommand*{\RS}{R_{k_1}S_{k_2}}
\newcommand*{\RI}{R_{k_1}I_{k_2}}

\newcommand*{\ddt}{\frac{d}{dt}}
\newcommand*{\bphy}{\beta_{\text{phy}}}
\newcommand*{\binfo}{\beta_{\text{info}}}
\newcommand*{\ba}{\beta_{\text{anti}}}
\newcommand*{\bp}{\beta_{\text{pro}}}
\newcommand*{\ga}{\gamma_{\text{anti}}}
\newcommand*{\gp}{\gamma_{\text{pro}}}
\newcommand*{\gphy}{\gamma_{\text{phy}}}
\newcommand*{\ginfo}{\gamma_{\text{info}}}
\newcommand*{\alphap}{\alpha_{\text{pro}}}
\newcommand*{\alphaa}{\alpha_{\text{anti}}}
\newcommand*{\USI}{U_{k_1}S_{k_2}\circ I_l}
\newcommand*{\USS}{U_{k_1}S_{k_2}\circ S_l}
\newcommand*{\USA}{U_{k_1}S_{k_2}\circ A_l}
\newcommand*{\USP}{U_{k_1}S_{k_2}\circ P_l}
\newcommand*{\USU}{U_{k_1}S_{k_2}\circ U_l}
\newcommand*{\UIA}{U_{k_1}I_{k_2}\circ A_l}
\newcommand*{\UIP}{U_{k_1}I_{k_2}\circ P_l}
\newcommand*{\UIU}{U_{k_1}I_{k_2}\circ U_l}
\newcommand*{\URA}{U_{k_1}R_{k_2}\circ A_l}
\newcommand*{\URP}{U_{k_1}R_{k_2}\circ P_l}
\newcommand*{\URU}{U_{k_1}R_{k_2}\circ U_l}
\newcommand*{\ASI}{A_{k_1}S_{k_2}\circ I_l}
\newcommand*{\ASS}{A_{k_1}S_{k_2}\circ S_l}
\newcommand*{\PSI}{P_{k_1}S_{k_2}\circ I_l}
\newcommand*{\PSS}{P_{k_1}S_{k_2}\circ S_l}
\newcommand*{\RSI}{R_{k_1}S_{k_2}\circ I_l}
\newcommand*{\RSS}{R_{k_1}S_{k_2}\circ S_l}
\newcommand*{\bhat}{\hat{\beta}}
\newcommand*{\pinfo}{\mathbb{P}_{\text{info}}}
\newcommand*{\pphy}{\mathbb{P}_{\text{phy}}}
\newcommand*{\Rphy}{R_{\text{phy}}}
\newcommand*{\Rinfo}{R_{\text{info}}}
\newcommand*{\Cm}{C_{k_1\,k_2}}

\title{\fontsize{10}{13}\bfseries\selectfont{A Multilayer Network Model of the Coevolution of the Spread of a Disease and
Competing Opinions}}

\author{\footnotesize{
	\parbox{0.9\linewidth}{\centering{
		Kaiyan Peng\footnote{Department of Mathematics, University of California, Los Angeles}, 
		Zheng Lu\footnote{Department of Mathematics, University of Wisconsin, Madison} \footnote{These authors contributed equally.}, 
		Vanessa Lin\footnote{University of North Carolina, Chapel Hill} \footnotemark[3], 
		Michael R. Lindstrom\footnotemark[1], 
		Christian Parkinson\footnote{Department of Mathematics, University of Arizona},  
		Chuntian Wang\footnote{Department of Mathematics, University of Alabama},  
		Andrea L. Bertozzi\footnotemark[1], 
		Mason A. Porter\footnotemark[1]}}}}

\date{}

\begin{document}

\maketitle

\begin{abstract}

During the COVID-19 pandemic, conflicting opinions on physical distancing swept across social media, affecting both human behavior and the spread of COVID-19. Inspired by such phenomena, we construct a two-layer multiplex network for the coupled spread of a disease and conflicting opinions. We model each process as a contagion. On one layer, we consider the concurrent evolution of two opinions --- pro-physical-distancing and anti-physical-distancing --- that compete with each other and have mutual immunity to each other. The disease evolves on the other layer, and individuals are less likely (respectively, more likely) to become infected when they adopt the pro-physical-distancing (respectively, anti-physical-distancing) opinion. We develop approximations of mean-field type by generalizing monolayer pair approximations to multilayer networks; these approximations agree well with Monte Carlo simulations for a broad range of parameters and several network structures. Through numerical simulations, we illustrate the influence of opinion dynamics on the spread of the disease from complex interactions both between the two conflicting opinions and between the opinions and the disease. We find that lengthening the duration that individuals hold an opinion may help suppress disease transmission, and we demonstrate that increasing the cross-layer correlations or intra-layer correlations of node degrees may lead to fewer individuals becoming infected with the disease.

\vspace{0.4cm}

\vspace{0.4cm}
\end{abstract}

\newpage
%
%
\section{Introduction}\label{intro}

Since the outbreak of coronavirus disease 2019 (COVID-19), researchers in numerous disciplines have used diverse approaches to analyze the spread of the disease, forecast its subsequent spread under many scenarios, and investigated strategies to mitigate it\cite{estrada2020covid, arino2021describing,bellomo2020multiscale}. As cases of COVID-19 escalated, collective compliance with non-pharmaceutical intervention (NPI) measures were vital for dealing with the COVID-19 pandemic in the absence of effective treatments, vaccines, and pharmacological interventions\cite{perra2021non}. As information --- some of which was accurate, and some of which was not --- flooded social media\cite{gallotti2020assessing, yang2020covid}, people adopted different opinions about the implementation of NPI measures\cite{Agusto2021.01.29.21250655}. These opinions affect human behavior and ultimately also the spread of diseases. Motivated by these observations, we build a multilayer network model to study disease spreading under the influence of the spread of competing information.

It is important to understand the influence of human behavior on the spread of diseases because these two processes are inextricably coupled\cite{verelst2016behavioural, Bedson2021}. The acquisition of prevalence-based and/or belief-based information from publicly available sources and/or individuals' social neighborhoods leads to changes of disease states, disease-state transition rates, and social contact patterns. 

A common approach to studying the spread of a disease is as a dynamical system on a network of individuals\cite{8955870}. This type of model emphasizes the importance of social contact patterns (typically in the form of physical contacts) and the heterogeneity of individuals. When disease and information spread through different venues --- such as face-to-face contacts versus online social platforms like Twitter and Facebook --- it is useful to use the formalism of multilayer networks\cite{kivela2014multilayer}, as one can encode different relations between people in a population in different layers of such a network. There have been many studies of spreading phenomena on multilayer networks and of how such network structures affect spreading processes\cite{de2016physics, wang2015coupled}. We discuss some of these in Section \ref{sec: model_background}. 

Past research has examined whether the spread of information can help contain an epidemic (e.g., through decreased transmission rate, fewer contacts, and/or acquired immunity) by leading to a smaller disease prevalence and/or a smaller basic reproduction number (and hence a reduced probability of a large outbreak of a disease)\cite{funk2010modelling, wang2015coupled}. However, as has been striking during the COVID-19 pandemic\cite{zarocostas2020fight}, how people act on information (and misinformation and disinformation) can also have a negative impact on disease propagation undermining the potential benefits of information. For example, there have been many anti-physical-distancing rallies in which protesters flout behavioral intervention measures such as wearing masks and practicing physical distancing. Moreover, such large gatherings can themselves cause surges in infections\cite{gabbatt_2020}. 

Motivated by the mixed effects of information and opinion spreading interacting with the spread of a disease, we study a model in which disease transmission is influenced by two opposing opinions: pro-physical-distancing (which we sometimes write simply as ``pro'' as a shorthand) and anti-physical-distancing (which we sometimes write as ``anti''). Following Ref.~\citen{granell2014competing}, we consider a two-layer multiplex network (a particular type of multilayer network\cite{kivela2014multilayer}) with interactions between the spread of opinions and a disease. We model the simultaneous evolution of two competing opinions on one layer of a multiplex network as a contagion process of either susceptible--infectious--recovered (SIR)\cite{kermack1927contribution} or susceptible--infectious--recovered--susceptible (SIRS) form, with opinion adoption that occurs between susceptible and infectious individuals. Similar models have been proposed for studying competing diseases\cite{karrer2011competing, miller2013cocirculation} and ideas\cite{wang2012dynamics}. The spread of the disease occurs on the other layer of the multiplex network, whose connectivity encodes in-person social contacts and whose connectivity strength depends on the opinions of the individuals in the first layer.

In our model, we investigate which opinion has greater influence, which we evaluate based on the disease's final epidemic size (i.e., the number of individuals who catch the disease during the disease outbreak).  We generate networks using configuration-model networks\cite{fosdick2018} and their extensions\cite{melnik2014dynamics}; we demonstrate complex interactions between the two opinions and (because of ensuing behavioral changes, which lead to changes in the network of in-person social contacts) between the opinions and the disease. We explore how the influence of opinions is affected by various factors, including opinion-contagion parameters and network structures. 

We derive a mean-field description, in the form of a pair approximation, of our multilayer dynamical system for the expected values of population-scale quantities, because it is costly to conduct direct simulations of the full stochastic model with a large population. Many approximation methods have been developed for dynamics on monolayer networks (i.e., ordinary graphs)\cite{kiss2017mathematics}, including edge-based compartmental modeling\cite{miller2012edge}, pair approximations\cite{keeling1997correlation}, effective-degree approximations\cite{lindquist2011effective}, and approximate master equations\cite{gleeson2011high, gleeson2013binary}. In the present paper, we use a degree-based pair approximation\cite{eames2002modeling} and generalize it to coupled dynamics on multiplex networks. In a pair approximation, one examines the various types of pairs in a system and approximates higher-order structures using moment closure\cite{kuehn2016moment}. Different pair approximations entail different choices when performing moment closure. Our approximation scheme incorporates dynamical correlations both within and across the layers of a multiplex network. To capture the influence of opinions on individuals' susceptibility to the disease, we also define effective transmission rates that take the form of time-dependent functions of the distribution of opinions in populations of interest. We develop approximations of the effective transmission rates based on the numbers of various types of pairs. Numerical simulations reveal that our approximate system is able to capture the influence of the spread of opinions on disease spread for population-scale quantities. We find that the time evolution of the expected numbers of individuals in different states in our pair approximation match very well with simulations on a variety of networks with different degree distributions and degree--degree correlations.

Our paper proceeds as follows. In Section \ref{sec: model}, we discuss prior work and relevant background information and then present our first model. In this model, opinions follow an SIR process. The disease follows an extended SIR process in which people who adopt the pro-physical-distancing opinion (respectively, anti-physical-distancing opinion) have a reduced (respectively, increased) disease transmission rate in comparison to a baseline. In Section \ref{sec: ODE}, we suppose that the SIR process occurs on a fully mixed population, yielding a description of our system in terms of a small set of coupled ordinary differential equations (ODEs). In this ODE system, we observe some influence of opinion dynamics on the spread of the disease, but this framework does not include the effects of social contacts. In Section \ref{sec: pwa}, we incorporate social contact structures between individuals to yield a dynamical system on a network. We derive a pair approximation for this network model. In Section \ref{sec: experiments}, we conduct stochastic simulations of this network model to investigate the influence of the contagion parameters and network structures on the dynamics of the system. In Section \ref{sec: SIRS-SIR}, we generalize the disease dynamics by considering an SIRS process for opinion spreading. 
We conclude in Section \ref{sec: conclusion}. 

%
%

\section{Background and Modeling}\label{sec: model}

We discuss prior work and background information in Section \ref{sec: model_background}, and we present our initial model in Section \ref{sec: model_model}. In this model, we consider a two-layer multiplex network with a disease that spreads on one layer and competing opinions that spread on the other layer. We specify network structures in Section \ref{sec: experiments}.

\subsection{Background}\label{sec: model_background}

A variety of past work has examined the influence of social behavior on disease transmission\cite{funk2010modelling, verelst2016behavioural}. We seek to do this in a way that incorporates the effects of network structures on the spread of disease \cite{kiss2017mathematics,porter2016}. We consider a social network of individuals (which are represented by nodes) and in-person social contacts (which are represented by edges) between them. The importance of such social contact patterns on dynamical processes has long been recognized\cite{durrett1994importance}, and there has been extensive research on the influence of network structures on the spread of infectious diseases\cite{pastor2015epidemic}. There is also a large body of work on the spread of social phenomena\cite{sune-yy2018}, including the dissemination of information\cite{bikhchandani1992theory} and the adoption of behaviors\cite{bass1969new}, which are often modeled as contagion processes that are similar to disease spread\cite{goffman1964generalization, bettencourt2006power}. (See \citen{weng2013, hodas2014} for discussions of when social contagions resemble and do not resemble contagions of infectious diseases.) In particular, interactions between peers strongly influence the dynamics that unfold on a network.

To give further context for our work, we briefly mention prior investigations on the co-evolution of diseases with behavior, awareness, and/or opinions on multilayer networks\cite{verelst2016behavioural,wang2015coupled}. In particular, many researchers have examined how the spread of awareness can suppress the spread of a disease. Funk et al.\cite{funk2009spread} developed a co-evolution model in which individuals acquire different levels of awareness of a disease either by becoming infected or by communicating with their neighbors. In their model, individuals are less susceptible to infection by a disease when they have a stronger awareness. Subsequent work by Funk et al.\cite{funk2010endemic} simplified the above model so that individuals are either aware or unaware of a disease, in analogy to the infectious and susceptible states (i.e., ``compartments'') of a traditional susceptible--infectious--susceptible (SIS) model of disease spread\cite{brauer2019mathematical}. A similar model was proposed by Granell et al.\cite{granell2013dynamical, granell2014competing}. Subsequent research has generalized these ideas by modeling the spread of the disease and information with other dynamics, such as modeling disease spread with an SIR model\cite{wang2014asymmetrically} and modeling information transmission with a threshold model\cite{guo2015two,guo2016epidemic} or a generalized Maki--Thompson rumor model \cite{da2019epidemic}. Other works have examined the influence of global information and mass media\cite{granell2014competing, qian2020modeling}, the relative speed of the dynamics for spreading information and disease\cite{velasquez2020disease, da2019epidemic}, and heterogeneous risk perceptions of disease spread\cite{pan2018impact, ye2020effect}. Researchers have also incorporated time-varying networks when studying the combined spread of disease and information, such as by coupling an activity-driven information layer with a time-independent disease layer\cite{guo2016epidemic} or a time-independent information layer with an adaptive physical layer\cite{peng2020contagion}. Additionally, evolutionary game theory has been used to study decision-making under government-mandated interventions, socioeconomic costs, perceived infection risks, and social influence\cite{ye2020modelling}.

During the COVID-19 pandemic, content and discussions on social media have played a prominent role. On one hand, social media can help rapidly disseminate transparent and accessible information about policy and scientific findings, and it gives crucial ways to advocate guidance such as mask-wearing and physical distancing\cite{lunn2020using}. On the other hand, social media also allows the pervasive spread of misinformation and disinformation, leading to so-called \textit{infodemics}, which refer to epidemics of information\cite{gallotti2020assessing, yang2020covid}. Distortion and inaccurate information can impair people's mental and physical health\cite{rapp2018can}, trigger anxiety and distrust, and ultimately lead to a worse situation for disease spread due to poor compliance with prevention measures. As information floods social media, opposing opinions about physical distancing and other intervention methods develop and propagate rapidly to many people. 

Several recent works have examined competing opinion dynamics. Johnson et al.\cite{johnson2020online} studied the evolution of anti-vaccine and pro-vaccine clusters of people on Facebook. She et al.\cite{she2021network} examined an opinion model with continuous-valued opinions to study the beliefs of different communities about the severity of disease spread with both cooperative and antagonistic opinion spreading. Epstein et al.\cite{epstein2021coupled} used compartmental models to study the fear of infection and the fear of vaccines, and Johnston et al.\cite{johnston2020dynamical} examined the fear of infection and frustration with physical distancing.  

\subsection{Our model}\label{sec: model_model}

We study the spread of a disease and two competing opinions on a two-layer multiplex network with one physical layer (where the disease spreads) and one information layer (where opinions spread). We assume that all individuals are present in both layers, and we ignore demographic processes such as birth, death, and migration. We model each layer as an undirected, unweighted, simple graph; we couple the two layers to each other by connecting nodes that correspond to the same individual. The edges within a layer are called ``intra-layer edges''; they encode in-person contacts in the physical layer and information-exchange channels (especially social-media) in the information layer. An individual can have different neighboring individuals (i.e., adjacent nodes) in the two layers, and the number of neighboring individuals (i.e., the degree) can also be different in the two layers. In Section \ref{sec: experiments}, we give further details about the structure of networks and we explore the influence of inter-layer and intra-layer structure on opinion and disease dynamics. In Section \ref{sec: model}--\ref{sec: experiments}, we use SIR dynamics\cite{kermack1927contribution} for each process and associate nodes in the physical layer with individuals' health states and nodes in the information layer with their opinions about physical distancing. In Section \ref{sec: SIRS-SIR}, we extend our model by modeling the spread of opinions as an SIRS process. In all versions of our model, we treat each process as a continuous-time Markov chain. We detail how the disease-spread and opinion-spread processes operate and interact in Sections \ref{sec: model_info} and \ref{sec: model_phy}. In Table \ref{table: param}, we summarize the key parameters of our model. 

\begin{table}[!h]
\caption{Key parameters in our model of coupled opinion spread and disease spread. In (a), the first column gives the parameters of the opinion dynamics that are related to pro-physical-distancing and the second column gives the parameters that are related to anti-physical-distancing. We use the subscript ``info" when the two opinions share parameters; we indicate these parameters in the third column. In (b), each column indicates the parameters of the disease dynamics when individuals adopt the corresponding opinions.}\label{table: param}
\begin{minipage}{1\textwidth}
\centering
\subcaption{Parameters for dynamics (of opinion adoption) on the information layer}
\begin{tabular}{l|c|c|c}
\hline
                                                                  & Pro   & Anti  & \multicolumn{1}{l}{Shared by pro and anti} \\ \hline
Transmission rate                                                 & $\bp$ & $\ba$ & $\binfo$                   \\ \hline
Recovery rate                                                     & $\gp$ & $\ga$ & $\ginfo$                   \\ \hline
\begin{tabular}[c]{@{}l@{}}Rate of losing\\ immunity\end{tabular} & \multicolumn{3}{c}{$\tau$}                 \\ \hline
\end{tabular}
\end{minipage}
\begin{minipage}{1\textwidth}
\centering
\subcaption{Parameters for dynamics (of disease spread) on the physical layer}
\begin{tabular}{l|c|c|c}
\hline
\backslashbox{Parameter}{Opinion}                  & $U$ or $R_{\text{info}}$ & $A$            & $P$            \\ \hline
Transmission rate & $\bphy$    & $\alphaa\bphy$ & $\alphap\bphy$ \\ \hline
Recovery rate     & \multicolumn{3}{c}{$\gphy$}                  \\ \hline
\end{tabular}
\end{minipage}
\end{table}

\subsubsection{Information layer}\label{sec: model_info}

\begin{figure}[!h]
    \centering
    \includegraphics[scale=0.4]{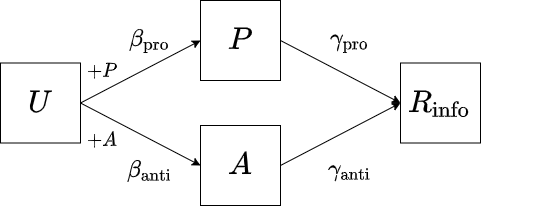}
    \vspace*{8pt}
    \caption{Schematic illustration of the (opinion-spreading) dynamics on the information layer of a two-layer multiplex network. There are four states in the information layer: uninformed ($U$), pro-physical-distancing ($P$), anti-physical-distancing ($A$), and recovered ($R_{\text{info}}$). Nodes in state $U$ transition to state $P$ (respectively, $A$) with a rate of $\bp$ (respectively, $\ba$) by communicating with neighbors in state $P$ (respectively, $A$). We use ``$+P$'' (respectively, ``$+A$'') to emphasize that state transitions occur under the influence of neighbors in $P$ (respectively, $A$). Nodes in state $P$ (respectively, $A$) transition to state $R_{\text{info}}$ at a rate of $\gp$ (respectively, $\ga$).
    }
    \label{fig: A}
\end{figure}

Two competing social contagions, which model pro-physical-distancing and anti-physical-distancing opinions, spread concurrently on the information layer. We use $P$ (respectively, $A$) to denote the pro-physical-distancing (respectively, anti-physical-distancing) state in which individuals both adopt the associated opinion and actively advocate the corresponding behavior. Uninformed ($U$) individuals are susceptible to both opinions and transition to the $P$ state or $A$ state with rates of $\bp$ and $\ba$, respectively, by communicating with neighbors in the corresponding states. We suppose that people who adopt either behavior can become weary of acting unusually in comparison with life without a disease epidemic, and they then become less passionate about maintaining their current conduct. We assume that individuals in either the $P$ state or the $A$ state transition to the recovered ($R_{\text{info}}$) state with rates of $\gp$ or $\ga$, respectively.  After this transition occurs, these individuals practice the same behavior as the uninformed group but are resistant to future influence from neighbors. When the two opinions share the same parameters, we use the subscript ``info". We make the assumption of permanent mutual immunity\cite{karrer2011competing}: once an uninformed node adopts one opinion, it can no longer be influenced by the other opinion. Therefore, upon recovery, it enters the $R_{\text{info}}$ state. Because pro- and anti-physical-distancing opinions are two opposing opinions, it is reasonable to assume that people do not adopt both behaviors simultaneously. We relax the assumption of permanent immunity in Section \ref{sec: SIRS-SIR}. In Figure \ref{fig: A}, we show the compartment flow diagram of the opinion dynamics.

\subsubsection{Physical layer}\label{sec: model_phy}

\begin{figure}[!h]
    \centering
    \vspace{0.1cm}
    \includegraphics[scale=0.4, trim=0 0 0 5, clip]{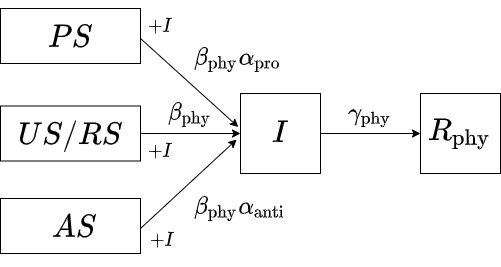}
    \caption{Schematic illustration of the (disease-spreading) dynamics on the physical layer of a two-layer multiplex network. There are three states in the physical layer
    : susceptible ($S$), infectious ($I$), and recovered ($R_{\text{phy}}$). Based on the opinion states of the node, we further divide the $S$ state into $PS$, $AS$, and $US/RS$. Nodes in state $S$ transition to state $I$ through in-person social contacts with infectious neighbors (which we emphasize with ``$+I$''), with rates that we mark close to the corresponding arrow. Nodes in state $I$ recover at a rate of $\gphy$.
    } \label{fig: B}
\end{figure}

We model the spread of a contagious disease on the physical layer as an SIR-like process. The key difference from a standard SIR contagion is that susceptible nodes have transmission rates that depend on their opinion states\cite{funk2009spread, granell2013dynamical}. We divide susceptible nodes into three types: (1) nodes that do not hold any opinion (i.e., their opinions are in the $U$ state or the $R_{\text{info}}$ state) experience the base transmission rate $\bphy$; (2) nodes that hold the pro-physical-distancing opinion experience a reduced transmission rate $\beta_{\text{phy,\,pro}} = \alphap \bphy$, with $\alphap \leq 1$; and (3) nodes that hold the anti-physical-distancing opinion experience an increased transmission rate $\beta_{\text{phy,\,anti}} = \alphaa\bphy$, with $\alphaa \geq 1$. We refer to $\alphap$ and $\alphaa$ as ``influence coefficients''. To model the effects of competing opinions on disease spread, it seems appropriate to study an adaptive network\cite{porter2016} in which structure coevolves with node states. For example, nodes with an anti-physical-distancing opinion may have more contacts than others. However, it is difficult to analyze such a model. Therefore, for simplicity, we assume that nodes with an anti-physical-distancing opinion have a higher risk of contracting the disease than the baseline through a higher transmission rate. We make an analogous assumption for nodes that hold the pro-physical-distancing opinion. Infected individuals (which we take to be the same as infectious individuals) recover at the rate $\gphy$. We show the compartment flow diagram of the disease dynamics in Figure \ref{fig: B}. 

\begin{figure}[!h]
    \centering
    \includegraphics[scale=0.5, trim = 0 50 0 50, clip]{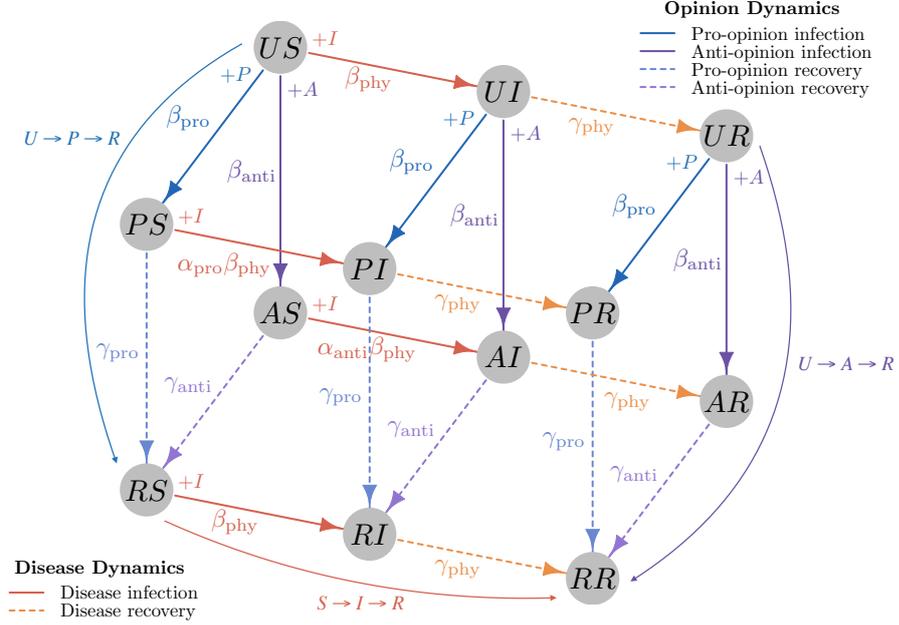}
    \caption{Schematic illustration of our model, with dynamics on both the information layer and the physical layer. The disks indicate the possible states (i.e., compartments) of a node. In each state, the first letter ($U$, $P$, $A$, or $R$) indicates the opinion state and the second letter ($S$, $I$, or $R$) indicates the disease state.  The arrows mark the possible state transitions.}
    \label{fig: C}
\end{figure}

Combining the dynamics on the two layers, we use two letters to describe the full profile of an individual; the first one indicates a node's opinion state, and the second one indicates its disease state. To simplify the notation, we also drop the subscript for the $R$ compartment, as the order of the two letters in a state already indicates whether we are referring to the opinion state or the disease state. There are a total of $12$ possible states (i.e., compartments). We show the complete compartment flow diagram for our model in Figure \ref{fig: C}. For convenience, we use the same notation for the state of a node and the set of nodes in the specified state throughout this paper. 

%
%
\section{Dynamics on a fully-mixed population}\label{sec: ODE}

We first study our model in a fully-mixed population, which yields a small set of coupled ODEs\cite{brauer2019mathematical}. We ignore contact patterns in both the information layer and the physical layer. Additionally, in every small time interval, we assume that each node interacts with other nodes in the same layer uniformly at random. We refer to this assumption as the ``random-mixing assumption''. In every time interval, we also assign each node in one layer to a counterpart node in the other layer uniformly at random (without replacement). We refer to this assumption as the ``random-recoupling assumption''. Throughout this paper, we use $[X] = \mathbb{E}[X]/N$ as the shorthand notation for the expectation of the random variable $X$ divided by the population size $N$. Using the law of mass action, we obtain the following population-level dynamics:
\begin{align}
    \ddt [\vert U\vert] &= -\frac{\bp}{N} [\vert U\vert \times \vert P\vert ] - \frac{\ba}{N} [\vert U\vert \times \vert A\vert]\notag\,,\\
    \ddt [\vert P \vert] &= \frac{\bp}{N} [\vert U\vert \times \vert P\vert]  - \gp [\vert P\vert]\,,\\
    \ddt [\vert A \vert] &= \frac{\ba}{N} [\vert U\vert \times \vert A\vert ] - \ga [\vert A\vert]\notag\,,\\
    \ddt [\vert R_{\text{info}}\vert] &= \gp [\vert P\vert ] + \ga [\vert A\vert]\notag\,,\\
    \ddt [\vert S \vert] &= -\frac{\beta^*}{N} [\vert S\vert \times \vert I\vert ] \notag\,,\\
   \ddt [\vert I\vert] &= \frac{\beta^*}{N} [\vert S\vert \times \vert I\vert ] - \gphy [\vert I\vert]\,,\\
    \ddt [\vert R_{\text{phy}}\vert] &=  \gphy [\vert I\vert] \,,\notag
\intertext{where we use $\vert \cdot \vert$ to denote cardinality and }
  \beta^*& = ([\vert P \vert]\alphap + [\vert A \vert]\alphaa + 1 - [\vert A \vert] - [\vert P \vert])\bphy\notag\,.
\end{align}
The first four equations describe the opinion dynamics. Uninformed individuals may adopt either pro- or anti-physical-distancing opinions by interacting with a node in the corresponding state. If a node adopts an opinion, it also becomes infectious and voices its opinion. The last four equations describe disease dynamics as a variant of the standard SIR model. The quantity $\beta^*$ is the effective transmission rate; it depends on the relative prevalence of nodes in states $P$ and $A$. To close the system, we approximate the expectations of products with the products of expectations. For example, 
\begin{equation*} 
\frac{ 1}{N}[\vert U\vert \times \vert P\vert ]\approx [\vert U \vert ]\times[\vert P \vert]\,.
\end{equation*}
This provides a good approximation when $N$ is large. Henceforth, we omit $|\cdot|$ to simplify our notation.

Consider the special case in which the transmission rate $\bp$, recovery rate $\gp$, and initial population proportion of the pro-physical-distancing opinion are the same as the corresponding parameters for the anti-physical-distancing opinion. In this case, the effective transmission rate is $\beta^* = ([P](\alphaa + \alphap - 2) + 1)\bphy$. Because $[P] \geq 0$, it follows that $\beta^* \geq \bphy $ if and only if $\alphaa + \alphap \geq 2$. Therefore, the spread of opinions always leads to more infections of the disease. If $\alphaa + \alphap = 2$, the opinion dynamics has no effect on disease spread. This conclusion relies on the random-recoupling assumption, which implies that the information-layer counterpart of any physical-layer node is equally likely to be in any given opinion state. Because individuals hold an opinion for some time, the sign of $\alphaa + \alphap - 2$ alone does not determine whether the influence of opinions leads to more infections or fewer infections when we consider the effects of network structures in Section \ref{sec: experiments}.

The epidemic threshold in a standard SIR model is characterized by the basic reproduction number \cite{diekmann1990definition,brauer2019mathematical} $R_0 = {\beta}/{\gamma}$, which is the mean number of secondary infections produced by a single infectious individual in a population in which everyone else is susceptible. An outbreak of the disease occurs if $R_0 > 1$. In our model, suppose that we start with a population in which most people are susceptible and uninformed about the disease. In this case, $\beta^*$ is close to $\beta$. In the limit in which the population becomes infinite with a vanishing fraction of people initially holding any opinion, the outbreak threshold is the same as in the standard SIR model and it is independent of the information layer. However, this conclusion does not hold if too many people hold some opinions about the disease at time $0$.

\begin{figure}[!h]
    \centering
        \begin{minipage}{.5\textwidth}
            \subcaptionbox{Final epidemic size\label{fig: ODEa}}
            {\includegraphics[scale=0.17, trim=12 0 0 0, clip ]{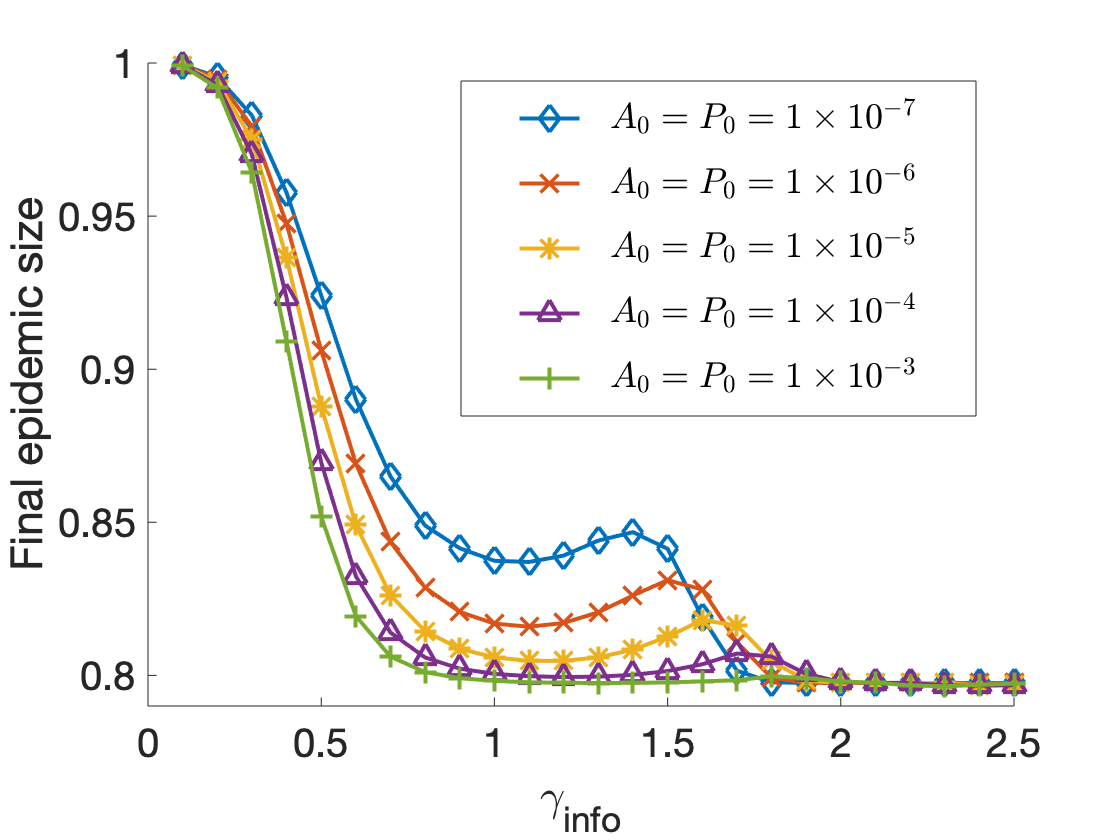}}
        \end{minipage}
        ~
        \begin{minipage}{.45\textwidth}
            \centering
            \subcaptionbox{Information layer\label{fig: ODEb}}
            {\includegraphics[width=\textwidth]{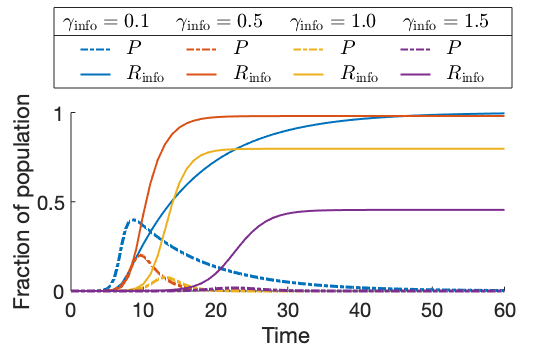}}
            \vfill
            \subcaptionbox{Physical layer\label{fig: ODEc}}
            {\includegraphics[width=\textwidth]{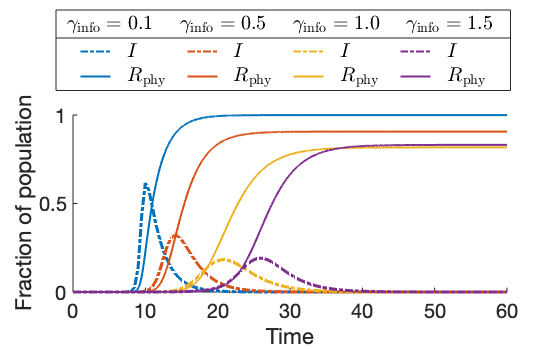}}
        \end{minipage}
    \caption{The influence of the information layer on the physical layer depends on the opinion recovery rates. (a) Effects of $\ginfo$ on the final epidemic size for different initial conditions. For simplicity, we suppose that the
    pro- and anti-physical-distancing opinions share the same contagion parameters for all examples in the paper. That is, $\ba=\bp$ (which we denote by $\binfo$), $\ga=\gp$ (which we denote by $\ginfo$), and $A_0=P_0$, where $A_0=[A](0)$ and $P_0=[P](0)$. The other parameters are $\bphy=1$, $\gphy=0.5$, $\alphap=0.1$, $\alphaa=10$, and $\binfo=2$. To help explain the non-monotonic curve in panel (a), we consider $P_0=A_0 =1\times 10^{-6}$ for different values of $\ginfo$. We show the ensuing dynamics of the fraction of the population in the $P$ and $R_{\text{info}}$ compartments in panel (b) and the fraction of the population in the $I$ and $R_{\text{phy}}$ compartments in panel (c).}
    \label{fig: ODE}
\end{figure}

Although an information contagion may not affect the epidemic threshold for the spread of a disease, it can still have a large impact on the disease's prevalence if a disease outbreak occurs. In Figure \ref{fig: ODE}, we show an illustrative example to demonstrate how the information layer can affect the spread of a disease. For simplicity, we suppose that the pro- and anti-physical-distancing opinions share the same contagion parameters. Figure \ref{fig: ODE}(a) shows an example in which we fix the parameters in the physical layer (on which the disease spreads) and investigate the effect of the opinion recovery rate on the final epidemic size (i.e., the total number of people who become infectious during the outbreak). Because we fix $\binfo=2$, the opinion contagion grows into an outbreak if $\ginfo$ is less than approximately $2$. Consequently, all curves for the final epidemic size converge to the same value when $\ginfo$ is at least approximately $2$. To assist exposition, we use the term ``basic size'' to indicate the final epidemic size when the disease spreads independently of opinions. Because the effective transmission rate $\beta^*$ satisfies $\beta^* \geq \bphy$, we expect the final epidemic size to be no smaller than the basic size.

The final epidemic size is affected by the prevalence of the nodes in states $P$ and $A$ and the relative spreading speeds of the opinions and the disease. As we increase $\ginfo$ in Figure \ref{fig: ODEa}, the final epidemic size tends to decrease, but it grows at first before decreasing to the basic size. To understand this, we compare the spreading dynamics on the two layers (see Figures \ref{fig: ODEb} and \ref{fig: ODEc}). Increasing $\ginfo$ leads to a reduction in the number of people in compartments $P$ and $A$, which reduces the negative influence from the information layer and results in fewer people infected. Increasing $\ginfo$ also postpones the time that it takes for the physical layer to achieve herd immunity. In other words, it takes longer for the $I$ compartment to reach its maximum size. We also see that increasing $\ginfo$ from $1$ to $1.5$ shortens the time difference between the opinion-prevalence peak and the disease-prevalence peak. As the two peaks become closer to each other, we observe transient growth in Figure \ref{fig: ODEa}. As we change the initial numbers of individuals in the $A$ and $P$ states while fixing the initial numbers of individuals in the $I$ compartment in Figure \ref{fig: ODEa}, we effectively change the relative starting times of the opinion dynamics versus the disease dynamics, leading to differences between the curves. 

%
%
\section{Our pair approximation (PA)}\label{sec: pwa}

The random-mixing assumption and the random-recoupling assumption in Section \ref{sec: ODE} ignore contact patterns and oversimplify the dynamics of the spread of opinions and diseases. In real life, both in-person contacts and online interactions have intricate structural patterns\cite{newman2018networks} that are far from homogeneous. Moreover, the random-recoupling assumption mixes the effects of the pro- and anti-physical-distancing opinions in a naive way and leads to features that contradict what we observe in individual-level simulations. Consequently, it is necessary to differentiate between different opinion states within the susceptible population and analyze the dynamics of the $12$ compartments in Figure \ref{fig: C}. Starting in this section, we incorporate network structure into our model and study the resulting dynamics in detail.

We develop a mean-field description of our system by generalizing the degree-based pair-approximation model of Eames and Keeling\cite{eames2002modeling} to coupled dynamics on multilayer networks. We assume that all nodes with the same degree are statistically equivalent, and we estimate the expected number of nodes and the expected number of dyads (i.e., pairs of nodes that are attached to the same edge) grouping both by degrees and by compartments\cite{kiss2017mathematics} using a closure model. We examine the dynamics of the spread of opinions and the disease using a mean of an ensemble of networks, rather than using a single realization with a generative network model\cite{pastor2015epidemic,porter2016}.  
We first develop an exact ODE system that involves single, pair, and triple terms based on the law of mass action. This system also depends on a set of time-dependent effective transmission rates. We close the system by approximating triple terms and the effective transmission rates using pair terms. The number of equations in this system depends on the number of distinct degrees and is independent of the population size. 

\subsection{Our dynamical system at the level of triples}

Recall that we use two adjacent letters $YX$ to describe the state (i.e., compartment) of an individual, where the first letter indicates the information layer and the second letter indicates the disease layer. The two intra-layer degrees of a node are its numbers of neighbors in the two layers. We are interested only in intra-layer degrees, so we treat our multiplex networks as edge-colored multigraphs\cite{kivela2014multilayer}. We use $Y_{k_1}X_{k_2}$ to refer to nodes in state $YX$ with degree $k_1$ in the information layer and degree $k_2$ in the physical layer. We write the expected density of these nodes as $[Y_{k_1}X_{k_2}]$. To simplify the notation, we drop subscripts to indicate summation over all possible degrees. For example, $[YX_{k_2}] = \sum_{k_1}[Y_{k_1}X_{k_2}]$. For ease of notation, we use one letter and thereby specify the state only in the other layer when the context is clear. For example, $S_{k_2}$ refers to susceptible nodes with degree $k_2$ in the physical layer and $[S_{k_2}]$ refers to the expected density of these nodes. 

To track the states of the neighbors of a node, we write the expected normalized count of the dyads of nodes with states ${\cal S}_1$ and ${\cal S}_2$ as $[{\cal S}_1\circ {\cal S}_2]$, where $\circ$ denotes an edge and the normalized count is the number of dyads divided by the population size. The layer of the dyad is clear from the context. For example, $[\US \circ I]$ represents the expected normalized count of dyads in the physical layer for which one end is attached to a $\US$ node and the other end is attached to an infectious node. 

Given the above notation and definitions, the time evolution of the expected density of each compartment as a function of their neighbors' states is 
\begin{align*}\label{equ: pair_approx_1}
        \ddt[\US] &= -[\US \circ I] \bphy - [\US \circ A]\ba - [\US \circ P]\bp\,,\\
    \ddt[\UI] &= [\US \circ I] \bphy - [\UI] \gphy- [\UI \circ A]\ba - [\UI \circ P]\bp\,,\\
    \ddt[\UR] &= [\UI] \gphy - [\UR \circ A]\ba - [\UR \circ P]\bp\,,\\
    \ddt[\AS] &= -[\AS \circ I] \bphy\times \alphaa + [\US \circ A]\ba -[\AS]\ga\,,\\
    \ddt[\AI] &= [\AS \circ I] \bphy\times \alphaa -[\AI]\gphy + [\UI \circ A]\ba - [\AI]\ga\,,\\
    \ddt[\AR] &= [\AI]\gphy + [\UR \circ A]\ba - [\AR]\ga\,,\\
    \ddt[\PS] &= -[\PS \circ I] \bphy\times \alphap + [\US \circ P]\bp - [\PS]\gp\,,\\
    \ddt[\PI] &= [\PS \circ I] \bphy\times \alphap -[\PI]\gphy + [\UI \circ P]\bp - [\PI]\gp\,,\\
    \ddt[\PR] &= [\PI]\gphy + [\UR \circ P]\bp - [\PR]\gp\,,\\
    \ddt [\RS] &= -[\RS\circ I]\bphy + [\AS]\ga + [\PS]\gp\,,\\
    \ddt[\RI] &= [\RS\circ I]\bphy - [\RI]\gphy + [\AI]\ga + [\PI]\gp\,.\tag{4.1}
\end{align*}
To illustrate the meaning of the equations in \eqref{equ: pair_approx_1}, we briefly go through one of them. In the first equation, the expected number of $\US$ nodes decreases as the nodes become infectious or adopt one of the two opinions. The infection rate is proportional to the number of infectious neighbors (or, equivalently, to the number of $\USI$ dyads). We do not track the opinion states of those infectious neighbors because we assume that those opinions do not affect the transmission rate of the $\US$ nodes. The same reasoning applies to the other dyads.

We expand the right-hand side of the system \eqref{equ: pair_approx_1} by tracking the dynamics of node pairs, which depend on the neighbors of both nodes and thus involve triples. Let $[{\cal S}_1 \circ {\cal S}_2 \circ {\cal S}_3]$ denote the expected normalized count of triples in which the center node ${\cal S}_2$ is adjacent to ${\cal S}_1$ and to ${\cal S}_3$. Analogously to the normalized count of a dyad, we define the normalized count of a triple to be the number of triples divided by the population size. The two edges may belong to the same layer or to different layers; this is clear from the context. For example, $\USS\circ I$ refers to triples in which the center node $S_l$ has two physical-layer neighbors in states $\US$ and $I$. Additionally, $P\circ \USI$ refers to triples in which one edge connects $\US$ and $P$ in the information layer and the other connects $\US$ and $I_l$ in the physical layer. We now write the evolution of the expected normalized count of the dyads in terms of triples terms:
\begin{align*}
    \ddt[\USI] &= [\USS \circ I]\bhat_{l,\,k_2}- [\USI]\bphy - [I\circ \USI]\bphy - [\USI]\gphy \\
    &\quad - [P\circ \USI]\bp - [A\circ\USI]\ba\,,\\
    \ddt[\USS] &= -[\USS \circ I]\bhat_{l,\,k_2}- [I\circ\USS]\bphy\\
    &\quad - [P\circ \USS]\bp - [A\circ\USS]\ba\,,\\
    \ddt[\USA] &= -[I\circ \USA]\bphy - [\USA]\ba - [A\circ \USA]\ba \\
    &\quad - [P\circ \USA]\bp + [\USU\circ A]\ba - [\USA]\ga\,,\\
    \ddt[\USP] &= -[I\circ \USP]\bphy - [\USP]\bp - [A\circ \USP]\ba \\
    &\quad - [P\circ \USP]\bp + [\USU\circ P]\bp - [\USP]\gp\,,\\
    \ddt[\USU] &= -[I\circ \USU]\bphy - [A\circ\USU]\ba - [P\circ\USU]\bp\\
    &\quad -[\USU\circ P]\bp - [\USU\circ A] \ba\,,\\
    \ddt[\UIA] &= [I \circ \USA]\bphy - [\UIA]\gphy + [\UIU\circ A]\ba - [\UIA]\ga\\& - [\UIA]\ba - [A\circ \UIA]\ba - [P\circ\UIA]\bp\,,\\
    \ddt[\UIP] &= [I \circ \USP]\bphy - [\UIP]\gphy + [\UIU\circ P]\bp - [\UIP]\gp\\& - [\UIP]\bp - [A\circ \UIP]\ba - [P\circ\UIP]\bp\,,\\
    \ddt[\UIU] &= [I \circ \USU]\bphy - [\UIU]\gphy - [\UIU\circ A]\ba\\& -[\UIU\circ P]\bp - [A\circ \UIU]\ba - [P\circ \UIU]\bp \,,\\
    \ddt[\URA] &=   [\UIA]\gphy + [\URU\circ A]\ba - [\URA]\ga\\& - [\URA]\ba - [A\circ \URA]\ba - [P\circ\URA]\bp\,,\\
    \ddt[\URP] &=  [\UIP]\gphy + [\URU\circ P]\bp - [\URP]\gp\\& - [\URP]\bp - [A\circ \URP]\ba - [P\circ\URP]\bp\,,\\
    \ddt[\URU] &=  [\UIU]\gphy - [\URU\circ A]\ba\\& -[\URU\circ P]\bp - [A\circ \URU]\ba - [P\circ \URU]\bp\,,\\
   \ddt[\ASI] &= [\ASS \circ I]\bhat_{l,\,k_2}- [\ASI]\bphy\times\alphaa - [I\circ \ASI]\bphy\times\alphaa \\
    &\quad -[\ASI]\gphy + [A\circ\USI]\ba - [\ASI]\ga\,,\\
    \ddt[\ASS] &= -[\ASS \circ I]\bhat_{l,\,k_2}- [I\circ\ASS]\bphy\times\alphaa\\
    &\quad + [A\circ\USS]\ba - [\ASS]\ga\,,\\
    \ddt[\PSI] &= [\PSS \circ I]\bhat_{l,\,k_2}- [\PSI]\bphy\times\alphap - [I\circ \PSI]\bphy\times\alphap \\
    &\quad -[\PSI]\gphy + [P\circ\USI]\bp - [\PSI]\ga\,,\\
    \ddt[\PSS] &= -[\PSS \circ I]\bhat_{l,\,k_2}- [I\circ\PSS]\bphy\times\alphap\\
    &\quad + [P\circ\USS]\bp - [\PSS]\gp\,,\\
    \ddt[\RSI] &= [\RSS \circ I]\bhat_{l,\,k_2}- [\RSI]\bphy - [I\circ \RSI]\bphy \\
    &\quad -[\RSI]\gphy + [\ASI]\ga + [\PSI]\gp\,,\\
    \ddt[\RSS] &= -[\RSS \circ I]\bhat_{l,\,k_2}- [I\circ\RSS]\bphy\\
    &\quad +[\ASS]\ga + [\PSS]\gp \,, \tag{4.2}
   \label{equ: pair_approx_2}
\end{align*}
where $\bhat_{l,\,k_2}$ is the expected transmission rate of the center nodes $S_l$ in triples of the form $YS_{k_2}\circ S_l\circ I$. 

One derives the system \eqref{equ: pair_approx_2} using the same reasoning as in \eqref{equ: pair_approx_1}. For example, consider the first equation in \eqref{equ: pair_approx_2}. The normalized count of the dyads $\USI$ decreases as $\US$ nodes adopt one of the two opinions at rate $[P\circ \USI]\bp + [A\circ\USI]\ba$, is infected by $I_l$ at rate $[\USI]\bphy$, or is infected by infectious neighbors other than $I_l$ at rate $[I\circ \USI]\bphy$. The normalized count of $\USI$ increases as susceptible neighbors of $\US$ are infected by their infectious neighbors at rate $[\USS \circ I]\bhat_{l,\,k_2}$. For each dyad in \eqref{equ: pair_approx_2}, there is one node for which we only track its status in one of the layers. For example, for the dyads $\USI$, we do not know the opinion states of node $I_l$. We need an approximation for the disease transmission rate $\bhat_{l,\,k_2}$ when node $I_l$ is in state $S$. In principle, one can track the states of both nodes on both layers and avoid the need for this approximation. However, doing this leads to an expanded system of higher dimension. We discuss the approximation of $\bhat_{l,\,k_2}$ in Section \ref{sec: bhat}. 

In principle, one can also work out the right-hand sides for the evolution of the expected normalized counts of the triple terms. These incorporate quadruple terms, and if we expand those terms and keep expanding expressions for the evolution of progressively larger network motifs (i.e., connected subgraphs), we eventually obtain an exact dynamical system. However, it is very high-dimensional and difficult to study. Therefore, we approximate the triple terms with pair terms on the right-hand sides of \eqref{equ: pair_approx_2} using the approach in Ref.~[\citen{eames2002modeling}].

\subsection{Closure of the triple terms}\label{sec: triple closure}

For a given type of triples $X\circ Y_k\circ Z$, we assume that the neighbors of all $Y_k$ nodes are interchangeable. Therefore, every neighbor has the same probability of being in a given state (e.g., state $X$).

If both edges are in the same layer, then for nodes $X$ and $Z$ that are adjacent to a center degree-$k$ node in state $Y$ in the same layer, it follows that 
\setcounter{equation}{2}
\begin{align}
    [X\circ Y_k \circ Z] &\approx k(k-1)[Y_k] \frac{[X\circ Y_k]}{k[Y_k]}\frac{[Y_k\circ Z]}{k[Y_k]}\notag\\
    &= \frac{k-1}{k}\frac{[X\circ Y_k][Y_k\circ Z]}{[Y_k]}\,.\label{equ:3}
\end{align}
Intuitively, nodes in state $Y_k$ have $k[Y_k]$ edges; an expected fraction $\frac{[X\circ Y_k]}{k[Y_k]}$ of these edges are attached to nodes in state $X$, and an expected fraction $\frac{[Y_k \circ Z]}{k[Y_k]}$ of these are attached to nodes in state $Z$.
 Therefore, if we choose a node in state $Y_k$ uniformly at random, the probability that two uniformly random neighbors of the $Y_k$ node are in states $X$ and $Z$ is approximately $\frac{[X\circ Y_k]}{k[Y_k]}\times\frac{[Y_k\circ Z]}{k[Y_k]}$ when $N$ is large. Because there are $k(k-1)$ ways to choose the two neighbors, we obtain expression \eqref{equ:3}. As concrete examples, 
\begin{align*}
    [\USS\circ I] &\approx \frac{l-1}{l}\frac{[\USS][S_l\circ I]}{[S_l]}\,,\\
    [I\circ \USS] &\approx \frac{k_2-1}{k_2}\frac{[I\circ \US][\USS]}{[\US]}\,,\\
    [A\circ \USA] &\approx \frac{k_1 - 1}{k_1} \frac{[A\circ \US][\USA]}{[\US]}\,.
\end{align*}

Now suppose that the two edges that connect the center node $Y_{1, k_1}Y_{2, k_2}$ to nodes in states $X$ and $Z$ are in different layers. If the node in state $X$ is in the information layer and the node in state $Z$ is in the physical layer, we obtain
\begin{align*}
    [X\circ Y_{1, k_1}Y_{2, k_2}\circ Z] &\approx k_1k_2[Y_{1,k_1}Y_{2,k_2}]\frac{[X\circ Y_{1, k_1}Y_{2, k_2}]}{k_1[Y_{1,k_1}Y_{2,k_2}]}\frac{[Y_{1, k_1}Y_{2, k_2}\circ Z]}{k_2[Y_{1, k_1}Y_{2, k_2}]}\\
    &= \frac{[X\circ Y_{1,k_1}Y_{2,k_2}][Y_{1, k_1}Y_{2, k_2}\circ Z]}{[Y_{1, k_1}Y_{2, k_2}]}\,.
\end{align*}
For example,
\begin{align*}
    [P\circ \USI] \approx \frac{[P\circ \US][\USI]}{[\US]}\,.
\end{align*}
One can work out approximations for the other triple terms similarly.

\subsection{Approximate transmission rate}\label{sec: bhat}

To close the dynamical system \eqref{equ: pair_approx_1}--\eqref{equ: pair_approx_2}, we need to find an approximation $\bhat_{l,\,k_2}$, the expected transmission rate of the center node for triples of the form $YS_{k_2}\circ S_l\circ I$. We need to approximate the opinion distribution in each population of interest. The random-recoupling assumption in Section \ref{sec: ODE} corresponds to setting
\begin{align}\label{equ:  0_approx}
    \hat\beta_{l,\,k_2} &\approx ([U] + [A]\alphaa + [P]\alphap + [R]) \bphy\,. 
\end{align}
However, now we can keep track of correspondence between the two layers. A naive approach is to weight the influence coefficients based on the densities of nodes with different opinion states among the $S_l$ nodes. That is,
\begin{equation}\label{equ:  half_approx}
    \bhat_{l,\,k_2} \approx \frac{[US_l] + [AS_l]\alphaa + [PS_l]\alphap + [RS_l]}{[S_l]}\bphy\,. 
\end{equation}
However, the approximation \eqref{equ:  half_approx} ignores the fact that the $S_l$ node of interest has both an intra-layer neighbor in state $S$ and an intra-layer neighbor in state $I$. Incorporating this neighborhood information yields the approximation
\begin{equation}
      \bhat_{l,\,k_2} \approx \frac{[S_{k_2}\circ US_l\circ I] + [S_{k_2}\circ AS_l\circ I]\alphaa + [S_{k_2}\circ PS_l\circ I]\alphap + [S_{k_2}\circ RS_l\circ I]}{[S_{k_2}\circ S_l\circ I]}\bphy\,.\label{equ:  full_approx}
\end{equation}
After inserting our pair approximation, we obtain 
\begin{equation*}
         \bhat_{l,\,k_2} = \frac{\bphy\sum_k\left(\frac{[U_{k}S_l\circ S_{k_2}][U_{k}S_l\circ I]}{[U_kS_l]} + \frac{[A_kS_l\circ S_{k_2}][ A_kS_l\circ I]}{[A_kS_l]}\alphaa + \frac{[P_kS_l\circ S_{k_2}][P_kS_l\circ I]}{[P_kS_l]}\alphap +\frac{[R_kS_l\circ S_{k_2}][ R_kS_l\circ I]}{[R_kS_l]}\right)}{\sum_k\left(\frac{[U_{k}S_l\circ S_{k_2}][U_{k}S_l\circ I]}{[U_kS_l]} + \frac{[A_kS_l\circ S_{k_2}][ A_kS_l\circ I]}{[A_kS_l]} + \frac{[P_kS_l\circ S_{k_2}][P_kS_l\circ I]}{[P_kS_l]} +\frac{[R_kS_l\circ S_{k_2}][ R_kS_l\circ I]}{[R_kS_l]}\right)}\,.
\end{equation*}

\begin{figure}
    \centering
    \includegraphics[scale=0.4]{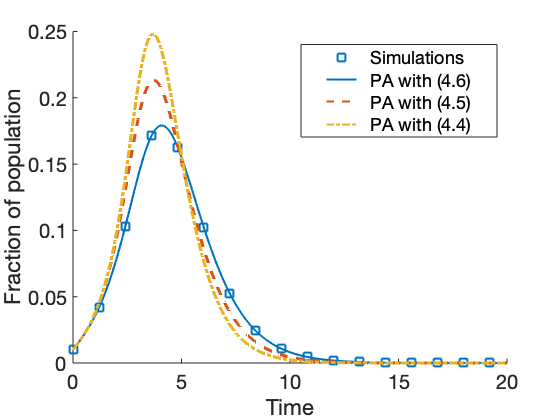}
    \caption{Comparison of our pair approximations (PAs) based on equations \eqref{equ:  0_approx}--\eqref{equ:  full_approx} (the curves) with direct numerical simulations (the markers). The trajectories show the time evolution of the infectious population. In each simulation, we generate networks with layers that consist of 5-regular configuration-model graphs (i.e., each node has degree 5). The results are a mean over $100$ simulations. The parameters are $\bphy =\binfo = 0.6$, $\gphy=\ginfo=1$, $\alphaa=10$, and $\alphap=0.1$.}
    \label{transmission_rate}
\end{figure}

We expect the value of $\bhat_{l,\,k_2}$ in equation \eqref{equ:  half_approx} to be smaller than its value in equation \eqref{equ:  0_approx}. This, in turn, leads to a smaller estimate of the disease prevalence from equation \eqref{equ:  half_approx} than from equation \eqref{equ:  0_approx}. Intuitively, because individuals who hold the anti-physical-distancing opinion become infected at a higher rate, a typical susceptible individual is less likely to have an anti-physical-distancing opinion than a member of the population selected uniformly at random. Therefore, we expect that $[A] \geq \frac{[AS_l]}{[S_l]}$. By applying analogous reasoning to individuals who hold the pro-physical-distancing opinion, we expect that $[P] \leq \frac{[PS_l]}{[S_l]}$. We do not have a mathematically rigorous understanding of how well the approximations $\eqref{equ:  half_approx}$ and $\eqref{equ:  full_approx}$ match the full stochastic system. (See Section \ref{sec: model} and our code in Ref.~\citen{code}.) We compare the pair approximation of the disease prevalence based on equations \eqref{equ:  0_approx}--\eqref{equ:  full_approx} and direct numerical simulations of the full stochastic system in Figure \ref{transmission_rate}. From this comparison, we see that the approximations \eqref{equ:  0_approx} and \eqref{equ:  half_approx} overestimate the infectious population and that the approximation \eqref{equ:  full_approx} matches the simulations very well. The numerical results indicate that it is essential to track the coupling of the nodes' states at both ends of inter-layer edges and intra-layer edges to ensure accurate estimations of time evolution of disease prevalence. We use the pair approximation (\ref{equ: pair_approx_1}, \ref{equ: pair_approx_2}, \ref{equ:  full_approx}) in subsequent experiments in Section \ref{sec: experiments}.

%
%
\section{Computational experiments}\label{sec: experiments}

We now investigate our full model by simulating the stochastic system \cite{code} and applying the pair approximation (\ref{equ: pair_approx_1}, \ref{equ: pair_approx_2}, \ref{equ:  full_approx}), which we refer to as ``PA". We explore the influence of competing opinion contagions on the spread of a disease for a variety of parameter values. We focus on examining different opinion contagion parameters (see Section \ref{sec: exp_contagion_params}) and network structures (see Sections \ref{sec: intra_layer} and \ref{sec: inter_layer}). For ease of comparison, we consider the special case in which the pro- and anti-physical-distancing opinions share the same contagion parameters for all examples in our paper. Additionally, we fix the disease contagion parameters to be $\bphy=0.6$ and $\gphy=1$. Unless we specify otherwise, we set the opinion influence coefficients to be $\alphap=0.1$ and $\alphaa=10$ to incorporate nontrivial influence from the corresponding opinion on the spread of the disease. This asymmetry between pro- and anti-physical-distancing opinions affects the dynamics in an interesting way, as we illustrate in this section. In many of the following examples, it is helpful to separate the influence of the two opinions to gain understanding of the overall behavior. To do this, we neutralize the influence from an opinion by setting its influence coefficient to be $1$. In Section \ref{sec: exp_contagion_params}, we do a parameter sweep for the opinion transmission parameters in the range $[0,\,2]$ to illustrate that the final epidemic size can change non-monotonically as we increase the recovery rate of an opinion. This feature occurs in networks with a variety of degree distributions. Due to the issuing or lifting of stay-at-home orders, people's contact patterns in the offline world can change a lot over the course of an epidemic (and especially a pandemic) \cite{zhang2020changes,feehan2021quantifying}. In Sections \ref{sec: intra_layer} and \ref{sec: inter_layer}, we show examples that illustrate that the influence from an opinion contagion on the spread of a disease can change in important ways when we change the intra-layer or cross-layer correlations of intra-layer degrees. 

In each computational experiment, we construct a network of $N = 10000$ nodes and simulate the dynamics on it using a Gillespie algorithm\cite{kiss2017mathematics}, which is a well-known approach for performing continuous-time simulations of Markovian processes. In all experiments in this section, the results are a mean over $200$ simulations. In each simulation, we generate new random graphs (of a few different types, which we specify below). We uniformly randomly infect $I_0=1\%$ of the nodes in the physical layer, and we independently and uniformly randomly choose $A_0=P_0=0.5\%$ nodes as anti- or pro-physical-distancing in the information layer. We set all remaining node states to $S$ in the physical layer and to $U$ in the information layer. 

In our initial experiments, we construct each network layer from a configuration model \cite{fosdick2018} and match the nodes from the two layers uniformly at random. Specifically, we specify degree distributions $\mathbb P_{\text{info}}$ and $\mathbb P_{\text{phy}}$, which need not be same. For each layer, we sample a degree sequence $\{k_i\}$ (where $i \in \{1,\ldots,N\}$ indexes the nodes) from the corresponding degree distribution; therefore, node $i$ has $k_i$ ends of edges (i.e., stubs). We match these stubs uniformly at random to form a network. Correspondingly, in the pair approximation, we have 
\begin{equation*}\begin{split}
    [Y_{k_1}X_{k_2}](0) &= Y_0X_0\pinfo(k_1)\pphy(k_2)\,,\\
    [Y_{k_1}X_{k_2}\circ Z_{k_3}](0) &= 
    \begin{cases}
    [Y_{k_1}X_{k_2}](0)\times Z_0\pinfo(k_3)k_1k_3/\langle k_{\text{info}}\rangle\,,
    ~Z\in\{U,\,P,\,A,\,\Rinfo\}\\
    [Y_{k_1}X_{k_2}](0)\times Z_0\pphy(k_3)k_2k_3/\langle k_{\text{phy}}\rangle\,,~Z\in\{S,\,I,\,\Rphy\} \,.
    \end{cases}
    \end{split}
\end{equation*}

\begin{figure}[!h]
    \centering
        \begin{minipage}{.4\textwidth}
            \subcaptionbox{Disease prevalence \label{fig:figA}}
            {\includegraphics[scale=0.183]{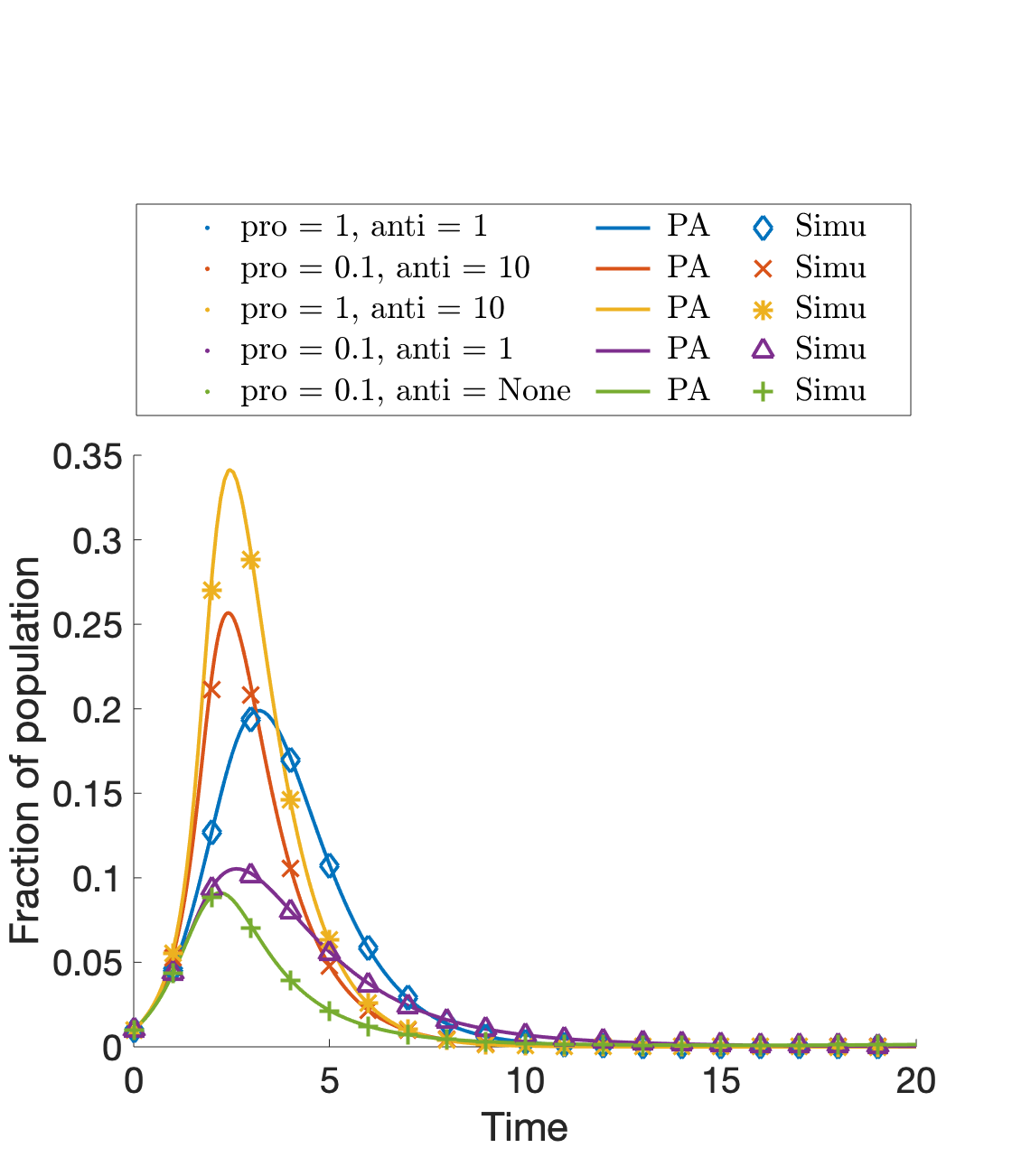}}
        \end{minipage}
        \hspace{1cm}~~~~~~
        \begin{minipage}{.45\textwidth}
            \centering
            \subcaptionbox{Information layer; only the pro-physical-distance opinion has influence.\label{fig:figB}}
            {\includegraphics[width=\textwidth,height=4cm]{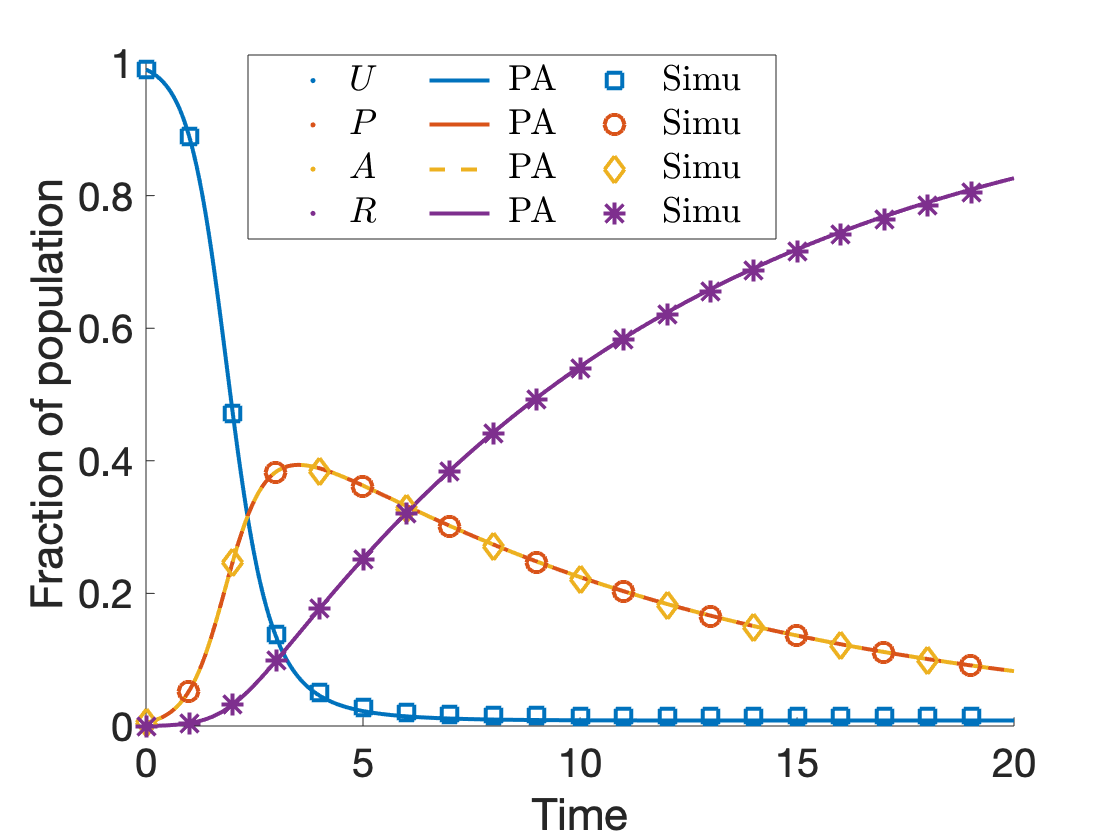}}
            \vfill
            \subcaptionbox{Information layer; no anti-physical-distance opinion\label{fig:figC}}
            {\includegraphics[width=\textwidth,height=4cm]{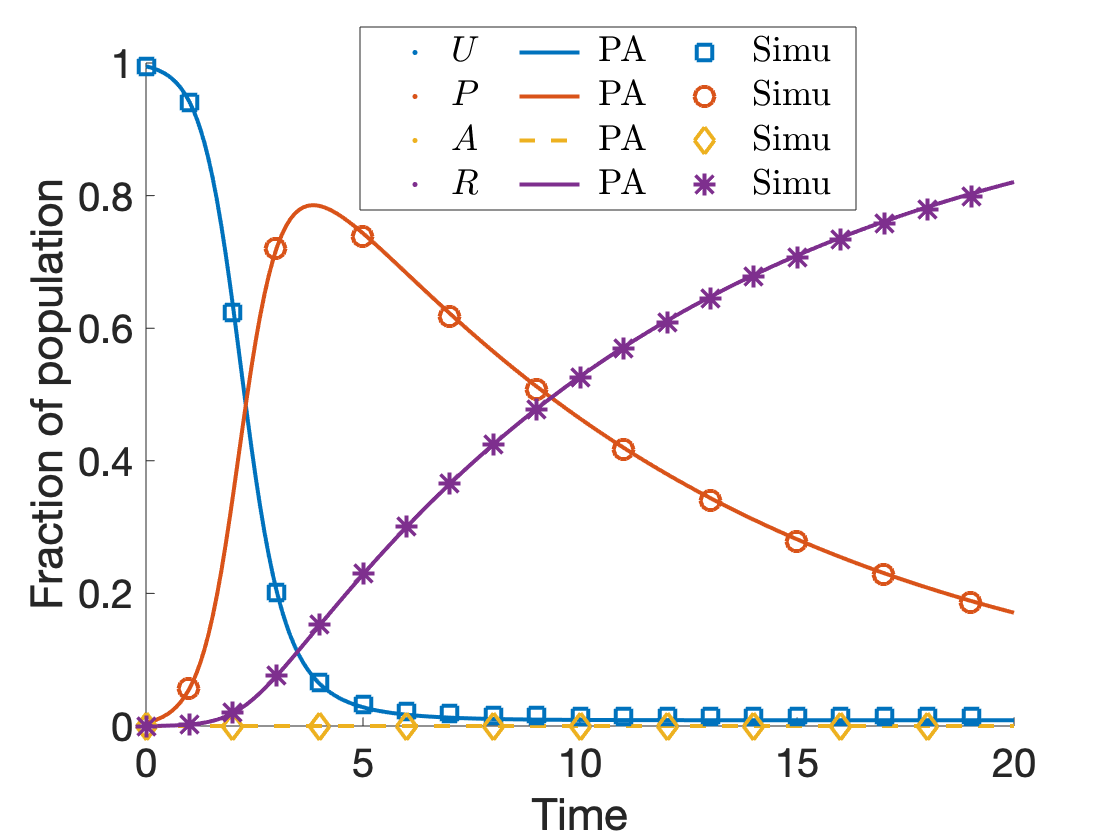}}
        \end{minipage}
    \caption{Dynamics for different opinion influence coefficients. (a) Disease prevalence curves for different influence coefficients, including when the anti-physical-distancing opinion is absent (which we denote by ``anti = None''). In (b, c), we show the dynamics on the information layer when (b) both opinions are present and (c) only the pro-physical-distancing opinion is present. The results are a mean over 200 simulations. We construct each layer from a configuration model with a degree sequence that we choose from a Poisson degree distribution with mean degree $5$. The other parameters are $\bphy =\binfo = 0.6$, $\gphy=1$, and $\ginfo=0.1$. The curves (respectively, markers) indicate results from the PA (respectively, direct simulations).}
    \label{fig: first_exp}
\end{figure}

In Figure \ref{fig: first_exp}, we compare typical disease prevalence curves (i.e., the time evolution of infectious populations) when we choose different coefficients for the influence of the opinions. The influence from the information layer changes a disease's prevalence, its peak value, and the time at which the peak number of infections occurs. Although an opinion does not alter the susceptibility of individuals when the corresponding influence coefficient is $1$, the spread of that opinion can still affect the overall disease dynamics, which thus can be different from what occurs in a system with only one opinion. For example, in Figure \ref{fig:figA}, the purple curve with triangle markers (for which $\alphaa=1$ and $\alphap=0.1$) has a higher disease prevalence than the green curve with plus signs (for which $\alphap=0.1$ and the anti-physical-distancing opinion is absent). The spread of the anti-physical-distancing opinion prevents some people from adopting the pro-physical-distancing opinion, although the anti-physical-distancing opinion has an influence coefficient of $1$. We show the corresponding dynamics on the information layer in Figures \ref{fig:figB} and \ref{fig:figC}. 

\subsection{Opinion contagion parameters}\label{sec: exp_contagion_params}

\begin{figure}[!h]
    \centering
    \begin{subfigure}{0.45\textwidth}
     \includegraphics[scale=0.16]{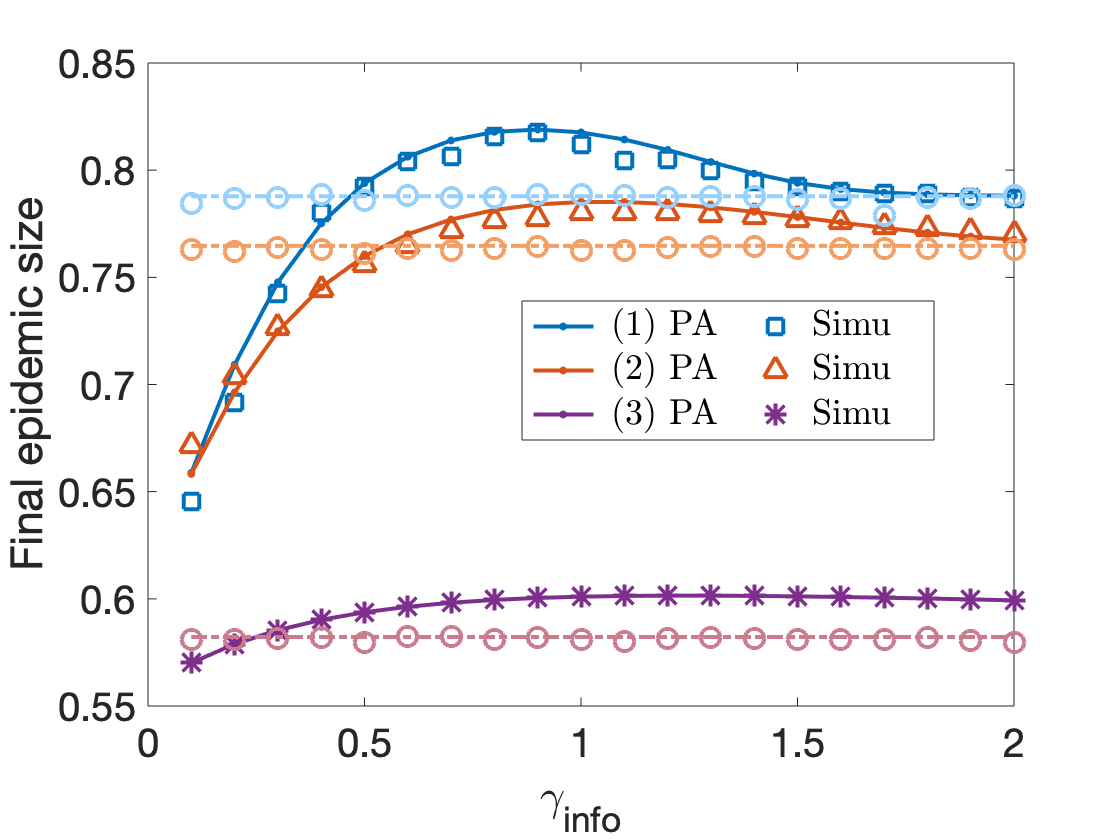}
     \caption{Final epidemic size}\label{fig: cont_param_a}
    \end{subfigure}
    ~~
    \begin{subfigure}{0.45\textwidth}
     \includegraphics[scale=0.16]{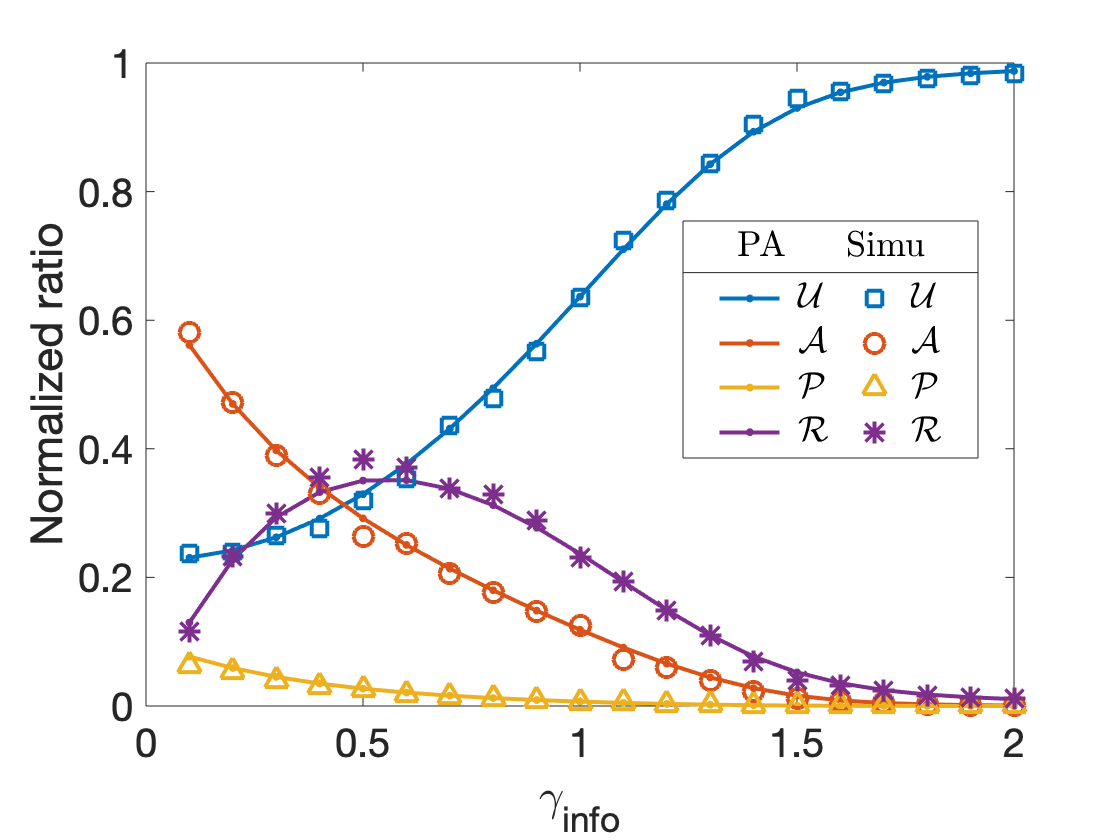}
     \caption{Decomposition of opinion states}\label{fig: cont_param_b}
    \end{subfigure}
    \caption{Influence of the opinion recovery rate $\gamma_{\text{info}}$ on disease prevalence and on the distribution of opinion states when people become infectious. (a) The final epidemic size for different values of $\gamma_{\text{info}}$. The solid curves and non-circle symbols mark the final epidemic sizes under influence from the information layer. The dashed curves and circles mark the basic size. We consider three situations: (1) each layer is a $5$-regular graph; (2) all node degrees follow a Poisson distribution with mean $5$; and (3) all node degrees follow a truncated power-law distribution with $\mathbb P(k=x) \propto x^{-1.32}e^{-x/35}$ for $x\leq 50$ and $\mathbb P(k=x) =0$ for $x>50$. We construct each layer from a configuration model with a degree sequence chosen from the specified degree distribution. The other parameters are $\bphy =\binfo = 0.6$, $\gphy=1$, $\alphaa=10$, and $\alphap=0.1$. (b) We group recovered people based on their opinion states when they become infectious (we use the notation $\cal{U}$, $\cal{P}$, $\cal{A}$, and $\cal{R}$ for these subpopulations) and plot the normalized size. We show results (which are means over 200 simulations) for $5$-regular configuration-model graphs. The curves (respectively, markers) indicate results from the PA (respectively, direct simulations).}
    \label{fig: non-monotone}
\end{figure}

Recall from Section \ref{sec: ODE} that in a fully-mixed population with $\alphap + \alphaa > 2$, the information layer leads to a larger epidemic size than when there is no influence from the information layer. We repeat the experiment in Figure \ref{fig: ODE}, but now we have a network structure and we employ our PA. We again consider the scenario in which the anti- and pro-physical-distancing opinions have the same contagion parameter values, denoted using the subscript ``info". Figure \ref{fig: cont_param_a} shows final epidemic sizes versus the recovery rate $\ginfo$ in the information layer for the following three situations: (1) all nodes have degree $5$, so we consider 5-regular graphs; (2) all node degrees follow a Poisson distribution with mean $5$; and (3) all node degrees follow a truncated power-law distribution with $\mathbb P(k=x) \propto x^{-1.32}e^{-x/35}$ for $x\leq 50$ and $\mathbb P(k=x) =0$ for $x>50$. In each situation, we generate both layers using configuration-model networks and we independently sample degrees for each layer from the same distribution. The mean degree is roughly $5$ in all three situations. In all three situations, the final epidemic size can be smaller than the corresponding basic size (i.e., without opinion spread) when $\ginfo$ is very small. As we increase $\ginfo$, the final epidemic size first increases and surpasses the basic size before reaching a peak; it subsequently decreases to the basic size. 

To explain this non-monotonic behavior, we decompose the recovered population at steady state into subpopulations based on their opinion states as they become infectious and plot the relative size of each subpopulation (with a sum that is normalized to $1$) in Figure \ref{fig: cont_param_b}. We use $\cal{U}$, $\cal{A}$, $\cal{P}$, and $\cal{R}$ to denote the subpopulations that become infectious when they are in the $U$, $A$, $P$, and $R_{\text{info}}$ states, respectively. We show results when both layers are $5$-regular graphs. Our results on networks with the Poisson and truncated power-law distribution are qualitatively the same. Because increasing opinion recovery rates results in fewer people adopting any opinions, the size of the $\cal{U}$ subpopulation increases, leading to more people becoming infectious while uninformed. For the same reason, the sizes of the subpopulations with the anti- and pro-physical-distancing opinions decrease with increasing opinion recovery rates. The size of the $\cal{R}$ subpopulation first increases as we increase $\ginfo$. This is because when $\ginfo$ is very small, many people keep the same opinion ($P$ or $A$)
until the disease dies out in the population. When we start to increase $\ginfo$, more people recover from either opinion when the disease is still actively spreading. Because people who give up pro-physical-distancing behavior increase their risk of becoming infectious, the overall epidemic size may increase when they become less cautious. As $\ginfo$ keeps growing, there is a decrease in the population that adopt either opinion; this, in turn, leads to a smaller $\cal{R}$ subpopulation and a drop in the overall epidemic size. 

The non-monotonic behavior described above suggests that if enough people practice pro-physical-distancing behavior for a sufficiently long time, the prevalence of a disease can be reduced, even for an arbitrarily large influence coefficient for the anti-physical-distancing opinion. When both opinions are present, the overall influence of opinions on disease spread is not determined by the two influence coefficients alone; instead, it arises from a complex interaction between the dynamics of the two opinions. 

\begin{figure}[!h]
\centering
     \includegraphics[scale=0.4]{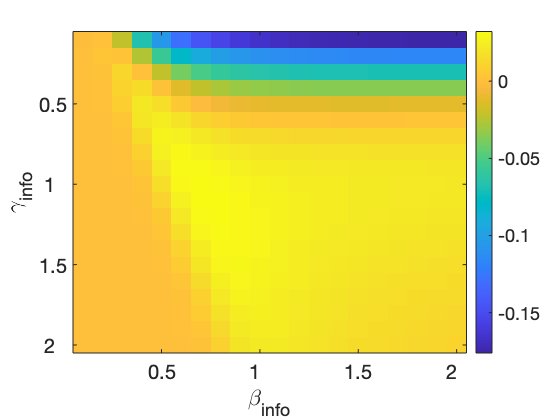}
    \caption{The final epidemic size minus the basic size for different values of the opinion contagion parameters for our PA on networks with layers that consist of $5$-regular configuration-model graphs. We fix the other parameters to be $\bphy = 0.6$, $\gphy=1$, $\alphaa=10$, and $\alphap=0.1$.}\label{fig: final_size_matrix}
\end{figure}

We plot the final epidemic size minus the basic size in Figure \ref{fig: final_size_matrix} for different values of the opinion transmission rates and opinion recovery rates. For fixed opinion recovery rates, the final epidemic size does not change much as we vary the opinion transmission rates if the information layer has an outbreak. Figure \ref{fig: final_size_matrix} shows results from using our PA on 5-regular configuration-model graphs. 

\subsection{Random graphs with intra-layer degree--degree correlations}\label{sec: intra_layer}

Nodes in a network with intra-layer edges to other nodes of similar degrees (i.e., degree assortativity\cite{newman2002assortative, newman2003mixing}, often called simply ``assortativity'') can have a strong impact on disease spread and other dynamical processes\cite{kiss2008effect,melnik2014dynamics}. Such assortative networks may have a core with large-degree nodes, so disease may spread faster but terminate with a smaller final epidemic size on such a network compared with a disassortative network\cite{newman2003mixing, moreno2003epidemic}. 
 In this subsection, we investigate how the change of assortativity structure in the form of an intra-layer degree--degree correlation can influence our model.

For all experiments in Section \ref{sec: intra_layer_2}, we generate networks with intra-layer degree--degree correlations using a model from Melnik et al.~\citen{melnik2014dynamics}. For each of the two layers in the network, we start with a mixing matrix $E$ that specifies the joint distribution of degrees at both ends of an edge chosen uniformly at random. The number of edges that connect nodes with degrees $k$ and $k'$ is ${\cal E}_{k,\,k'} = E_{k,k'}\sum_k (k p_k)N/2$, where $p_k = \left(\sum_{k'} \left.\frac{E_{k,\,k'}}{k}\right)\right/\left(\sum_{k,\,k'} \frac{E_{k,\,k'}}{k}\right)$ specifies the degree distribution and we recall that $N$ is the number of nodes. We first create the required number of edges that connect node pairs with specified combinations of the degrees $k$ and $k'$. We then generate nodes by collecting $k$ ends of edges that we attach uniformly at random to nodes with degree $k$. We obtain networks with the desired degree--degree correlation when we finish attaching all ends of edges to nodes.

Because we track the expected number of edges with all possible degree combinations explicitly and separately in the PA system \eqref{equ:  pair_approx_1}--\eqref{equ: pair_approx_2}, we only need to modify the initialization step to encode the desired intra-layer degree--degree correlation. For example, we let $[U_{k}\circ P_{l}](0) =  {\cal E}_{\text{info},\,k,\,l} \times (1-P_0-A_0)P_0$. We initialize the physical layer and the information layer independently, so $[\USP](0) = [U_{k_1}\circ P_l](0)\times \mathbb{P}_{\text{phy}}(k_2)\times (1-I_0)$. We initialize the other dyads similarly.

\subsubsection{Pedagogical example: Networks whose nodes have one of two different degrees}\label{sec: intra_layer_2}

To illustrate the importance of intra-layer degree--degree correlations, we consider a simple example of a network whose nodes have one of two different degrees, with the degree distribution $\mathbb{P}(k = k_1 ) = p_1$ and $\mathbb{P}(k=k_2) = p_2$, where $p_1 + p_2 = 1$. The mixing matrix is then
\begin{equation} \label{equ: intra}
    E = 
    \begin{bmatrix}
    a & \frac{k_1p_1}{\langle k\rangle } - a \\  \frac{k_1p_1}{\langle k\rangle } - a & \frac{k_2p_2 - k_1p_1}{\langle k\rangle }+ a
    \end{bmatrix}\,,
\end{equation}
where $a\in \left[\max\left\{0, \frac{k_1p_1 - k_2p_2}{\langle k\rangle }\right\}, \,\frac{k_1p_1}{\langle k\rangle }\right]$ and $\langle k\rangle$ denotes the mean degree. We calculate the assortativity coefficient $r_{\text{intra}}$, which is given by the Pearson correlation coefficient of the degrees at the two ends of an edge that we choose uniformly at random. Given the mixing matrix \eqref{equ: intra}, the assortativity coefficient $r_{\text{intra}}$ is linear in $a$ and is given by
\begin{equation*}
\begin{split}
    r_{\text{intra}} &= \frac{a - k_1^2p_1^2/\langle k\rangle ^2}{k_1k_2p_1p_2/\langle k\rangle ^2}\,.
\end{split}
\end{equation*}

\begin{figure}[!h]
    \centering
    \begin{subfigure}{\textwidth}
    \centering
    \hspace{0.8cm}
        \includegraphics[scale=0.16, trim = 433 1460 785 0, clip]{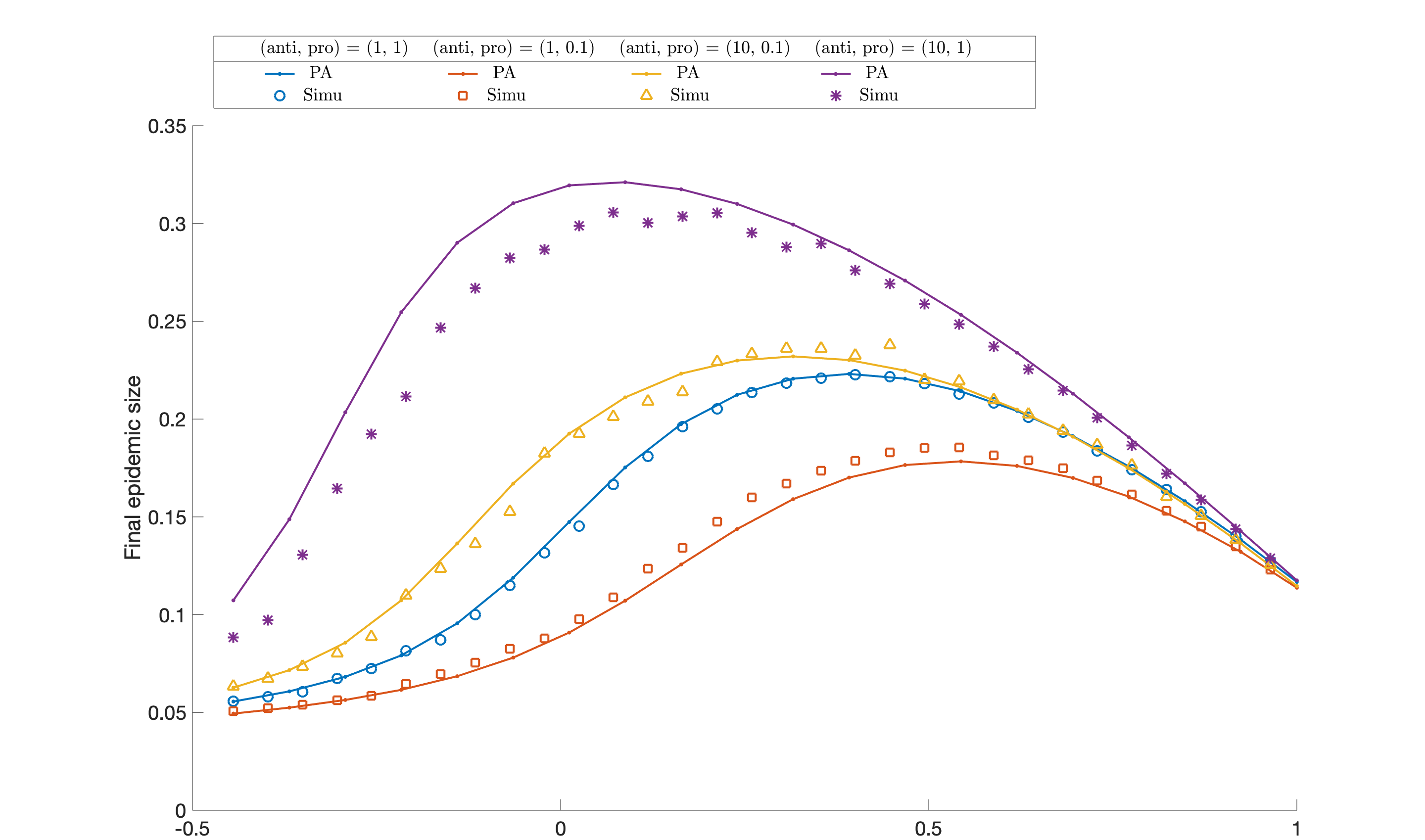}
    \end{subfigure}
    
    \begin{subfigure}{0.45\textwidth}
         \includegraphics[scale=0.16, trim = 0 0 0 30, clip]{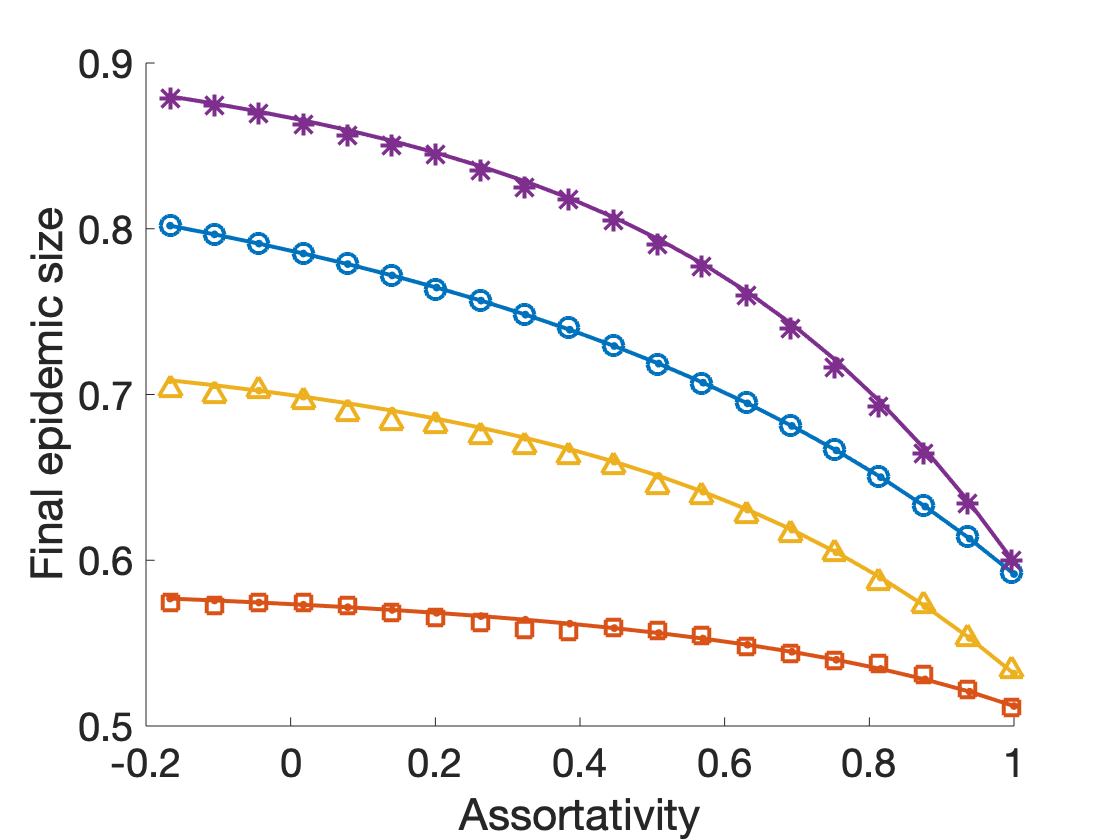}
         \caption{$\pinfo(k=2)=\pphy(k=2)=0.4$}
    \end{subfigure}
    ~~~
    \begin{subfigure}{0.45\textwidth}
    \includegraphics[scale=0.16, trim = 0 0 0 30, clip]{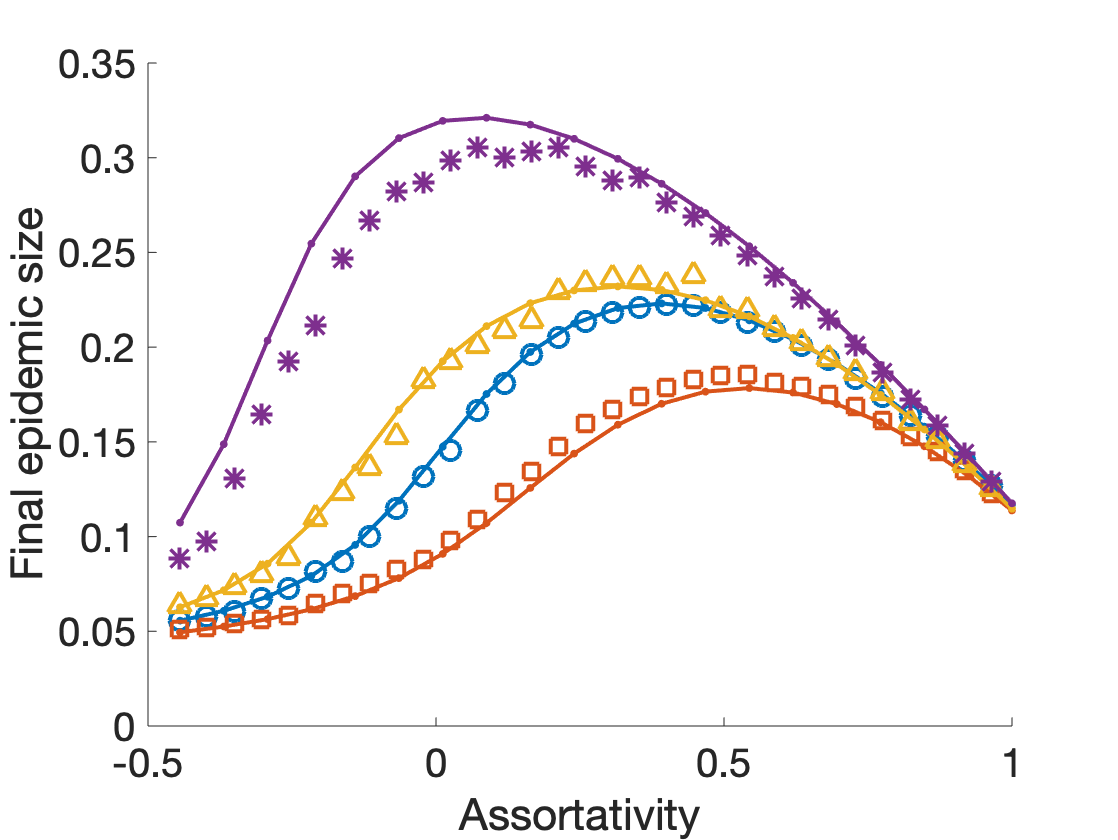}
    \caption{$\pinfo(k=2)=\pphy(k=2)=0.9$}
    \end{subfigure}
    \caption{The final epidemic size depends on the intra-layer degree--degree correlations. We generate each layer independently using a generalization of a configuration-model network with the procedure described in the text. Each layer consists of nodes with degrees $2$ and $8$, and the intra-layer degree--degree correlation is the same in the two layers. The other parameters are $\bphy =\binfo = 0.6$, $\gphy=1$, and $\ginfo=0.1$. The curves (respectively, markers) indicate results from our PA (respectively, direct numerical simulations averaged over 200 simulations).}
    \label{fig: correlated vary distribution}
\end{figure}

\begin{figure}[!h]
    \centering
    \fbox{
        \begin{subfigure}{0.4\textwidth}
        \includegraphics[scale=0.14]{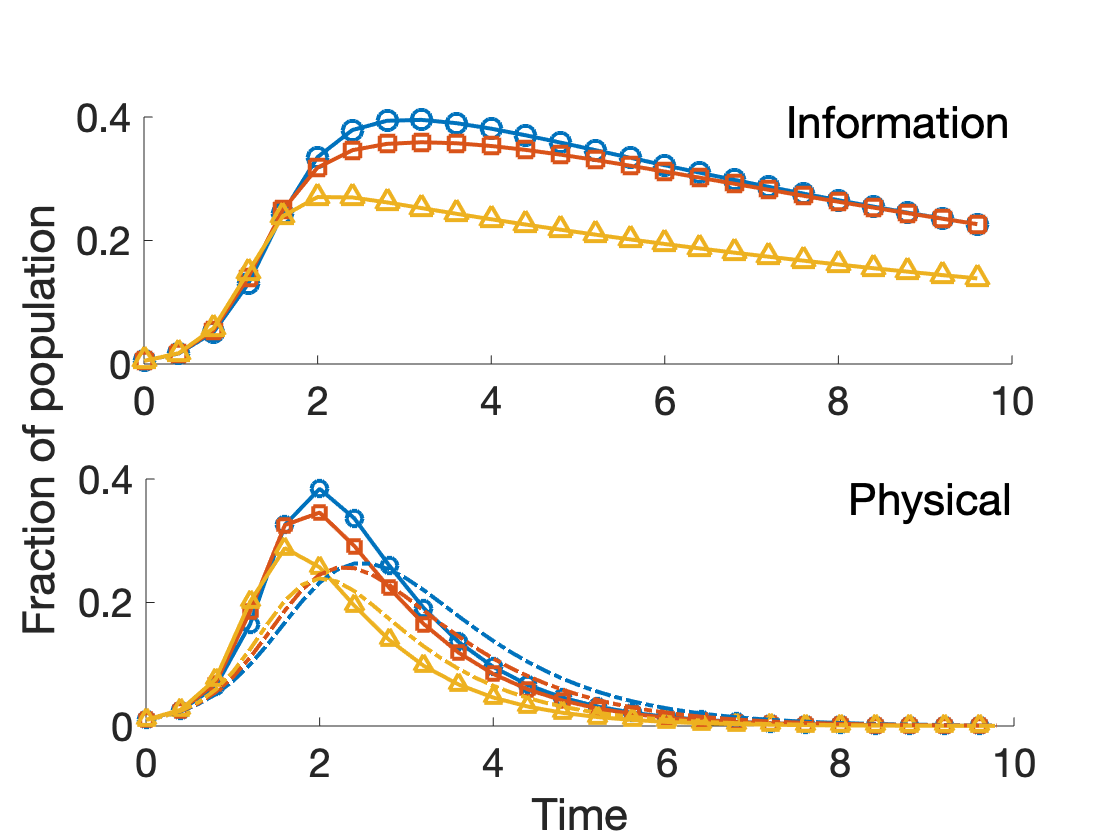}
        \caption{$\alphaa=10\,,\,\alphap=1$}
    \end{subfigure}
    ~~
    \begin{subfigure}{0.4\textwidth}
        \includegraphics[scale=0.14]{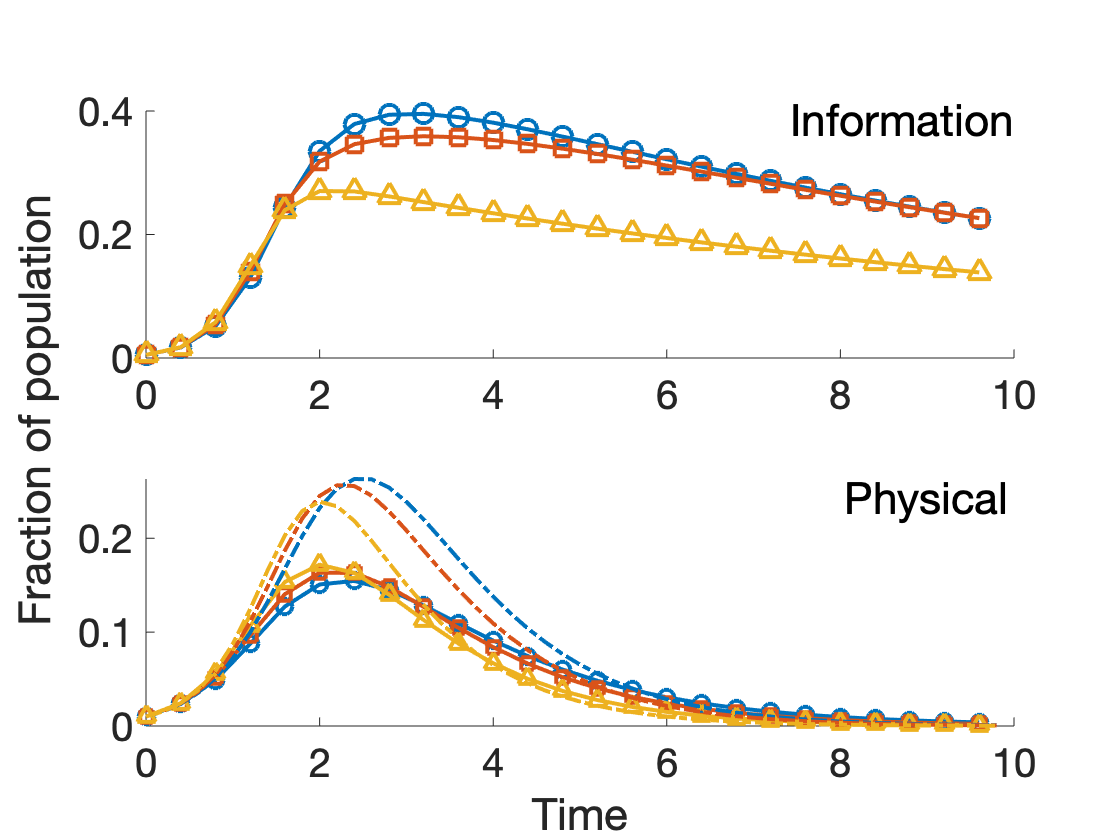}
        \caption{$\alphaa=1\,,\,\alphap=0.1$}
    \end{subfigure}
    }
    \begin{subfigure}{0.1\textwidth}
    \includegraphics[scale=0.15, trim=1050 0 0 100, clip]{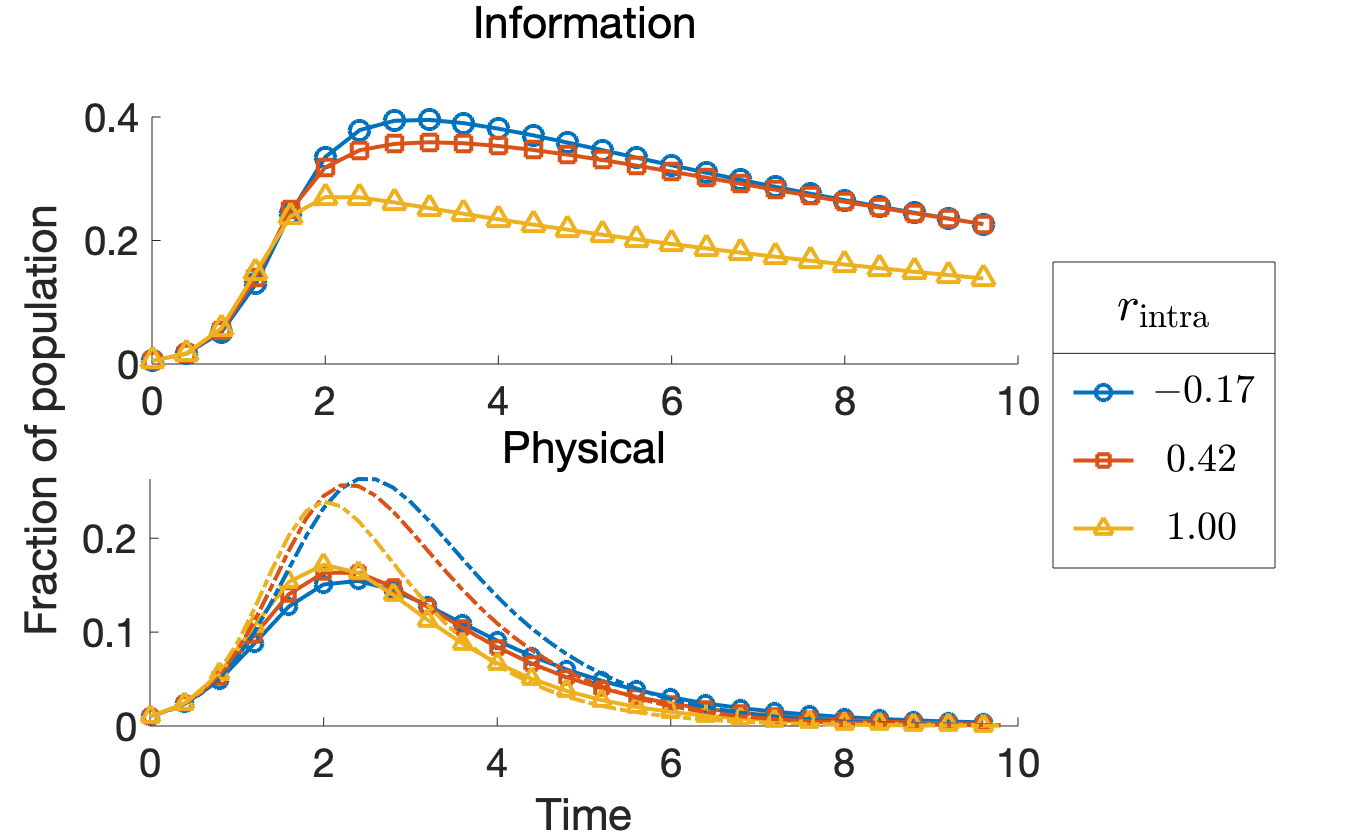}
    \end{subfigure}
    \caption{Incidence curves from our PA for different intra-layer degree--degree correlations in networks with $\pinfo(k=2)=\pphy(k=2)=0.4$ and $\pinfo(k=8)=\pphy(k=8)=0.6$. We generate each layer independently using a generalization of a configuration-model network with a procedure described in the text. The upper row shows the (identical) dynamics of the fraction of individuals in the $P$ and $A$ compartments. The lower row shows the dynamics of the population in the $I$ compartment. The solid curves show results when (left) $\alphaa=10$ and $\alphap=1$ and (right) $\alphaa=1$ and $\alphap=0.1$. The dashed curves indicate results without opinion contagions. Curves with the same color share all parameters except opinion influence coefficients. The other parameters are $\bphy =\binfo = 0.6$, $\gphy=1$, and $\ginfo=0.1$. }\label{fig: intra_dynamics}
\end{figure}

Figure \ref{fig: correlated vary distribution} shows two typical sets of curves for the final epidemic size for different values of the intra-layer degree--degree correlation. We fix $k_1=2$ and $k_2=8$, and we assign $40\%$ nodes to have degree $2$ in Figure \ref{fig: correlated vary distribution}(a) and $90\%$ to have degree $2$ in Figure \ref{fig: correlated vary distribution}(b). We set the intra-layer degree--degree correlations to be the same in the two layers. The blue circles indicate the influence of degree assortativity on disease spreading when the disease spreads independently of opinions. The decreasing trend in Figure \ref{fig: correlated vary distribution}(a) and in the right part of Figure \ref{fig: correlated vary distribution}(b) is consistent with the known result\cite{moreno2003epidemic, kiss2008effect} that  the infection tends to affect a smaller fraction of a population in an assortative network than in a disassortative network when a disease outbreak occurs. The increasing trend in the left part of Figure \ref{fig: correlated vary distribution}(b) arises from the fact that the disease is initially impeded from spreading because of disassortative structures and a denser network helps the disease to spread and persist. Similar trends also occur in the information layer, so if outbreaks do occur on both layers, an opinion contagion has a smaller impact when networks have a larger degree assortativity (as we see in both panels of Figure \ref{fig: correlated vary distribution}). In Figure \ref{fig: intra_dynamics}, we compare the disease prevalence curves as we turn on and off the influence from a specific opinion on networks with $\pinfo(k=2)=\pphy(k=2)=0.4$. The results demonstrate that disassortative structures tend to enhance the influence of both pro- and anti-physical-distancing opinions. When both opinions have nontrivial effects on the transmission of a disease, the overall effect of the opinion dynamics on the disease dynamics is a complicated combination of the dynamics of the opinions; in this situation, it is unclear whether an assortative or a disassortative structure is better for the spread of the disease.   

\begin{figure}[!h]
    \centering
    \begin{subfigure}{\textwidth}
      \includegraphics[scale=0.21]{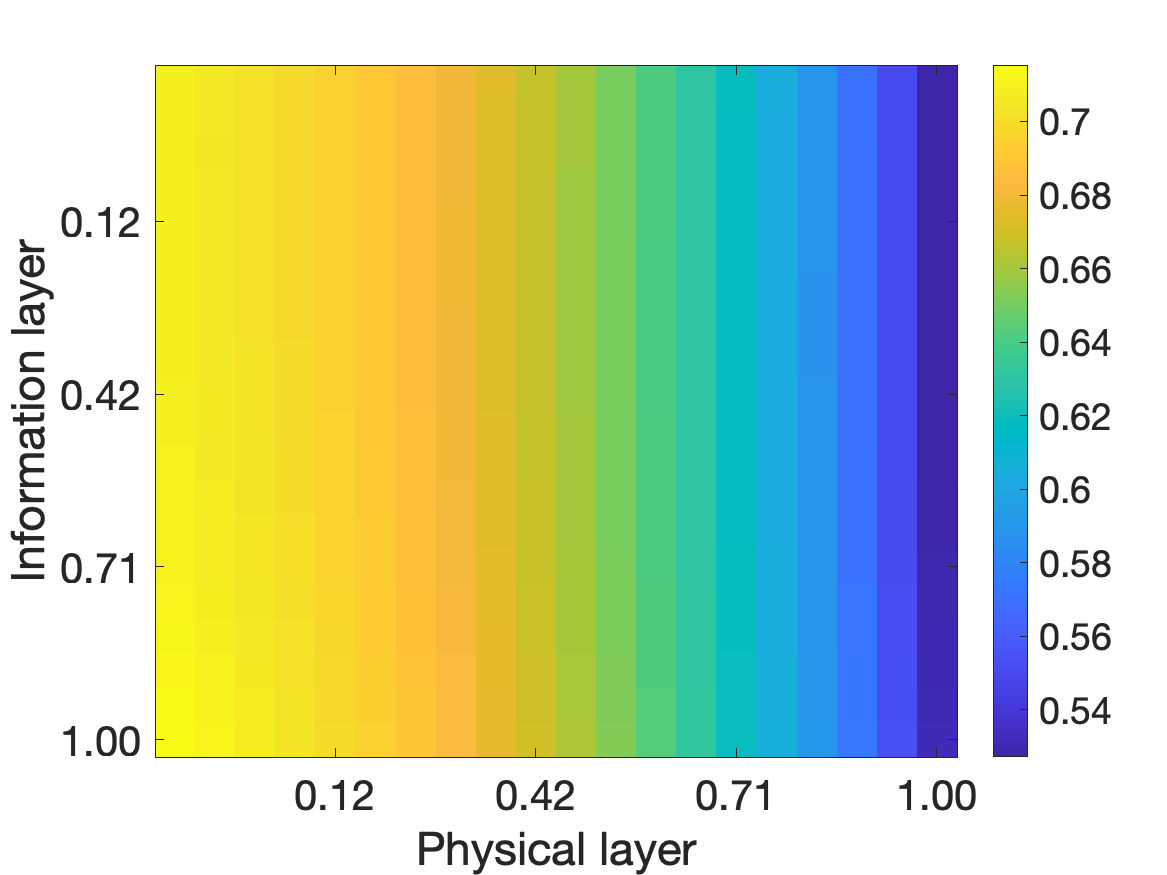}
    \includegraphics[scale=0.21]{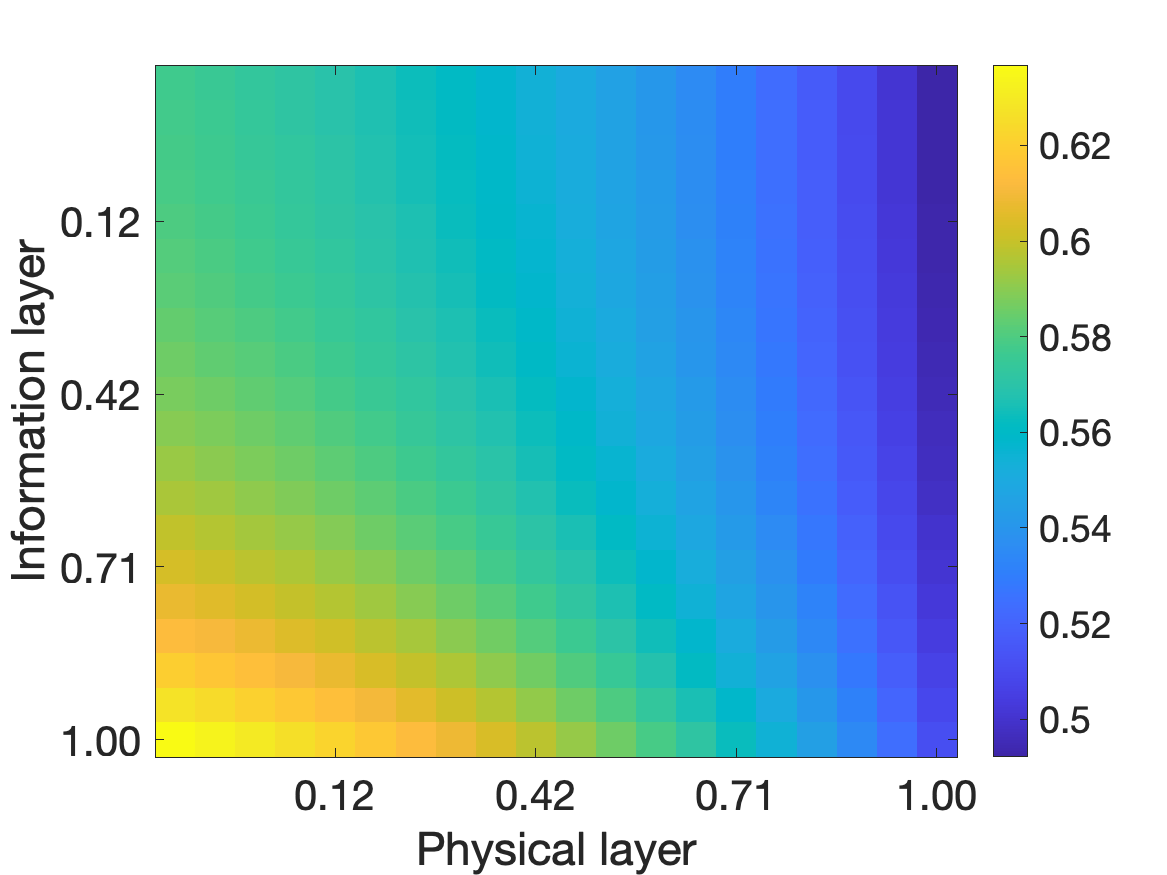} \includegraphics[scale=0.21]{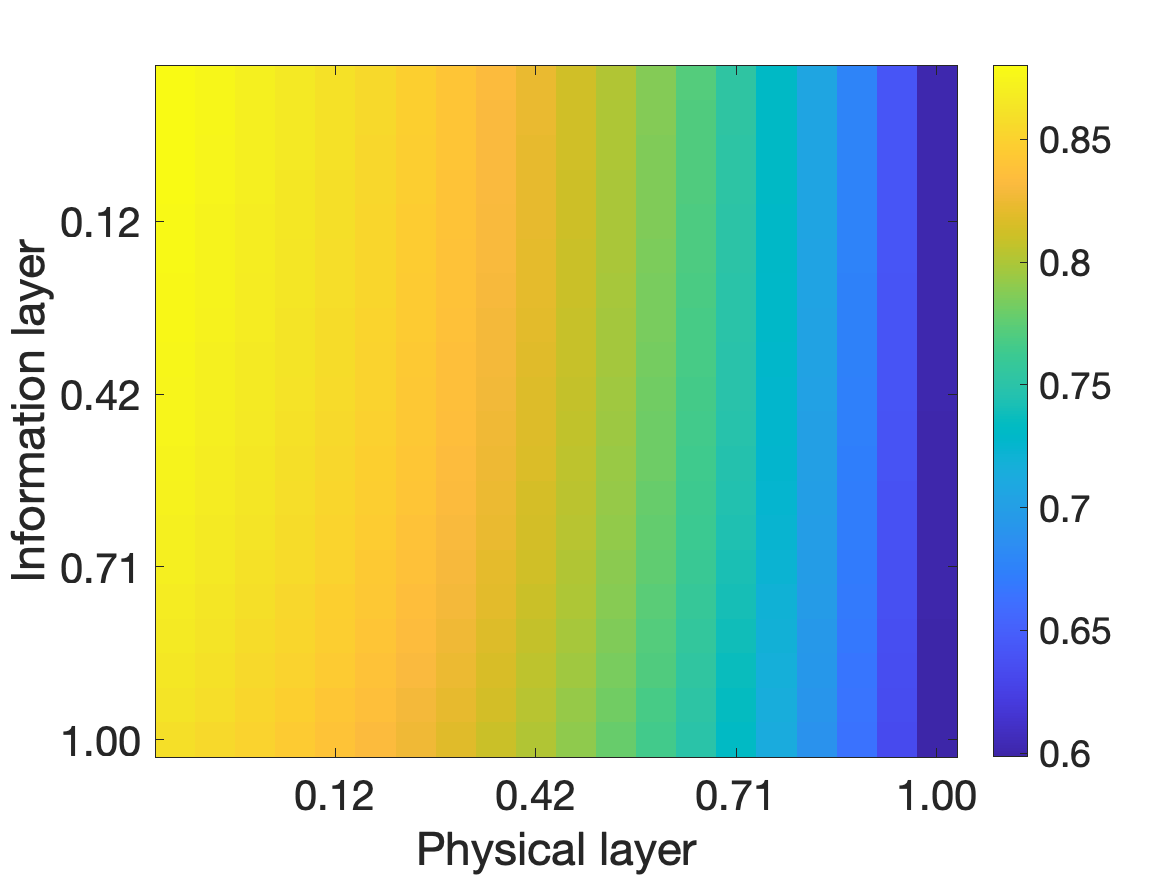}
    \caption{$\pinfo(k=2)=\pphy(k=2)=0.4$}
    \end{subfigure}
    \begin{subfigure}{\textwidth}
       \includegraphics[scale=0.21]{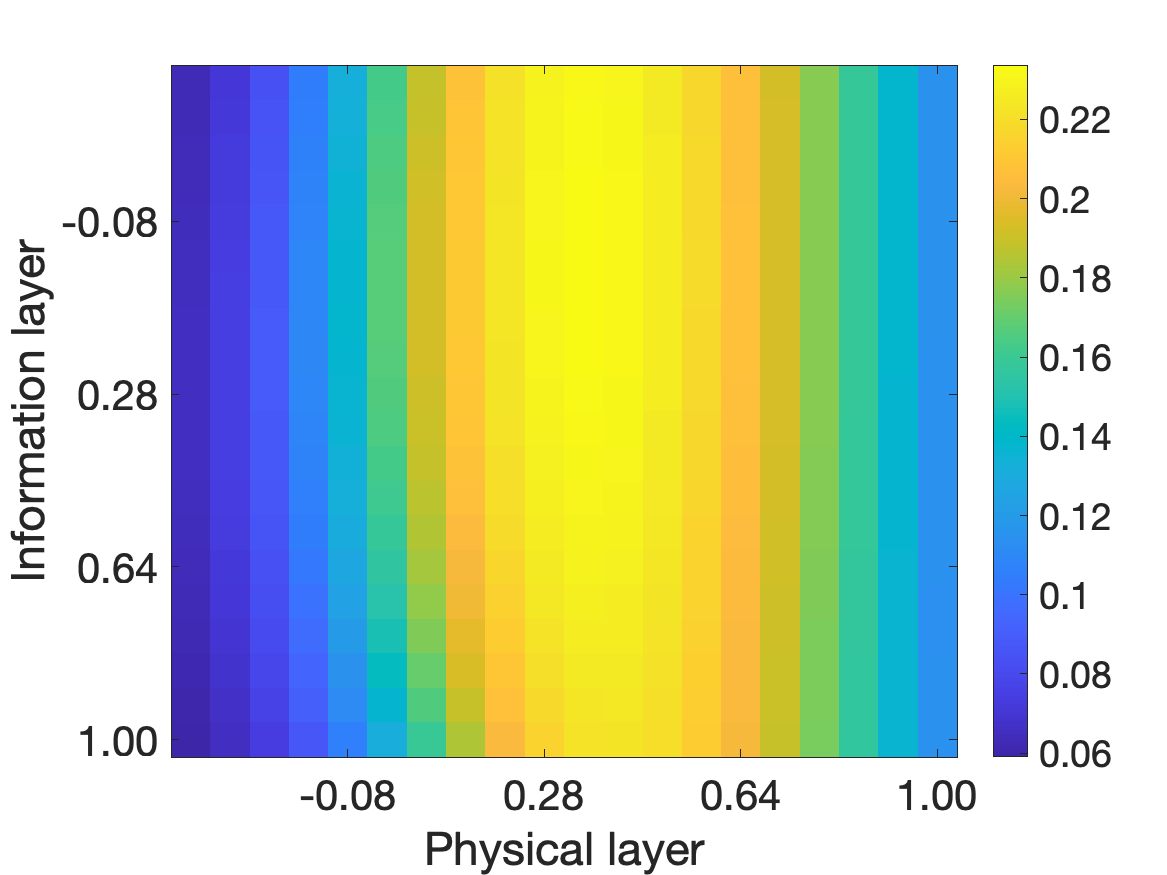} \includegraphics[scale=0.21]{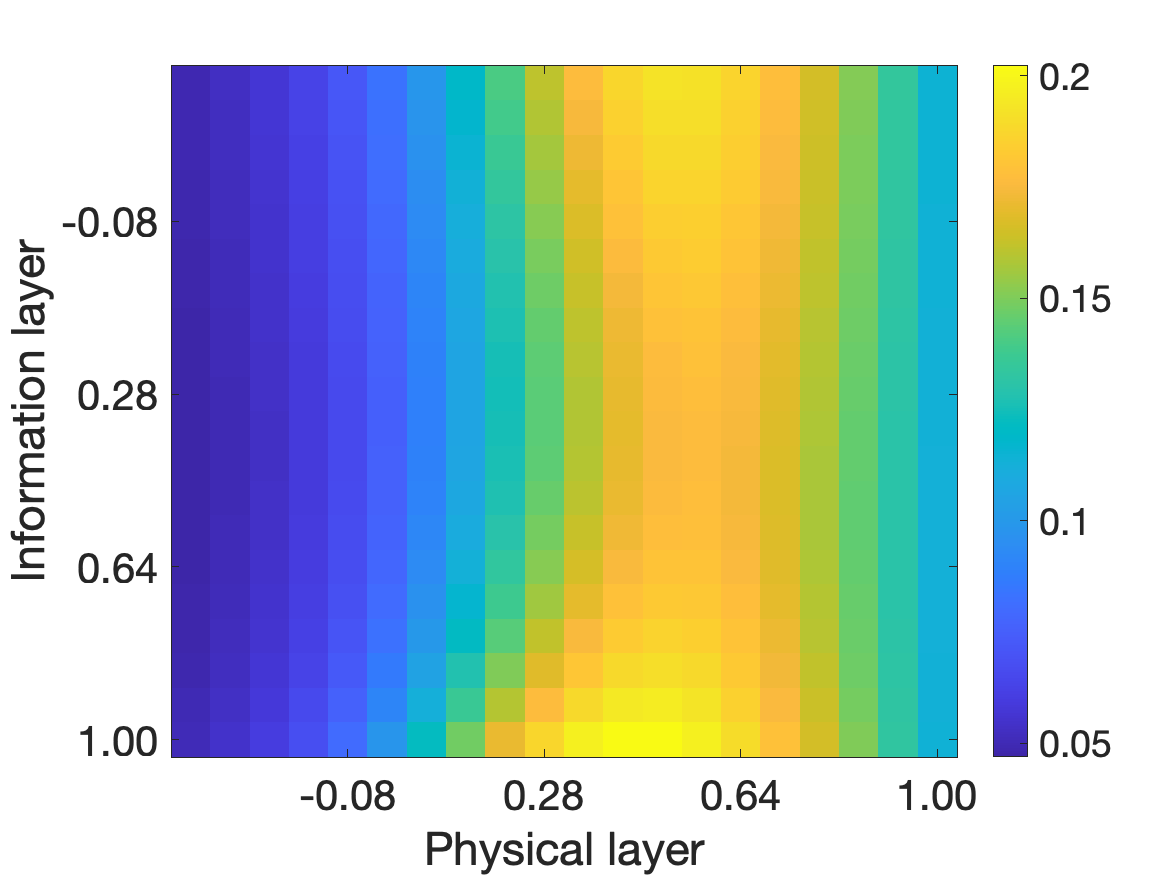} \includegraphics[scale=0.21]{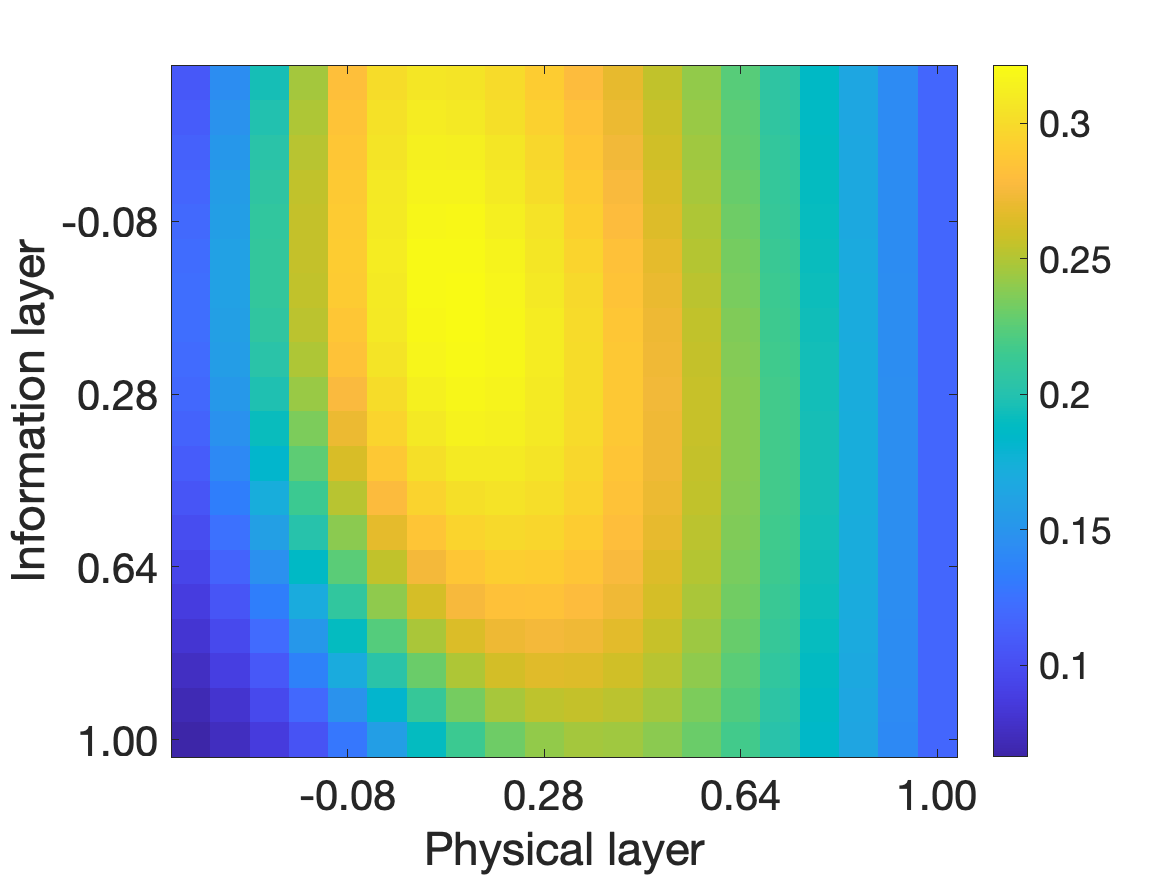}
    \caption{$\pinfo(k=2)=\pphy(k=2)=0.9$}
    \end{subfigure}
    \begin{subfigure}{\textwidth}
         \includegraphics[scale=0.21]{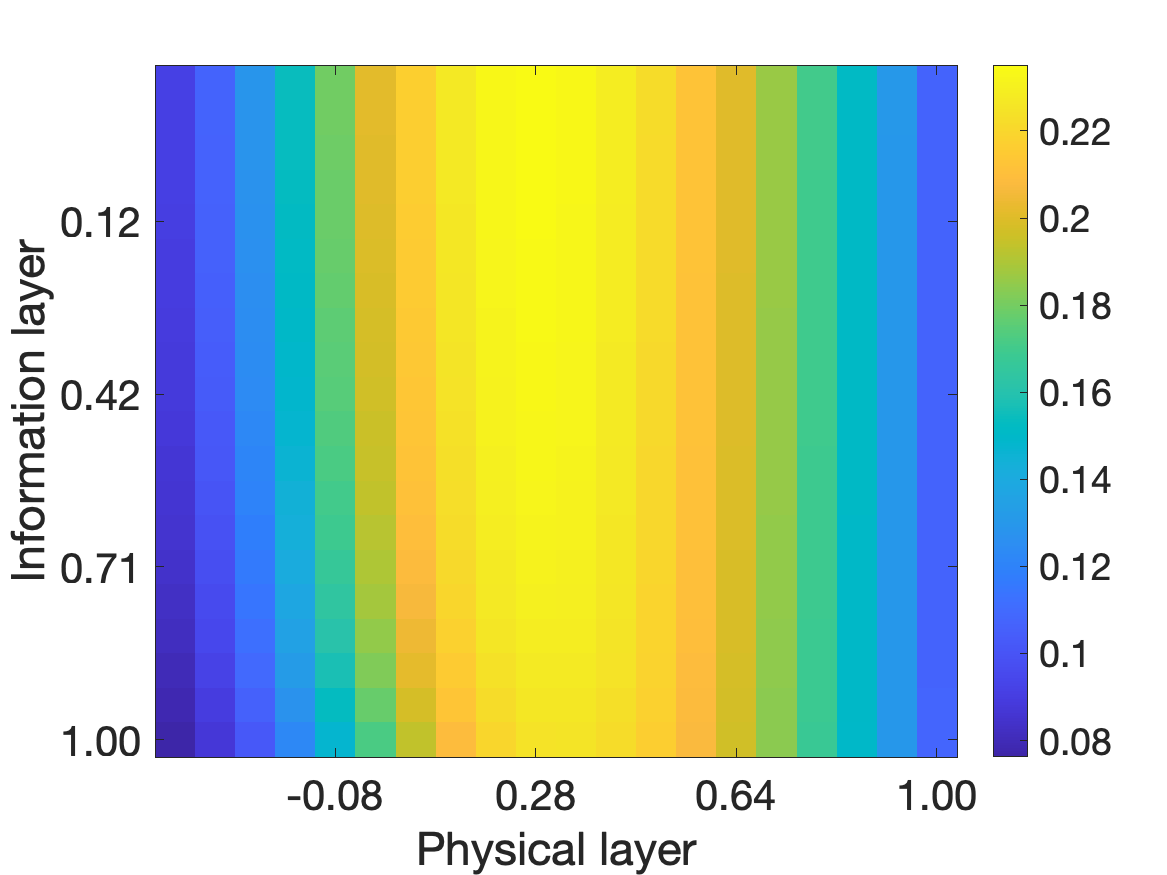} \includegraphics[scale=0.21]{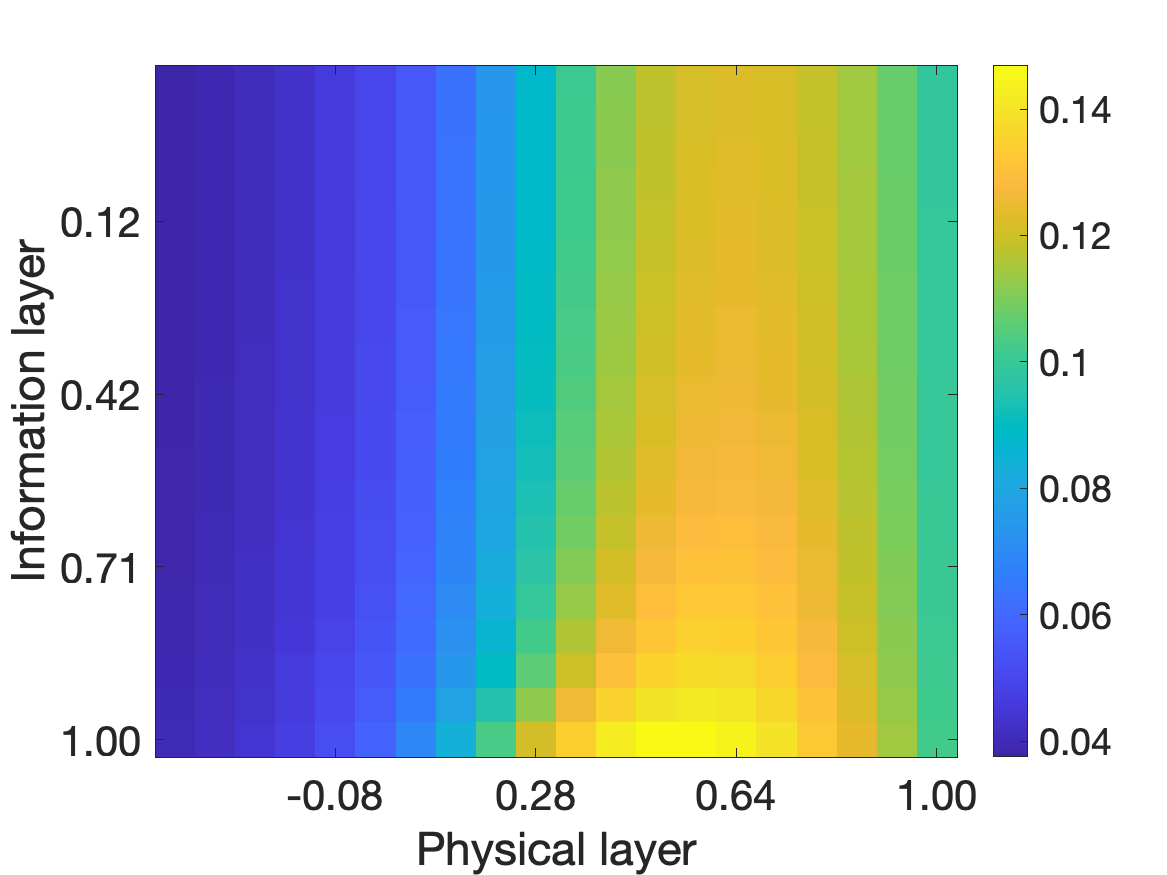} \includegraphics[scale=0.21]{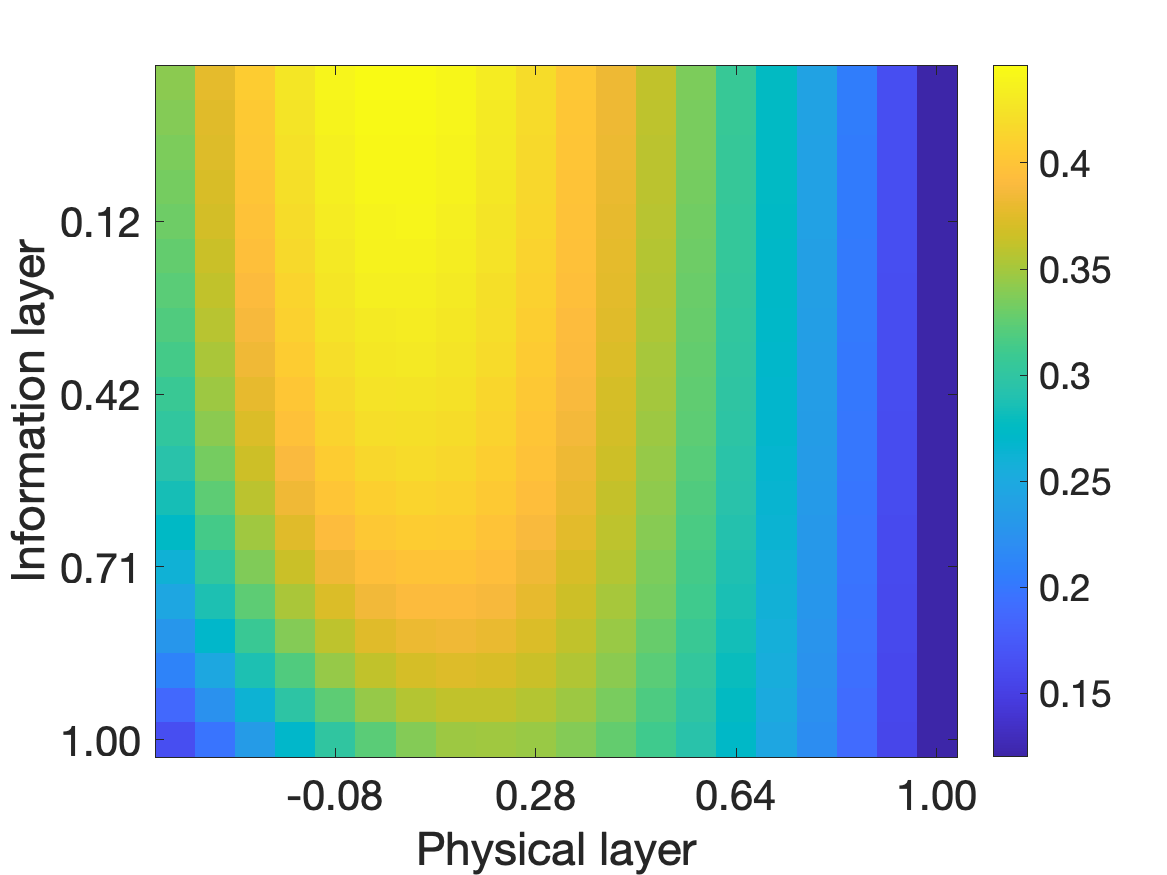}
    \caption{$\pinfo(k=2)=0.4$, $\pphy(k=2)=0.9$ }
    \end{subfigure}
    \caption{Heat maps of the final epidemic size from our PA as we vary the assortativities in the two layers. The three columns have parameter values of (left) $\alphap = 0.1$ and $\alphaa= 10$, (center) $\alphap =0.1$ and $\alphaa= 1$, and (right) $\alphap =1$, and $\alphaa= 10$. The other parameters are $\bphy =\binfo = 0.6$, $\gphy=1$, and $\ginfo=0.1$. }\label{fig: intra_final_size_matrix}
\end{figure}

The intra-layer degree--degree correlations in the two layers need not be the same. Figure \ref{fig: intra_final_size_matrix} shows heat maps of the final epidemic size for different values of the two degree--degree correlations, which we vary independently in each layer. The issuance of a stay-at-home order may lead to a physical layer with many small-degree nodes, and such an order is not likely to affect the information layer (which may describe online contacts). Therefore, we also consider the case with $\pinfo(k=2)=0.4$ and $\pphy(k=2)=0.9$. We show the simulation results in the third row of Figure \ref{fig: intra_final_size_matrix}. In this example, when both physical-distancing opinions have a nontrivial influence as specified by the influence coefficients, the physical-layer network structures dominate the effect on disease dynamics.

\subsection{Random graphs with cross-layer correlations of intra-layer degrees}\label{sec: inter_layer}

We also investigate the influence of cross-layer correlations of intra-layer degrees on the dynamics. We refer to such correlations as ``inter-layer degree--degree correlations''. People who are active on social-media platforms may also have frequent offline social contacts, and vice versa\cite{acar2008antecedents}. Let $C$ denote the inter-layer degree--degree correlation matrix, so $C_{k_1,\, k_2}$ is the probability that a node that we choose uniformly at random has degree $k_1$ in the information layer and degree $k_2$ in the physical layer. We say that these nodes are ``of type $(k_1,\,k_2)$''. An uncorrelated model corresponds to $C_{k_1,\, k_2} = \pinfo(k_1) \pphy(k_2)$. We uniformly randomly pair $N\times C_{k_1,\,k_2}$ degree-$k_1$ nodes from the information layer with the same number of degree-$k_2$ nodes from the physical layer to construct a desired network with $N$ nodes and a specified inter-layer degree--degree correlation.

As with the situation in Section \ref{sec: intra_layer}, a PA can deal with inter-layer degree--degree correlations properly as long as we build them into the initial values. The modification is straightforward. For example, we write
\begin{equation*}
    \begin{split}
        [\US](0) &= \Cm(1 - A_0 - P_0)(1 - I_0)\,,\\
        [\USI](0) &= [S_{k_2}\circ I_l](0)\times \frac{\Cm}{\pphy(k_2)}\times (1 - A_0 - P_0)\,.
    \end{split}
\end{equation*}
We use similar formulas for the other pairs.

\subsubsection{Pedagogical example: Networks whose nodes have one of two different degrees}

We again suppose that nodes have one of two different degrees in each layer. These degrees are $k_{\text{info},\, 1}$, $k_{\text{info},\,2}$, $k_{\text{phy},\,1}$, and $ k_{\text{phy},\,2 }$, where ${\mathbb P}_{\text{info}}(k= k_{\text{info},\, 1}) = q_1$ and ${\mathbb P}_{\text{phy}}(k= k_{\text{phy},\, 1}) = q_2$. The correlation matrix $C$ is
\begin{equation}
    \begin{bmatrix}
        a & q_1 - a\\
        q_2 - a & 1 - q_1 - q_2 + a
    \end{bmatrix}\,,\label{equ: inter}
\end{equation}
where $a \in [\min\{0,\, q_1 + q_2 -1\},\, \min\{q_1, \,q_2\}]$. The Pearson correlation coefficient is 
\begin{equation*}
    r_{\text{inter}} = \frac{(k_{\text{info},\, 1} - k_{\text{info},\, 2})(k_{\text{phy},\, 1} - k_{\text{phy},\, 2})(a-q_1q_2)}{\sigma_{\text{info}}\sigma_{\text{phy}}}\,,
\end{equation*}
where $\sigma_{\text{info}}$ and $\sigma_{\text{phy}}$ denote the standard deviations of the degrees in the two layers.

\begin{figure}[!h]
\centering
\begin{subfigure}{\textwidth}
\hspace{0.5cm}
        \includegraphics[scale=0.15, trim = 250 1410 0 0, clip]{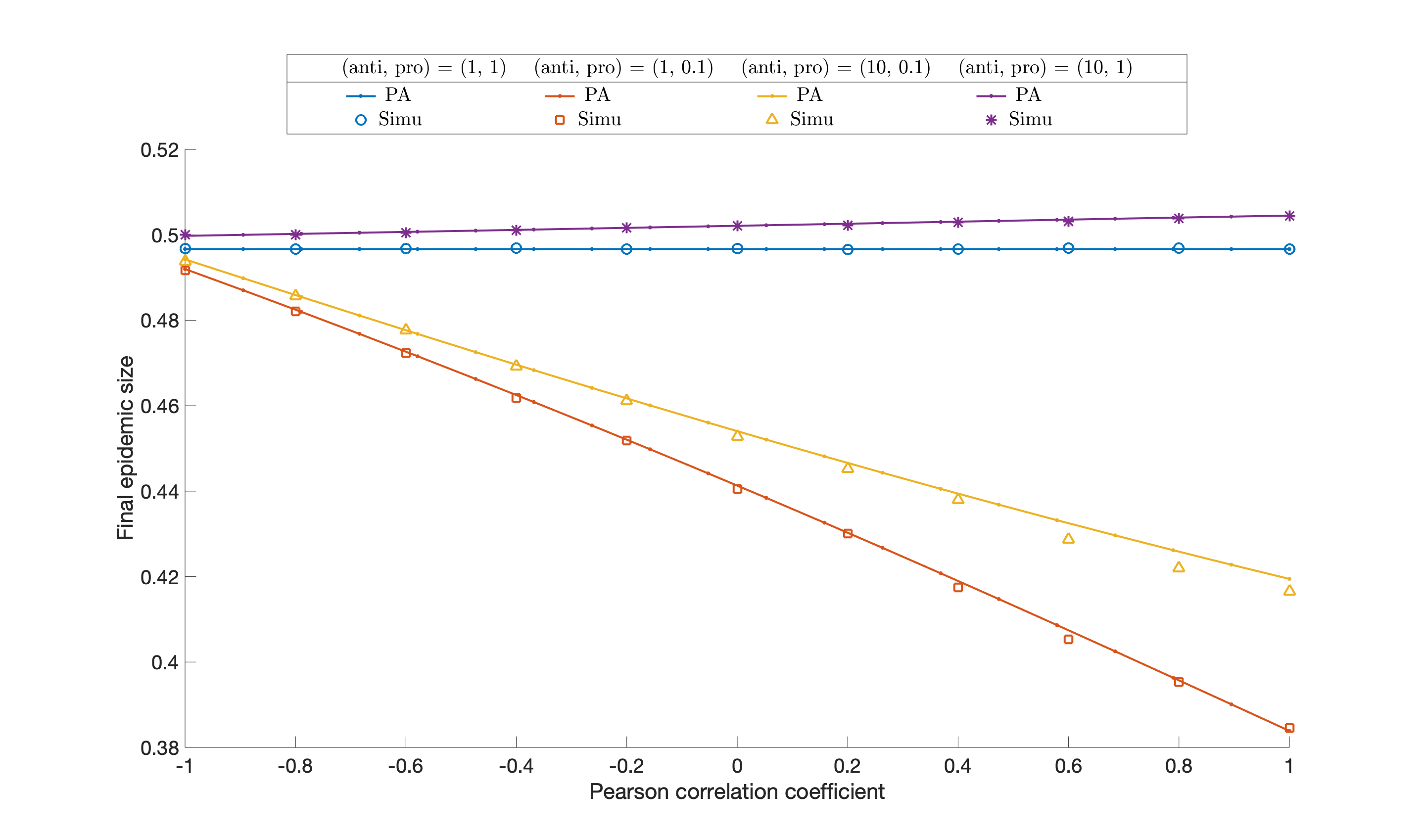}
    \end{subfigure}
    \begin{subfigure}{0.45\textwidth}
        \includegraphics[scale=0.15,trim=0 0 0 0, clip]{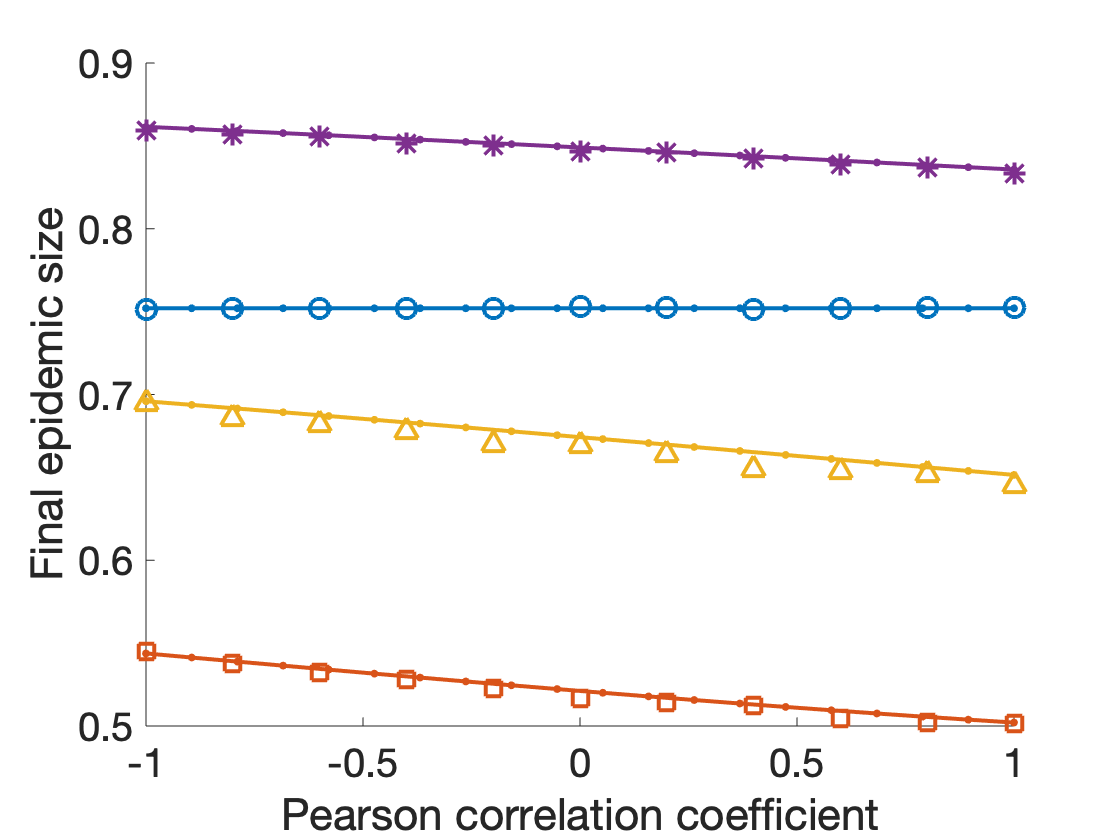}
        \caption{$r_\text{intra} = -0.25$}\label{fig: inter_correlation_a}
    \end{subfigure}
    ~~~
    \begin{subfigure}{0.45\textwidth}
        \includegraphics[scale=0.15]{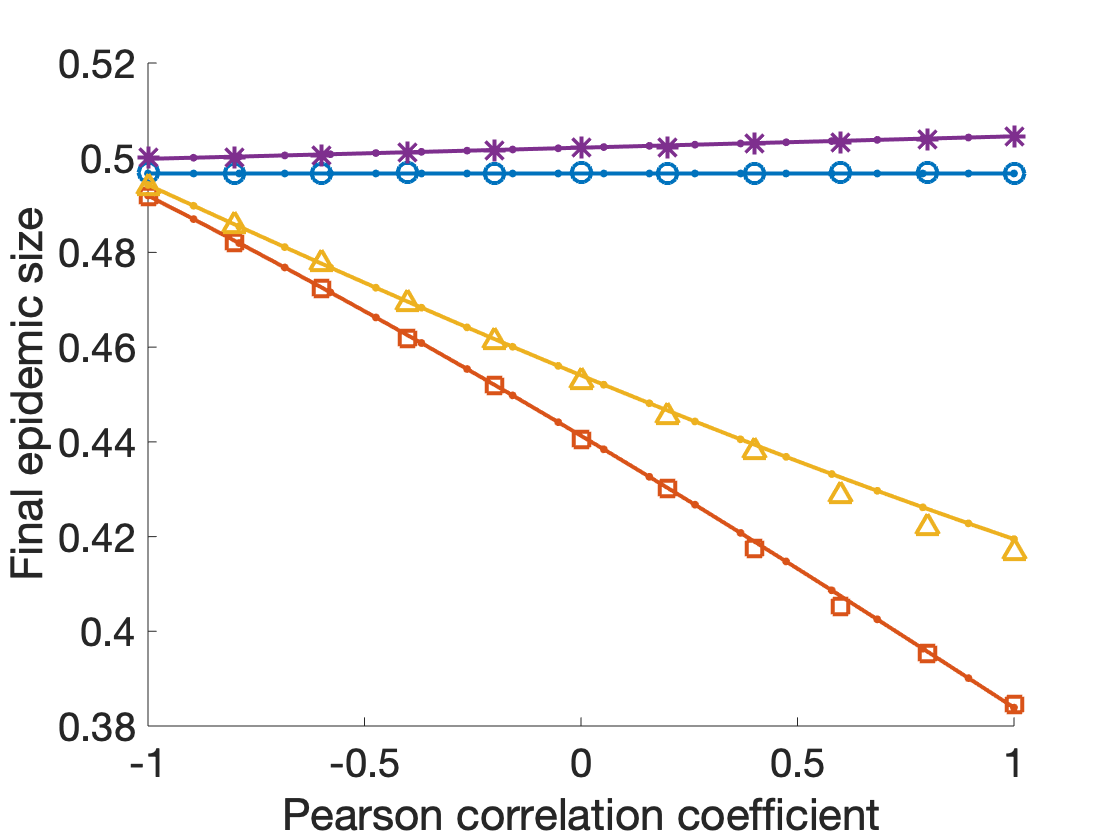}
        \caption{$r_\text{intra} = 1$}\label{fig: inter_correlation_b}
    \end{subfigure}
    \caption{The final epidemic size depends on the inter-layer degree--degree correlation. Both layers have nodes of degrees $2$ and $8$, and we set $\pinfo(k=2)=\pphy(k=2)=0.5$. We generate each layer using the procedure in Section \ref{sec: intra_layer}. We couple the two layers following the approach in Section \ref{sec: inter_layer}. We set the intra-layer degree--degree correlation of both layers to be (a) $-0.25$ and (b) $1$. The other parameters are $\bphy =\binfo = 0.6$, $\gphy=1$, and $\ginfo=0.1$. The curves (respectively, markers) show results from our PA (respectively, direct simulations averaged over 200 simulations).}
    \label{fig: inter-correlation}
\end{figure}

Figure \ref{fig: inter-correlation} shows the dependence of the final epidemic size on inter-layer degree--degree correlations. Each of the two layers has nodes of degrees $2$ and $8$, and we set $\pinfo(k=2)=\pphy(k=2)=0.5$. We generate each layer independently with a generalization of a configuration-model network following the procedure in Section \ref{sec: intra_layer}. We consider cases in which the intra-layer degree--degree correlation is $-0.25$ and $1$. We couple the two layers as described above. The pro-physical-distancing opinion has a larger influence when the two layers are more positively correlated. (See the red curves with square markers.) However, the anti-physical-distancing opinion's influence can either decrease or increase as we increase the inter-layer degree--degree correlation. (See the purple curves with asterisk markers.)

\begin{figure}
    \centering
    \fbox{
    \begin{subfigure}{0.4\textwidth}
          \includegraphics[scale=0.13]{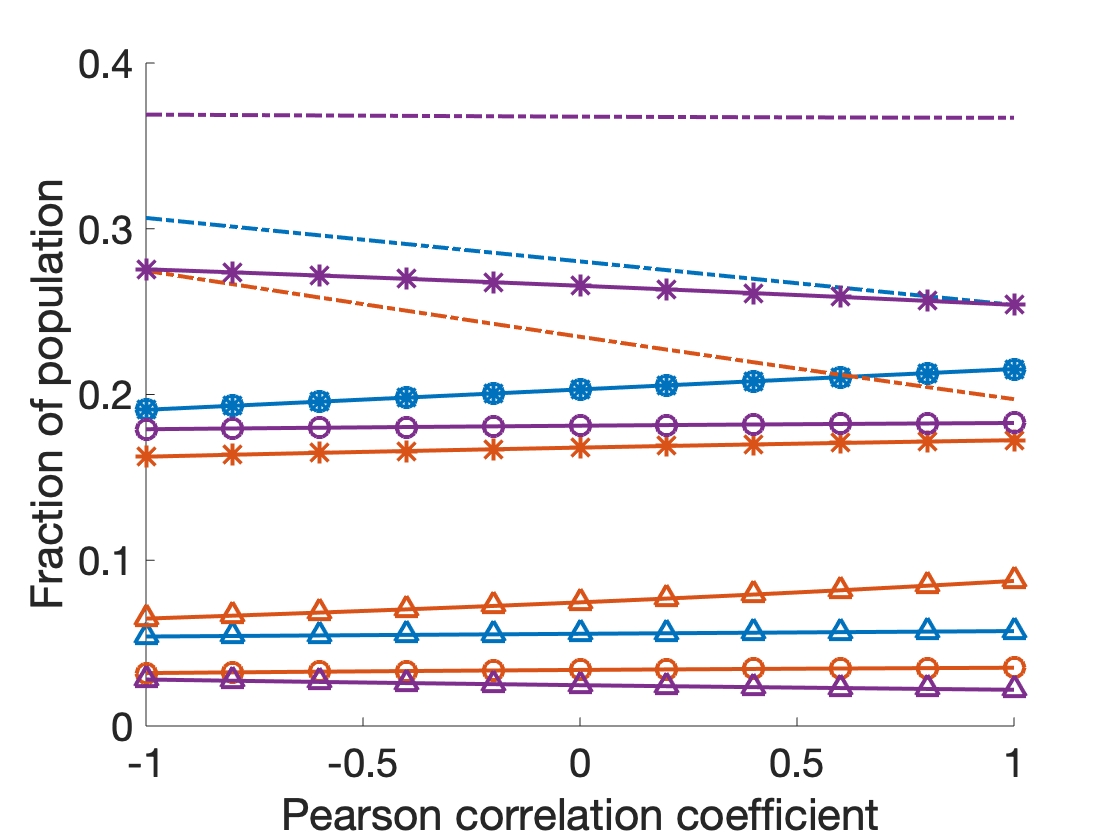}
          \caption{$r_\text{intra} = -0.25$}\label{fig: inter_awareness_a}
    \end{subfigure}
    \begin{subfigure}{0.4\textwidth}
         \includegraphics[scale=0.13]{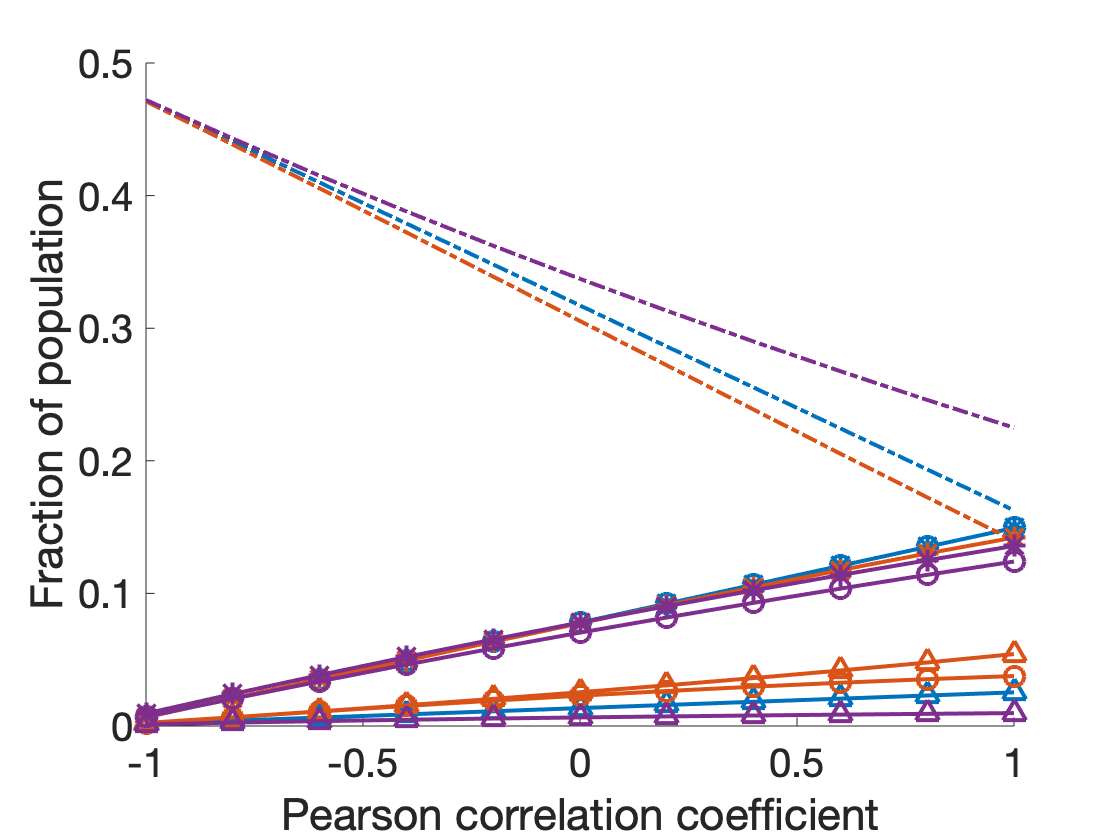}
         \caption{$r_\text{intra} = 1$}\label{fig: inter_awareness_b}
    \end{subfigure}
    }
    \hspace{-0.2cm}
    \begin{subfigure}{0.15\textwidth}
    \includegraphics[scale=0.16, trim = 600 60 70 30, clip]{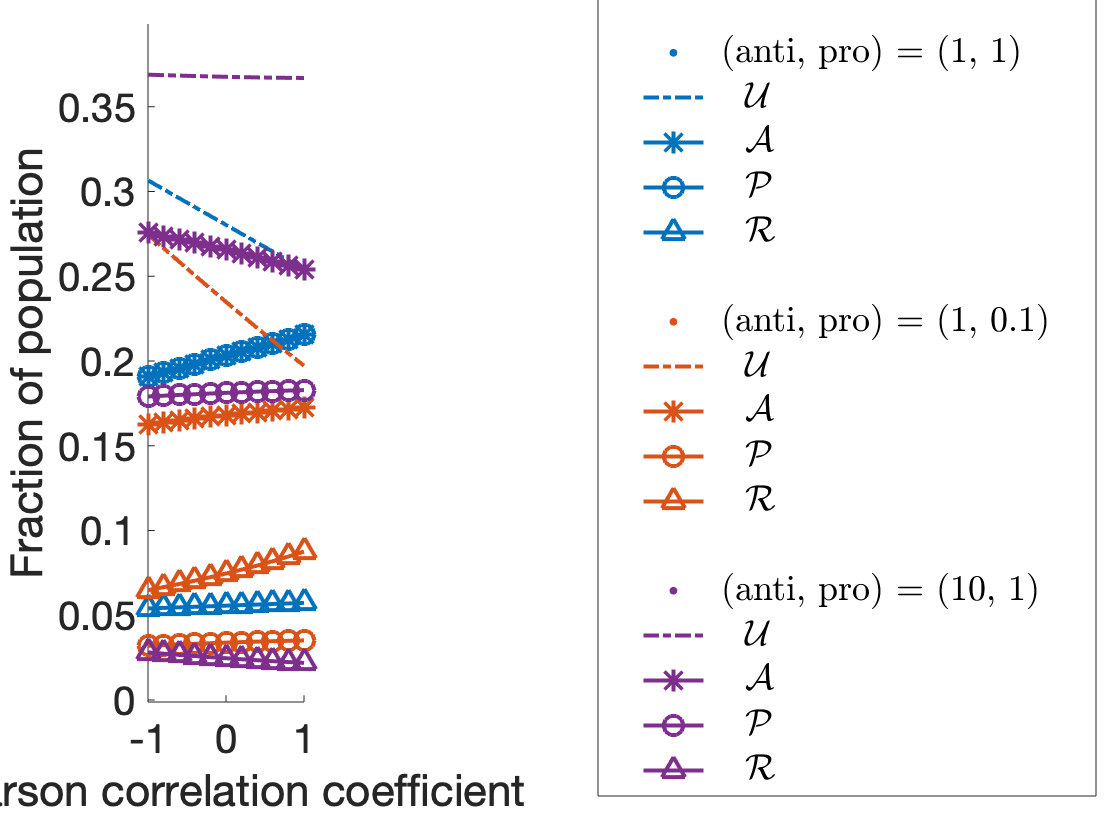}
    \end{subfigure}
    \caption{Decomposition of opinion states for nodes that eventually become infectious and recover. We group the recovered population based on their opinion states when they become infectious. Recall that the corresponding subpopulations are $\cal{U}$, $\cal{P}$, $\cal{A}$, and $\cal{R}$. The vertical axis indicates the fraction of the population in each of these subpopulations. We plot the subpopulation sizes versus the inter-layer degree--degree correlation from our PA. The intra-layer degree--degree correlation is (a) $-0.25$ and (b) $1$. The other parameters are $\bphy =\binfo = 0.6$, $\gphy=1$, and $\ginfo=0.1$.}\label{fig: inter_awareness}
\end{figure}

To understand the trends in Figure \ref{fig: inter-correlation}, we again decompose the population based on their opinion states. Recall that $\cal{U}$, $\cal{A}$, $\cal{P}$, and $\cal{R}$ denote the subpopulations that become infectious when they are in the $U$, $A$, $P$, and $R_{\text{info}}$ states, respectively. It is instructive to consider the case of two independent layers (i.e., $\alphap = \alphaa=1$). The inter-layer degree--degree correlation changes the opinion distributions, but the final epidemic size remains constant. We investigate the influence of opinions and how such influence depends on the inter-layer degree--degree correlation.  

\begin{figure}[t!h]
    \centering
    \begin{minipage}{0.87\textwidth}
     \begin{subfigure}{0.43\textwidth}
      \includegraphics[scale=0.12]{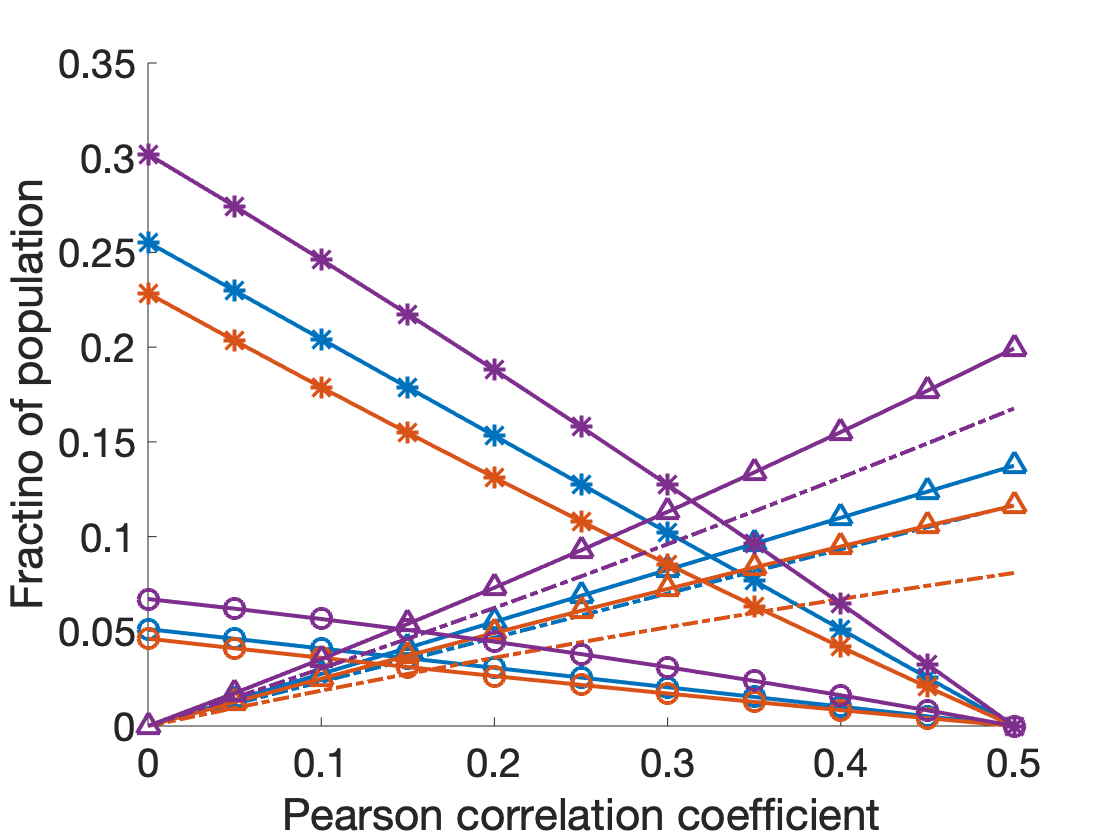}
      \caption{Decomposition of the $\cal{U}$ subpopulation when the intra-layer degree--degree correlation is $-0.25$}\label{fig: inter_degree_a}
     \end{subfigure}
        ~~~
     \begin{subfigure}{0.43\textwidth}
      \includegraphics[scale=0.12]{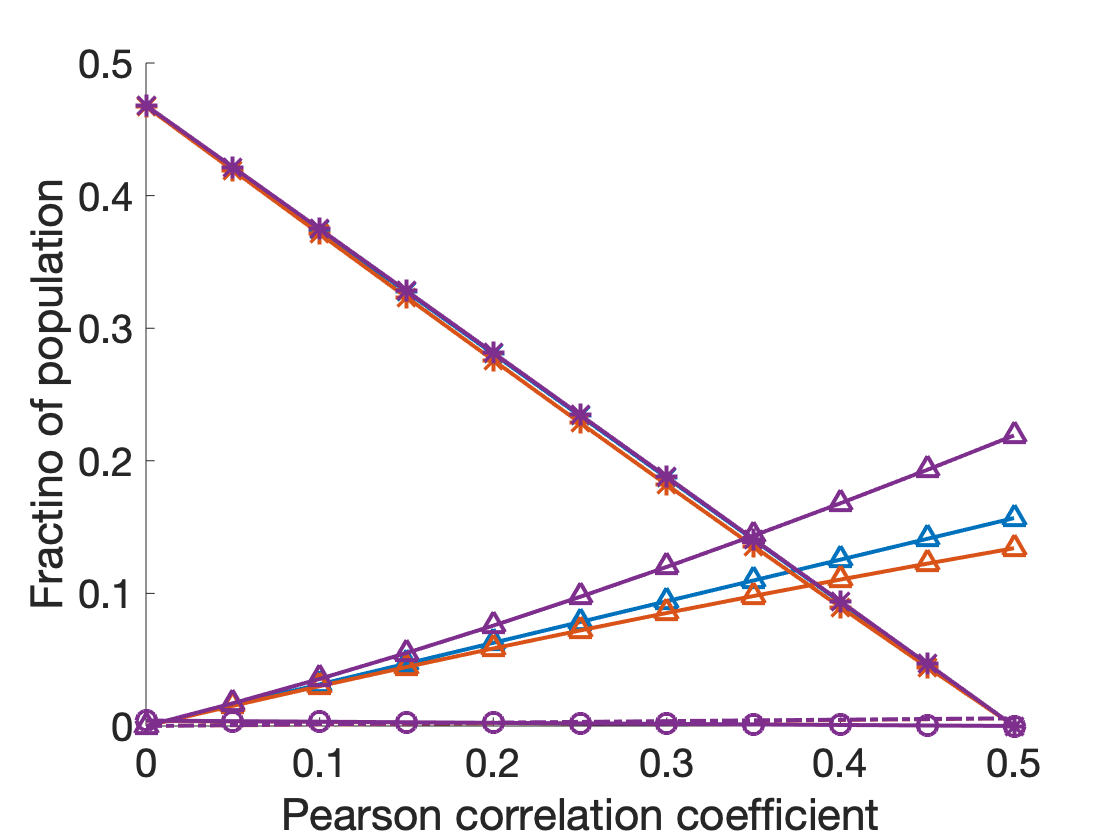}   
      \caption{Decomposition of the $\cal{U}$ subpopulation when the intra-layer degree--degree correlation is $1$}\label{fig: inter_degree_b}
     \end{subfigure}
    
     \begin{subfigure}{0.43\textwidth}
      \includegraphics[scale=0.12]{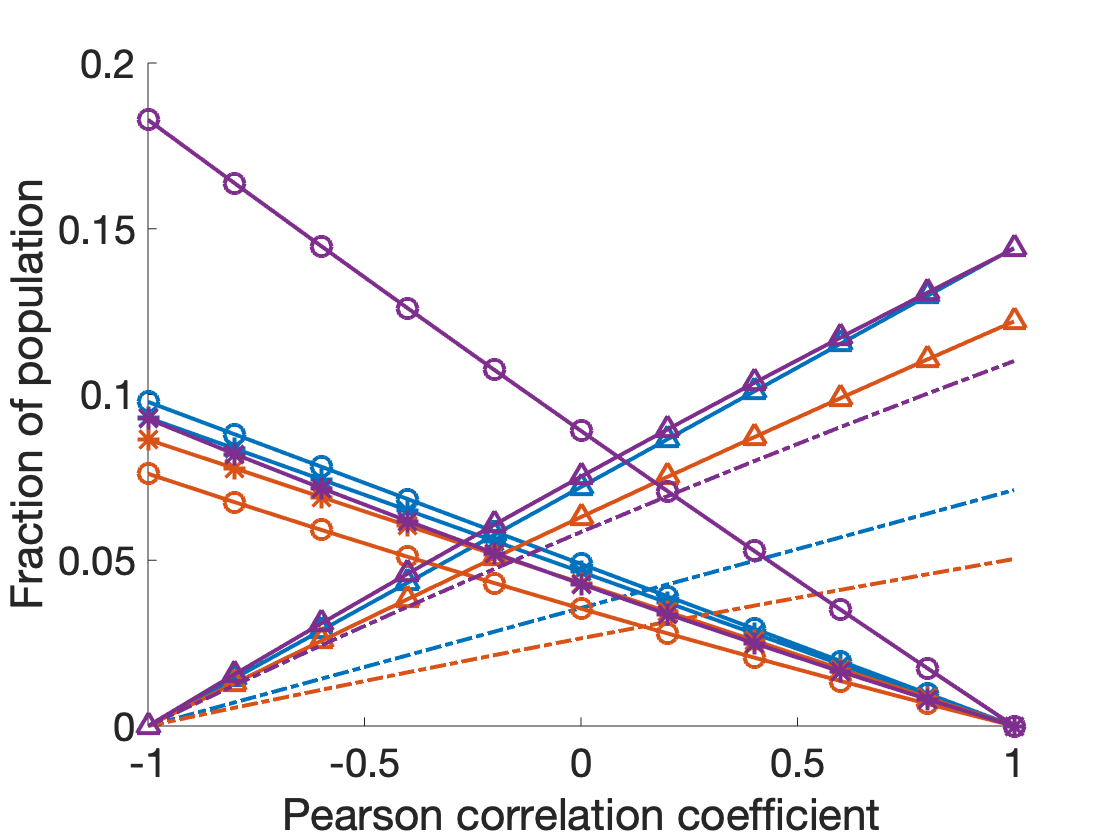}
      \caption{Decomposition of the $\cal{A}$ subpopulation when the intra-layer degree--degree correlation is $-0.25$}\label{fig: inter_degree_c}
     \end{subfigure}
        ~~~
     \begin{subfigure}{0.43\textwidth}
      \includegraphics[scale=0.12]{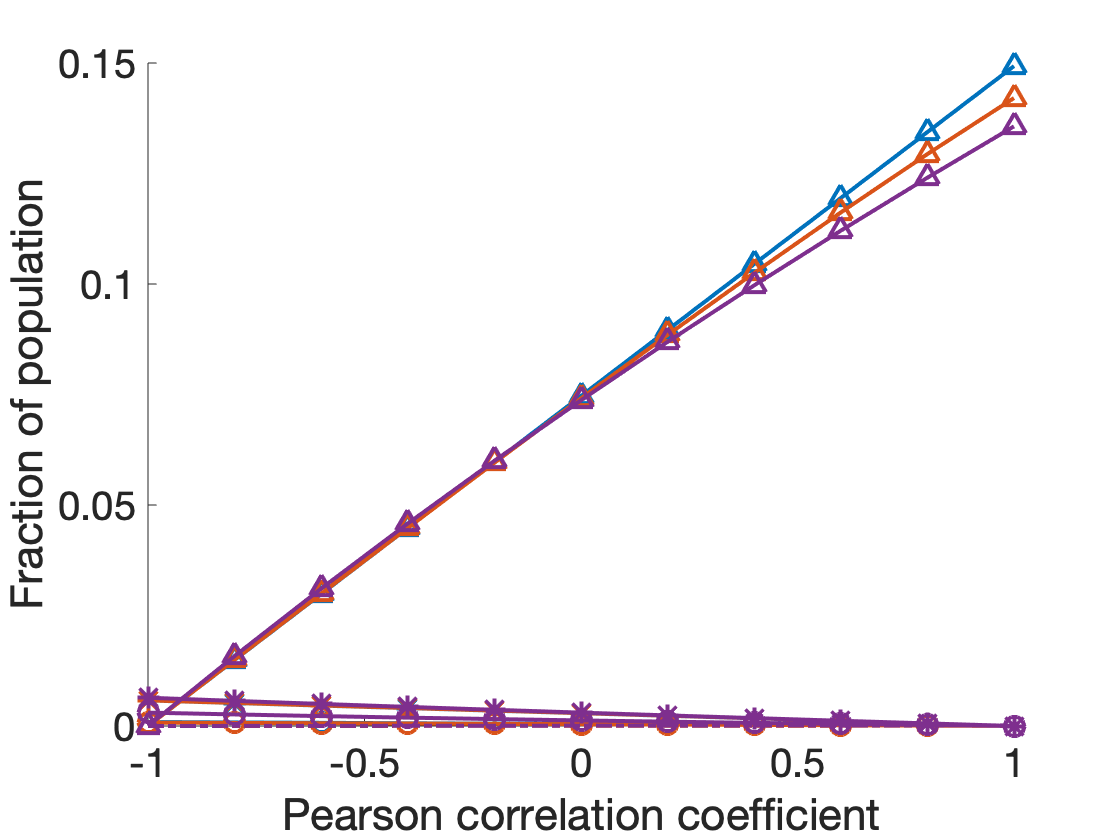} 
      \caption{Decomposition of the $\cal{A}$ subpopulation when the intra-layer degree--degree correlation is $1$}\label{fig: inter_degree_d}
     \end{subfigure}
    \end{minipage}
    \hspace{-1.3cm}
    \begin{minipage}{0.12\textwidth}
     \vspace{-0.6cm}
     \includegraphics[scale=0.15, trim=1900 160 400 120,clip]{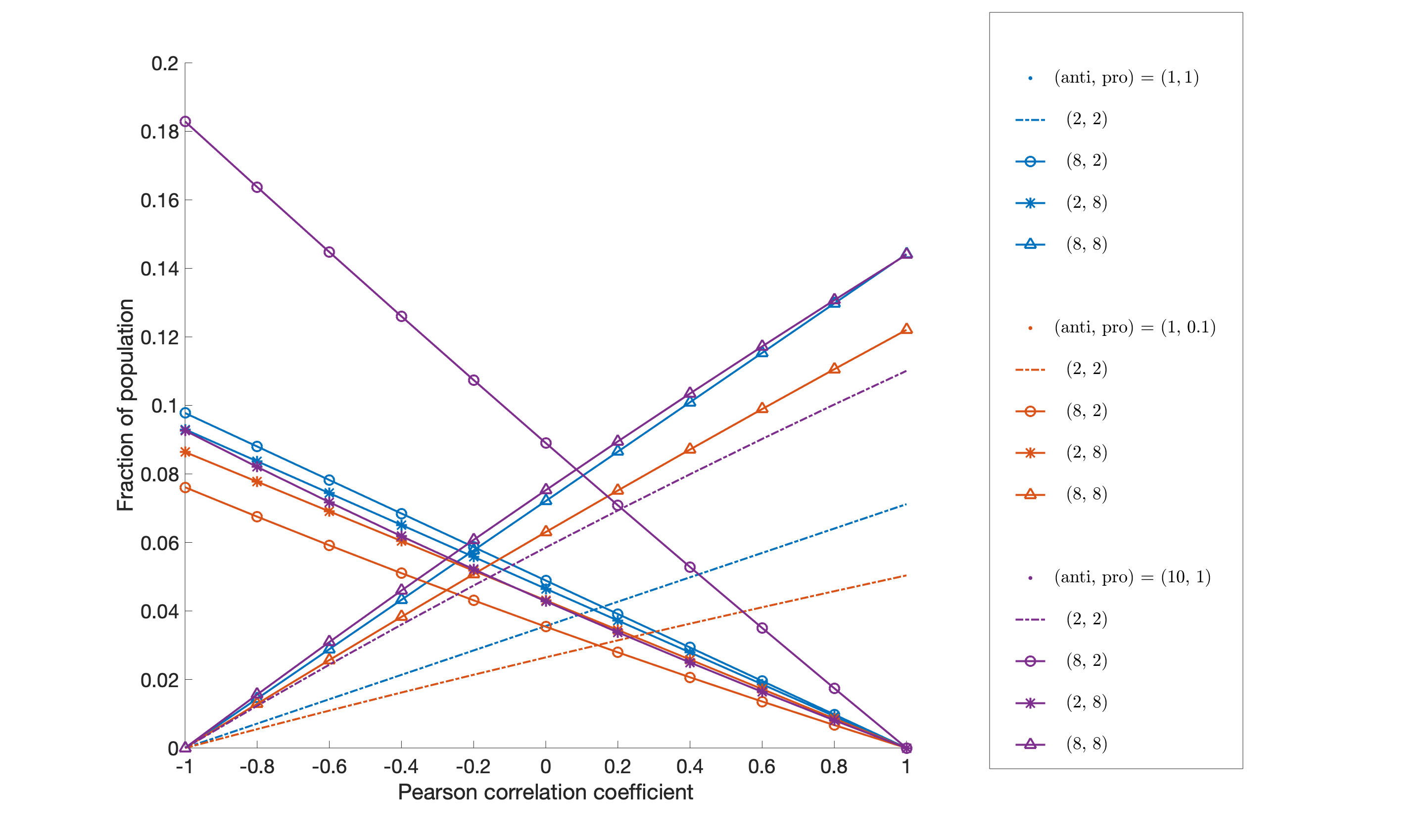} 
    \end{minipage}
    \caption{Decomposition based on degrees of the nodes that eventually become infectious and recover. We decompose (a, b) the $\cal{U}$ subpopulation and (c, d) the $\cal{A}$ subpopulation by degree. The vertical axes indicate the fraction of the population in each subpopulation. We show the changes in subpopulation size versus the inter-layer degree--degree correlation from our PA. The other parameters are $\bphy =\binfo = 0.6$, $\gphy=1$, and $\ginfo=0.1$.}
    \label{fig: inter_degree}
\end{figure}

We first examine the case in which the intra-layer degree--degree correlation is $1$ and node opinions do not affect the spread of the disease. A positive inter-layer degree--degree correlation encourages the coupling of large-degree nodes in the two layers; these nodes have a larger probability than small-degree nodes of becoming infectious or forming an opinion. Therefore, as we increase the inter-layer degree--degree correlation, fewer nodes are uninformed when they catch the disease. Figure \ref{fig: inter_degree_b} shows the decomposition of the $\cal{U}$ subpopulation based on the degrees of its nodes. Because the degree-$2$ nodes are adjacent only to other degree-$2$ nodes in each layer, the degree-$2$ nodes are rarely infectious or form an opinion. Therefore, the $\cal{U}$ subpopulation consists primarily of nodes of types $(2,\,8)$ or $(8,\,8)$. As we increase the inter-layer degree--degree correlation, there are more $(8,\,8)$-type nodes and fewer $(2,\,8)$-type nodes in the network. Because $U_8$-nodes (i.e., nodes that are uninformed and have degree $8$ in the information layer) have a larger probability of forming an opinion than $U_2$-nodes, a smaller number of people become infectious while still uninformed. 

If we perturb the influence coefficients $\alphaa$ and $\alphap$ from $1$, the opinions on the information layer directly affect the $\cal{A}$ and $\cal{P}$ subpopulations, respectively, through modified infection risks. These modified risks then influence the speed of disease spread and affect the other subpopulations. Therefore, the information layer has a larger affect on the disease dynamics when more people become infected while holding some opinion. Based on our discussion above, we expect that increasing the inter-layer degree--degree correlation amplifies the influence of opinion spread.

We now examine the case in which either the pro-physical-distancing opinion or the anti-physical-distancing opinion has a nontrivial influence on disease dynamics. 
For simplicity, we suppose that only one opinion is effective. When $\alphap < 1$, the spread of the pro-physical-distancing opinion protects people who adopt that opinion because it suppresses the spread of the disease. In Figure \ref{fig: inter_awareness_b}, we see that the $\cal{U}$ subpopulation tends to decrease faster 
and that the $\cal{P}$ and $\cal{A}$ subpopulations tend to increase slower in this situation than when the disease spreads independently of opinions. Therefore, the final epidemic size decreases as we increase the inter-layer degree--degree correlation. When $\alphaa >1$, the anti-physical-distancing opinion accelerates the spread of the disease. Therefore, more people become infected before their opinions change; this, in turn, leads to a larger $\cal{U}$ subpopulation and smaller $\cal{A}$, $\cal{P}$, and $\cal{R}$ subpopulations. Overall, the growing gap between the dashed purple curve and the dashed blue curves (i.e., the $\cal{U}$ subpopulation for different parameter values) in Figure \ref{fig: inter_awareness_b} illustrates the increase in the epidemic size as we increase the inter-layer degree--degree correlation.

The situation is more intricate when the intra-layer degree--degree correlation is $-0.25$. Because the intra-layer edges now connect degree-$2$ nodes to degree-$8$ nodes, the former are more likely to become infectious or adopt an opinion than when the intra-layer degree--degree correlation is $1$. In Figure \ref{fig: inter_degree_a}, we see that when $\alphaa=\alphap=1$, nodes of types $(8,\,2)$ and $(2,\,2)$ constitute a larger proportion of the $\cal{U}$ subpopulation than in Figure \ref{fig: inter_degree_b}. As we increase the inter-layer degree--degree correlation, the $\cal{U}$ subpopulation has progressively more nodes with degree $2$ in the physical layer because there are gradually fewer $(8,\,2)$-type nodes and gradually more $(2,\,2)$-type nodes and it is more difficult for the $(2,\,2)$-type nodes to form an opinion. Consequently, the decreasing trend in the $\cal{U}$ subpopulation (see the dashed blue curve) in Figure \ref{fig: inter_awareness_a} is less drastic than in Figure \ref{fig: inter_awareness_b}. 

When $\alphaa >1$, the $\cal{A}$ subpopulation in Figure \ref{fig: inter_awareness_a} (see the purple curve with asterisk markers) has qualitatively different dynamics than in Figure \ref{fig: inter_awareness_b}. This is due to the influence of opinions on degree-$2$ nodes in the physical layer. Nodes that adopt the anti-physical-distancing opinion now have a larger infection risk than when $\alphaa=1$, so we expect more nodes to become infected while holding the anti-physical-distancing opinion. In Figure \ref{fig: inter_degree_c}, we see that the anti-physical-distancing opinion leads to an increase in the numbers of nodes of types $(8,\,2)$ and $(2,\,2)$ in the $\cal{A}$ subpopulation as we increase $\alphaa$ from $1$ to $10$. This, in turn, leads to the growth of the $\cal{A}$ subpopulation. Moreover, as we increase $\alphaa$ from $1$ to $10$, the increase in the number of $(8,\,2)$-type nodes when the inter-layer degree--degree correlation is $-1$ is larger than the increase in the number of $(2,\,2)$-type nodes when the inter-layer degree--degree correlation is $1$. This phenomenon arises because degree-$8$ nodes are more likely than degree-$2$ nodes to adopt an opinion. This, in turn, leads to a decrease of the $\cal{A}$ subpopulation and ultimately to a decrease of the total epidemic size as we increase the inter-layer degree--degree correlation. The anti-physical-distancing opinion does not lead to a clear increase in the number of the degree-$8$ nodes in the $\cal{A}$ subpopulation. One plausible explanation is that the degree-$8$ nodes are already very likely to become infected at the baseline transmission rate for the disease. Finally, the anti-physical-distancing opinion does not trigger an increase in the number of nodes of types $(8,\,2)$ or $(2,\,2)$ in Figure \ref{fig: inter_degree_d}. We conjecture that this is because the positive intra-layer degree--degree correlation imposes sufficiently strong constraints so that it is difficult for nodes with degree $2$ in the physical layer to become infected even with the higher risk of infection. 

%
%
\section{Temporary immunity to opinions}\label{sec: SIRS-SIR}

\begin{figure}[!h]
    \centering
    \includegraphics[scale=0.45, trim=0 0 0 0, clip]{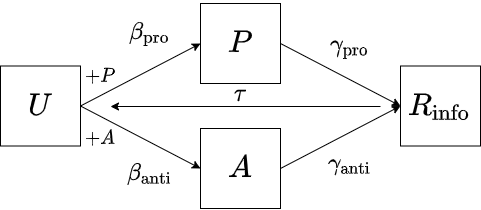}
    \caption{Schematic illustration of the dynamics on the information layer with temporary immunity to opinions. The transitions are the same as in Figure \ref{fig: A}, except that nodes in state $R_{\text{info}}$ transition to state $U$ at rate $\tau$.}
    \label{fig:SIRS}
\end{figure}

In previous sections, we assume that people who adopt a pro- or anti-physical-distancing opinion develop immunity to both opinions after they recover. More generally, individuals' ideas can change back and forth\cite{glaubitz2020oscillatory}. In this section, we extend the opinion dynamics to an SIRS process (see Figure \ref{fig:SIRS}). With conversion rate $\tau$, people in the $R_{\text{info}}$ compartment return to the $U$ compartment and again become susceptible to the pro- and anti-physical-distancing opinions. When $\tau = 0$, this refined model reduces to the model in Section \ref{sec: model} (See Figure \ref{fig: A}).

\begin{figure}[!h]
    \centering
    \begin{subfigure}[t]{0.32\textwidth}
    \includegraphics[scale=0.12]{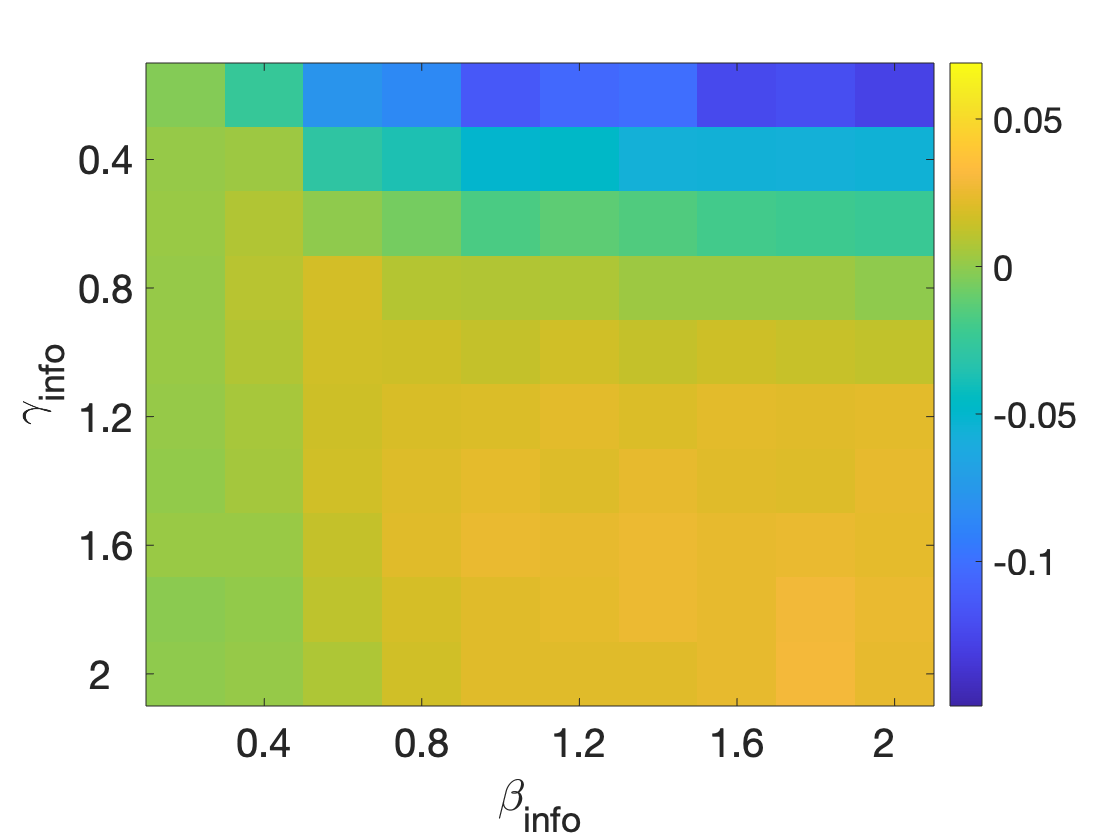}
    \caption{$\tau=0$}
    \end{subfigure}
    \begin{subfigure}[t]{0.32\textwidth}
    \includegraphics[scale=0.12]{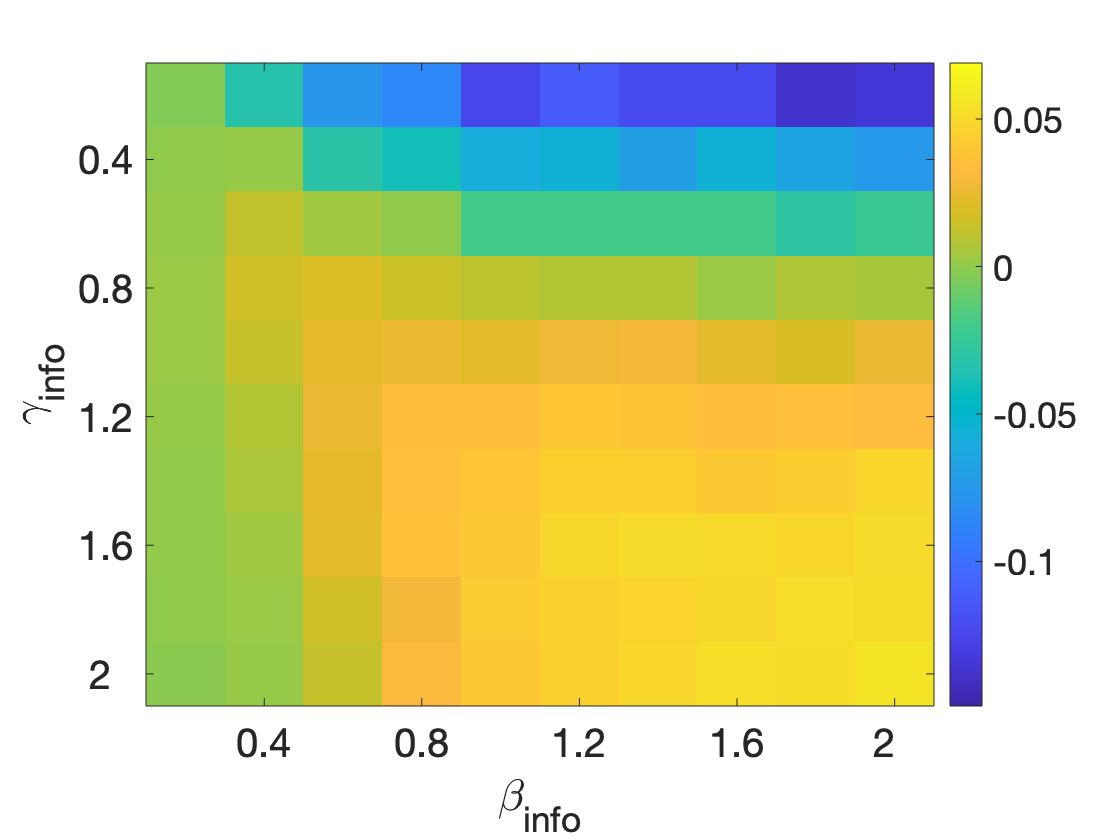}
    \caption{$\tau=1$}
     \end{subfigure}
    \begin{subfigure}[t]{0.32\textwidth}
    \includegraphics[scale=0.12]{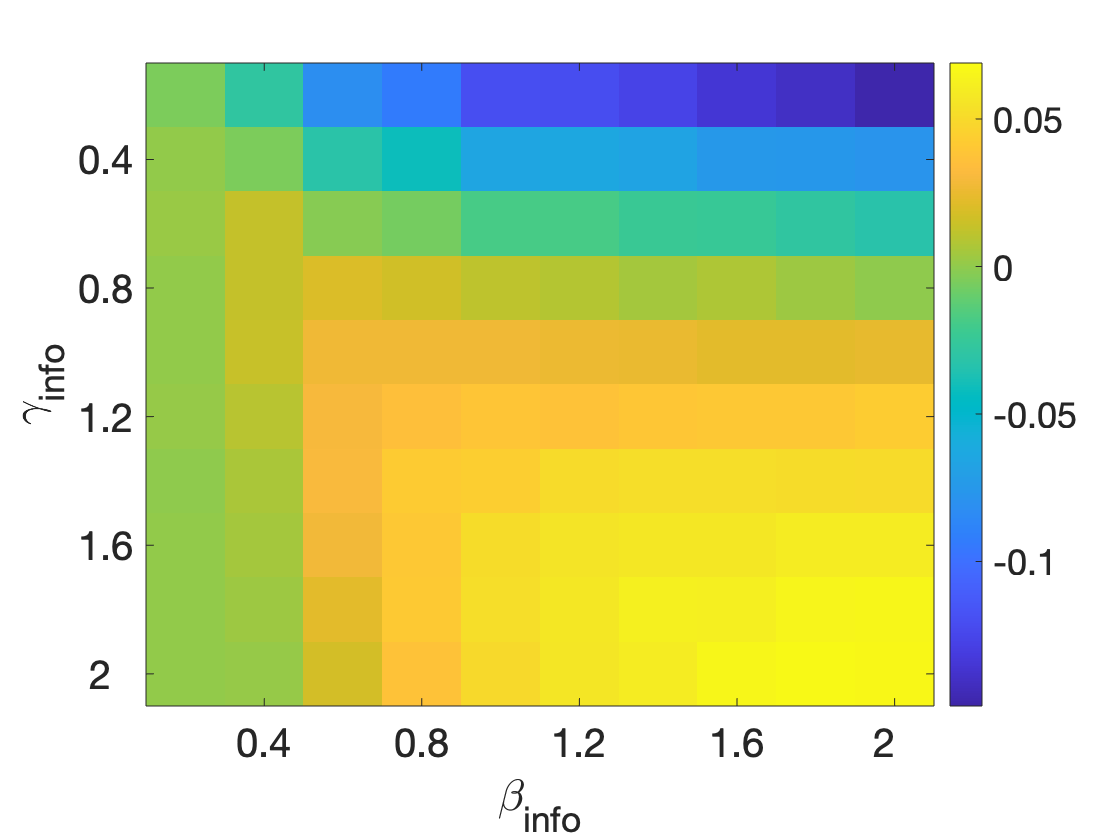}
    \caption{$\tau=2$}
     \end{subfigure}
    \caption{The final epidemic size minus the basic size as we vary $\ginfo$, $\binfo$, and $\tau$. We construct both layers with configuration-model networks with Poisson intra-layer degree distributions with a mean degree of $5$. The other parameters are $\bphy = 0.6$, $\gphy=1$, $\alphaa=10$, and $\alphap=0.1$. Each panel is a mean over 600 simulations.}
    \label{fig: SIRS_matrix}
\end{figure}

Figure \ref{fig: SIRS_matrix} shows the final epidemic size minus the basic size as we vary the contagion parameters $\tau$, $\ginfo$, and $\binfo$. For a fixed value of $\tau$, we obtain a heat map similar to that in Figure \ref{fig: final_size_matrix}. As we increase $\tau$, the simulations suggest that the overall influence from the information layer increases. For fixed values of $\ginfo$ and $\binfo$, the colors in the heat map become brighter (respectively, darker) from left to right because the spread of opinions leads to a decrease (respectively, increase) in the final epidemic size in comparison to the basic size as we increase $\tau$.

\begin{figure}[t!h]
    \centering
    \begin{subfigure}[t]{0.5\textwidth}
        \centering
        \includegraphics[scale=0.16]{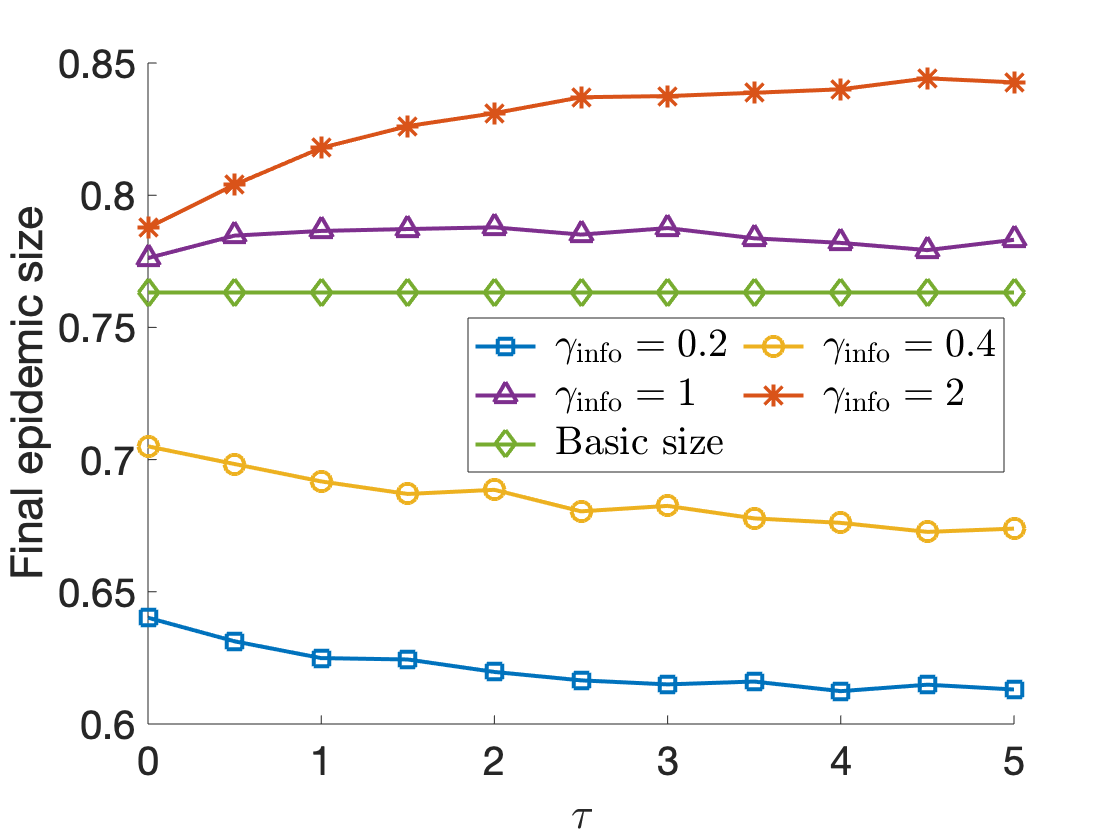}
        \caption{Final epidemic size}\label{fig: SIRS_final_size_tau}
    \end{subfigure}
     ~ 
    \begin{subfigure}[t]{0.45\textwidth}
        \centering
        \includegraphics[scale=0.16]{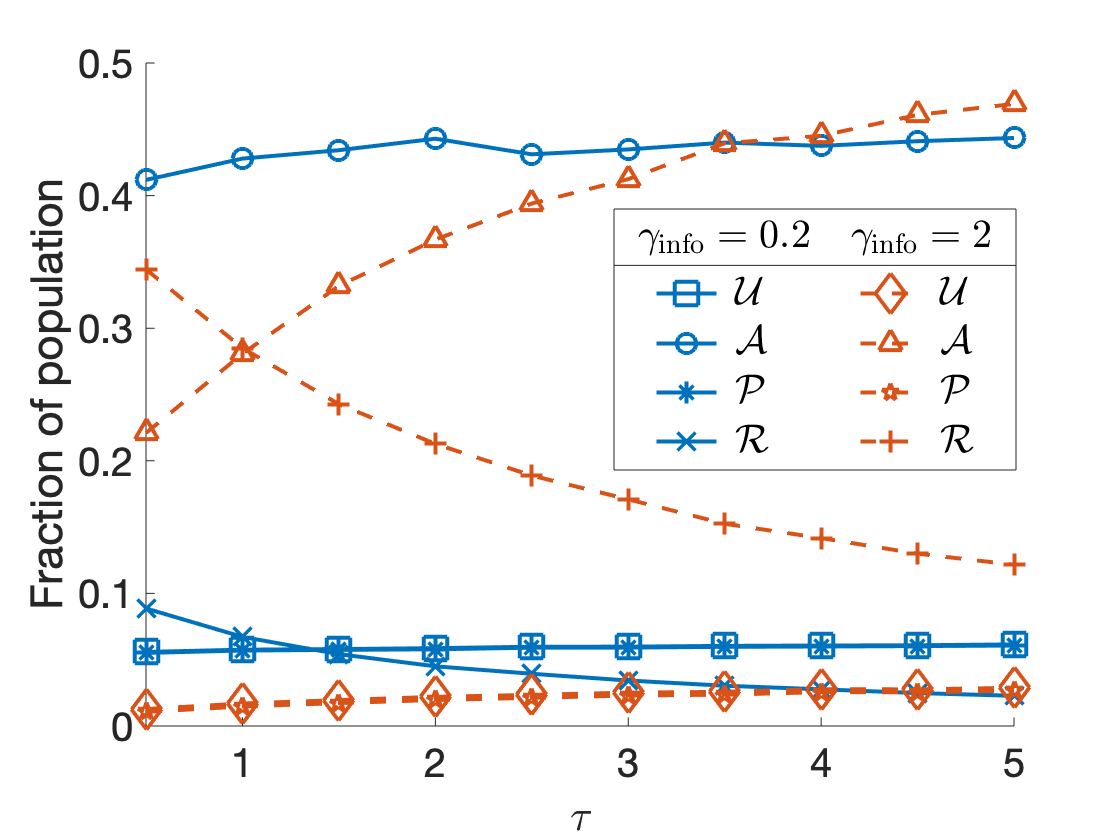}
        \caption{Decomposition of opinion states}\label{fig: SIRS_awareness}
    \end{subfigure}
    \begin{subfigure}[t]{0.5\textwidth}
        \centering
        \includegraphics[scale=0.16]{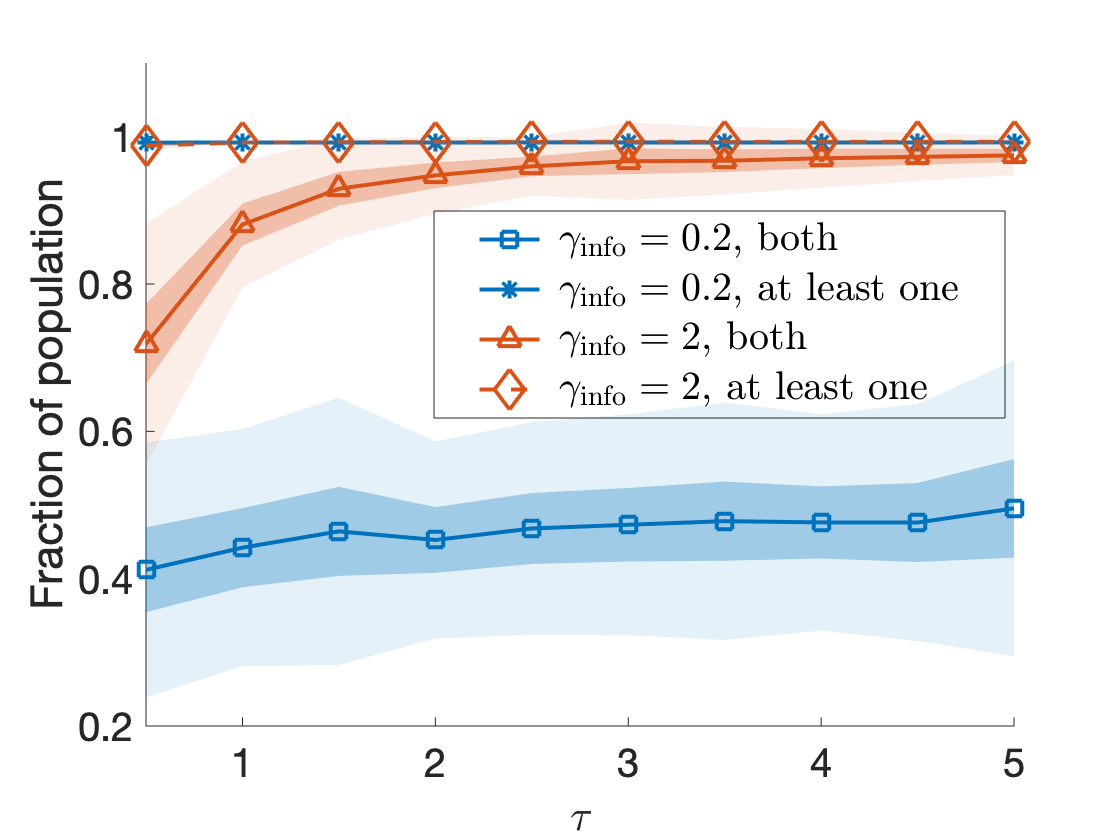}
        \caption{Fractions of the population that adopt at least one or both opinions}\label{fig: SIRS_Jaccard}
    \end{subfigure}
    ~
    \begin{subfigure}[t]{0.45\textwidth}
        \centering
        \includegraphics[scale=0.16]{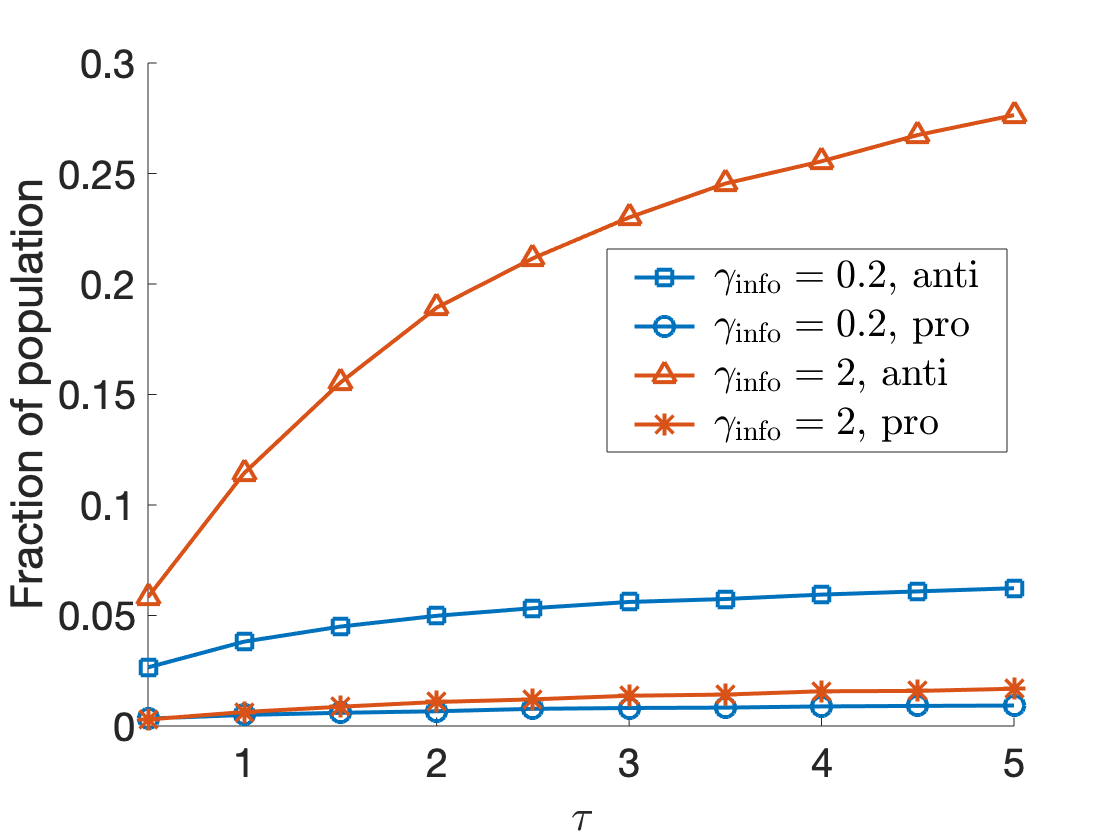} 
        \caption{Fractions of the population infected during their second or subsequent time that they adopt an opinion (The second and subsequent opinions need not be the same as prior opinions.)}\label{fig: SIRS_second_infection}
    \end{subfigure}
    \caption{Influence of the opinion recovery rate $\ginfo$ and the conversion rate $\tau$ on the final epidemic size and the opinion distribution. (a) The influence of $\tau$ on the final epidemic size. We decompose the recovered population at steady states based on their opinion states when they become infected and plot the size of these subpopulations in (b). In (c, d), we show additional statistics of the opinion distribution, which we define in the subtitles. We construct each layer using a configuration model with a Poisson intra-layer degree distribution with a mean degree of $5$. The curves show means of $200$ simulations with $\bphy =0.6$, $\gphy=1$, $\binfo= 2$, $\alphap=0.1$, and $\alphaa=10$.}\label{fig: Final_size_selected}
\end{figure}

Consider the case $\binfo=2$, where the effect of $\tau$ is particularly evident. Figure \ref{fig: SIRS_final_size_tau} shows the influence of $\tau$ on the final epidemic size for a few values of $\ginfo$. It is hard to see the trend when $\ginfo=1$ because of the stochasticity of the simulations, but the behavior of the other three curves is consistent with that of Figure \ref{fig: SIRS_matrix}. As we increase $\tau$, the expected duration that individuals stay in the $R_{\text{info}}$ state decreases. Consequently, as we increase $\tau$ in Figure \ref{fig: SIRS_awareness}, the size of the $\cal{R}$ subpopulation decreases and sizes of the other subpopulations ($\cal{U}$, $\cal{A}$, and $\cal{P}$) increase. 

We conjecture that the overall influence (i.e., whether increasing $\tau$ leads to more or fewer disease infections) depends on whether people tend to adopt an opinion that departs from their earlier opinion(s). If individuals tend to adopt different opinions over time, increasing $\tau$ makes the model with SIRS opinion dynamics exhibit behavior like what we observed from the random-recoupling assumption (see Section \ref{sec: ODE}) and leads to more people adopting the anti-physical-distancing opinion at an earlier time. Susceptible individuals are likely to become infected when they adopt the anti-physical-distancing opinion, regardless of whether they have previously adopted the pro-physical-distancing opinion. Consequently, increasing $\tau$ leads to more disease infections. However, if people tend to adopt the same opinion over time, susceptible individuals who adopt the pro-physical-distancing opinion also are more likely to avoid future infections. In this case, enforcing a faster reversion to the $U$ state has a similar effect to increasing $\ginfo$, which (as we showed in Section \ref{sec: exp_contagion_params}) may help suppress disease spreading. Figure \ref{fig: SIRS_Jaccard} shows the fractions of individuals who adopt at least one opinion and both opinions within the time frame of our experiments for $\ginfo=0.2$ and $\ginfo=2$. Many fewer people adopt both opinions when $\ginfo=0.2$ than when $\ginfo=2$. Additionally, as we increase $\tau$, there is only a slightly increasing trend in the fraction of people who adopt both opinions for $\ginfo=0.2$, in contrast to the rapid growth of the fraction for $\ginfo=2$. Intuitively, a node that adopts one opinion can influence more neighbors when $\ginfo$ is smaller. Therefore, it is more likely to adopt the same opinion later. Suppose an individual holds an opinion when it becomes infected on the physical layer, and suppose that this is not the first opinion that it has held (i.e., it previously returned to the $U$ state in the information layer). In Figure \ref{fig: SIRS_second_infection}, we examine the counts of these individuals for both opinions as a function of $\tau$. Consistent with our conjecture, as we increase $\tau$, we observe a larger increase of the fraction of the population that holds the anti-physical-distancing opinion when $\ginfo = 2$ than when $\ginfo = 0.2$. 

%
%
\section{Conclusions and discussion}\label{sec: conclusion}

We studied the influence of the spread of competing opinions on the spread of a disease. We assumed that pro- and anti-physical-distancing opinions circulate within a population and affect the spread of the disease. We developed a degree-based pair approximation for the time evolution of the expected number of individuals in different compartments and is applicable to dynamics on heterogeneous networks with specified inter-layer and intra-layer degree--degree correlations. We examined different approximation schemes for the effective transmission rate of susceptible individuals in the physical layer. We found that the distribution of opinions in nodes in a given disease state is correlated with both its own disease state and the disease states of their neighbors in the physical layer.

Through extensive numerical simulations, we showed that the opinion contagions in our model can either increase or decrease the disease transmission speed, the peak infection counts, and the number of people who become infected. We demonstrated that the overall impact of the opinion dynamics on the disease prevalence depends not only on their influence coefficients, but also on the network architecture and how the opinions couple to the spread of the disease.

We found that lengthening the duration time (through decreasing the opinion recovery rate) over which people adopt opinions --- whether in favor of or against physical distancing --- may help suppress disease transmission. We also saw that physically distancing for too short a time period may still place people at high infection risk; this is well-known for epidemic models in a fully mixed population.\cite{bootsma2007effect,bertozzi2020challenges} We observed that the benefit of a long opinion-adoption period is reinforced when we let the spread of opinions follow SIRS dynamics instead of SIR dynamics. Allowing people to become susceptible to opinions after having a previous opinion helps create neighborhoods in a network's information layer in which adjacent nodes tend to adopt the same opinion over time. Consequently, people who adopt the pro-physical-distancing opinion are more likely to adopt it again later. Although the same phenomenon applies to the spread of the anti-physical-distancing opinion, the difference in the influence of the two opinions on disease transmission rates leads to the asymmetry of their influence on disease prevalence.
  
Our work examines both beneficial and harmful effects of the spread of opinions on other dynamical processes (such as the spread of a disease). There are many ways to build on our research. Although the two opinions can have different contagion parameters, we only showed results in which these parameters are identical. One can study this model when the two competing opinions spread asymmetrically.
We also assumed a unidirectional influence from the information layer to the physical layer, but disease states can also influence opinion states\cite{funk2010modelling, wang2015coupled}, and one can incorporate such coupling. Time-dependent network structures in which node states coevolve with network structures \cite{sayama2013modeling,porter2016} are relevant for behavioral changes when individuals adopt opinions about physical distancing. Such time-dependent networks model changes in contact patterns due to lockdowns and stay-at-home orders. Additional opinions (e.g., opinions on vaccines) can be considered in conjunction with more complex disease dynamics due to vaccines \cite{stella2021} and variants \cite{bellomo2021multiscale}.

%
%
\section*{Acknowledgements}
This research is supported by NSF's Rapid Response Research (RAPID) grant DMS-2027438 and by NSF grants DMS-1922952 and DMS-1737770 through the Algorithms for Threat Detection (ATD) program. It is also supported by Simons Foundation Math + X Investigator Award number 510776.




\end{document}